\documentclass[a4paper,11pt]{article}
\pdfoutput=1 

\usepackage{jcappub} 

\usepackage[T1]{fontenc} 

\title{Exploring Cosmic Origins with CORE: The Instrument}

\author[1,2,a] {P. de Bernardis,\note[$a$]{Corresponding author}}
\author[3] {P.~A.~R. Ade,}
\author[4] {J.J.A. Baselmans,}
\author[1,2] {E.~S. Battistelli,}
\author[5] {A. Benoit,}
\author[6,7] {M. Bersanelli,}
\author[5] {A. Bideaud,}
\author[5] {M. Calvo,}
\author[8] {F.~J. Casas,}
\author[9] {G. Castellano,}
% ...

\affiliation[1]{ Dipartimento di Fisica, Universit\'a di Roma  La Sapienza , P.le A. Moro 2, 00185 Roma, Italy }
\affiliation[2]{ INFN, Sezione di Roma, P.le A. Moro 2, 00185 Roma, Italy}
\affiliation[3]{ School of Physics and Astronomy, Cardiff University, The Parade, Cardiff CF24 3AA, UK }
\affiliation[4]{ SRON - Netherlands Institute for Space Research, Sorbonnelaan 2, 3584 CA - Utrecht, The Netherlands }
\affiliation[5]{ Institut N\'eel, CNRS and Universit\'e Grenoble Alpes, F-38042 Grenoble, France }
\affiliation[6]{ Dipartimento di Fisica, Universit\'a  degli Studi di Milano, Via Celoria 16, I-20133 Milano, Italy }
\affiliation[7]{ INAF IASF, Via Bassini 15, I-20133 Milano, Italy }
\affiliation[8]{ Instituto de Fisica de Cantabria (CSIC-UC), Avda. los Castros s/n, 39005 Santander, Spain }
\affiliation[9]{ Istituto di Fotonica e Nanotecnologie - CNR, Via Cineto Romano 42, I-00156 Roma, Italy }
% ...

\emailAdd{paolo.debernardis@roma1.infn.it}

\abstract{We describe a space-borne, multi-band, multi-beam polarimeter aiming at a precise and accurate measurement of the polarization of the Cosmic Microwave Background. The instrument is optimized to be compatible with the strict budget requirements of a medium-size space mission within the Cosmic Vision Programme of the European Space Agency. The instrument has no moving parts, and uses arrays of diffraction-limited Kinetic Inductance Detectors to cover the frequency range from \SI{60}{GHz} to \SI{600}{GHz} in 19 wide bands, in the focal plane of a \SI{1.2}{m} aperture telescope cooled at \SI{40}{K}, allowing for an accurate extraction of the CMB signal from polarized foreground emission. The projected CMB polarization survey sensitivity of this instrument, after foregrounds removal, is \SI{1.7}{\micro\kelvin}$\cdot$arcmin. The design is robust enough to allow, if needed, a downscoped version of the instrument covering the \SI{100}{GHz} to \SI{600}{GHz} range with a \SI{0.8}{m} aperture telescope cooled at \SI{85}{K}, with a projected CMB polarization survey sensitivity of \SI{3.2}{\micro\kelvin}$\cdot$arcmin.}

\usepackage{float}
\restylefloat{table}
\usepackage{multirow}
\usepackage{graphicx}
\usepackage[table,xcdraw]{xcolor}
\usepackage{rotating}
\usepackage{adjustbox}
\usepackage{subfig}
\def\lsim{\raise0.3ex\hbox{$<$\kern-0.75em\raise-1.1ex\hbox{$\sim$}}}
\usepackage[load-configurations=abbreviations]{siunitx}
\DeclareSIUnit\mK{\milli\kelvin}
\DeclareSIUnit\mbar{\milli\bar}
\DeclareSIUnit\micron{\micro\meter}

\usepackage{wasysym}
\newcommand{\coremfive}{\textit{\negthinspace CORE}}
\newcommand{\MiniCORE}{\textit{\negthinspace MiniCORE}}

 \newcommand{\planck}{\textit{\negthinspace Planck\/}}
\newcommand{\Herschel}{\textit{\negthinspace Herschel\/}}

\newcommand*\He[1]{$^{#1}$He}
\newcommand*\PQ[2]{\ensuremath{#1_{\text{#2}}}}

\newcommand*\dQ[1]{\ensuremath{\dot{Q}_{\text{#1}}}}
\newcommand*\dn[1]{\ensuremath{\dot{n}_{\text{#1}}}}

\newcommand*{\dg}{\nobreak\ensuremath{^{\circ}}}

\begin{document}
\maketitle
\flushbottom





\newpage
\section{Introduction}
\label{sec:intro}

The measurement of the polarization state of the CMB promises to shed light on the earliest phases of the evolution of the Universe, testing the cosmic inflation scenario, but poses difficult experimental challenges, requiring a polarimeter with very high sensitivity, exquisite control of systematic effects, and the ability to extract the tiny inflation-related signal from overwhelming polarized foregrounds. 

In this paper we describe a space-based instrument, optimized to represent the \emph{ultimate} experiment for probing cosmic inflation measuring CMB polarization (an objective which cannot be achieved by ground-based experiments alone, nor by simpler space-missions), while fulfilling the stringent requirements of a medium-class mission of the European Space Agency. We underline that also the lack of detection of primordial B-modes by CORE would have an impact on the current cosmological paradigm, since a large part of inflation scenarios would be ruled out. 

\noindent This paper is part of a series describing the \coremfive \ (Cosmic ORigins Explorer) mission and its scientific context. Here we aim at an instrument baseline description with a level of detail adequate for starting a Phase-A study.

The \coremfive \ instrument inherits the legacy of Planck, as well as our previous proposals COrE and COrE+. 

The goal of \coremfive \ is to detect unambiguously a tensor to scalar ratio $r$ as small as 1$\times$10$^{-3}$ (3$\sigma$ CL), even in the presence of complex polarized foregrounds. In fact, the Starobinsky model, R$^2$ (Higgs) inflation, foresees $r > 2\times10^{-3}$. With this level of sensitivity, a null result would basically disfavour most of the large-field inflation models allowed by Planck (see the companion paper on inflation \cite{2016arXiv161208270C}).

This ultimate measurement is possible only from space and requires a mission that will map the CMB B-modes polarization over most of the sky, with an angular resolution of a few arcminutes and with a polarization survey sensitivity better than \SI{2.5}{\micro\kelvin} per square arcminute pixel, after foreground removal (20 times better than the aggregated CMB polarization sensitivity of the entire set of \planck\ polarized detectors). This mission will collect virtually all the information about the Early Universe encoded in the CMB polarization.

The \coremfive \ instrument builds on the success of \planck\ and \Herschel, re-using many of the subsystems and methods developed by the mm/submm community. The sensitivity requirements described above, combined with the requirement of internal control of polarized foregrounds, drives the dimension of the focal plane of the instrument. The proposed \coremfive \ baseline instrument is based on an array of 2100 cryogenically cooled, linearly polarized detectors at the focus of a 1.2 meter aperture telescope. The entire sky will be surveyed with 19 frequency bands, spanning the range 60 to 600 GHz. The spacecraft will be located in a large Lissajous orbit around the Sun-Earth L2 
Lagrange point to keep the Sun, Earth and Moon well away from the line of sight at all times, and thus avoid that they pollute the scientific signal with unwanted far sidelobe contamination. The combination of three rotations of the spacecraft at different timescales provides an observation pattern such that each sky pixel is crossed frequently, and  from many different directions. This scan strategy provides for a strong mitigation of systematic effects and would thus ensure optimal use of the high inherent sensitivity, especially for extracting the large angular scale signals. 

With this implementation, \coremfive \ also addresses a broad range of other questions of prime scientific importance that cannot be answered by any other means than very accurate observation of the CMB polarization on all angular scales and over the near full sky. It will probe the distribution of clustered mass in the Universe through the observation of the lensing of CMB polarization due to dark matter structures between our telescopes and the last scattering surface.
The reconstruction of the CMB lensing potential will provide high signal-to-noise-ratio maps of the distribution of dark matter at redshifts $z=1-3$ without recourse to biased baryonic tracers. 
In addition to providing a map of the dark matter integrated along the line of sight up to high redshift, this measurement, combined with cosmological constraints from Euclid, will constrain the sum of the 3 light neutrino masses with a statistical error of 3 meV, 5 times better than any single cosmological probe alone and sufficient to distinguish unambiguously between a normal neutrino hierarchy ($m_1, m_2 \ll m_3$) with a mass sum of approximately at least 60 meV, and an inverted hierarchy ($m_3 \ll m_1, m_2$) with a minimal mass sum of about 100 meV (see the companion paper on cosmological parameters \cite{2016arXiv161200021D}).

\coremfive \ will also probe the distribution of hot gas up to redshifts $z=2-3$ by measuring the thermal Sunyaev-Zel'dovich effect, the inverse Compton scattering of CMB photons by energetic electrons. It will detect $\sim 50,000$ galaxy clusters extending to high redshift, more than 300000 clusters in combination with high-resolution ground-based surveys, and part of the hot baryons in the cosmic web. Combined with high resolution (2-3$^\prime$) ground-based CMB data in atmospheric windows between 90 and \SI{250}{GHz} \coremfive \ will also detect the individual peculiar motions of $\sim 30,000$ galaxy clusters, thus directly measuring the cosmic velocity field at large redshift, a measurement that cannot be performed by any other means (see the companion paper on clusters science \cite{Meli17}). 

At frequencies above 350 GHz, where sky signals are dominated by emission from thermal dust and point sources, \coremfive \ will for the first time provide full sky, high quality polarization maps. These maps will provide astrophysicists with the most 
detailed view yet at the Galactic magnetic field, unveiling its role in creating the  filamentary, web-like, structures where stars form. Magneto-hydrodynamical turbulence will be revealed with unprecedented statistical information characterizing the energy injection and energy transfer down to dissipation scales. The spectral dependence of the polarized signal from dust will be measured with high accuracy across the sky, furthering our understanding of the nature of interstellar dust. 
Moreover, together with the corresponding high sensitivity intensity maps, these observations will discover and characterize a large number of new galactic and extragalactic point sources and also measure their polarization properties.

Achieving the \coremfive \ cosmological science programme will require accurate separation of the many astrophysical foregrounds as well as exquisite control and assessment of systematic errors. The \coremfive \ instrument configuration and calibration procedure are designed to generate all the data needed for this assessment. In particular, the \coremfive \ array will include a large number of closely packed spectral bands for optimal monitornig of polarized foreground components. Simulations based on the full presently available information, as summarized in the Planck Sky Model \cite{PSM2013}, and analyzed using state-of-the-art component separation algorithms show that \coremfive \ will achieve its science objectives and that the design includes redundancy and margin for error. This is described in detail in the ``mission'' paper of this series \cite{missionpaper}.

The \coremfive \ ultra-high sensitivity maps of the three Stokes parameters I, Q, and U in 19 frequency bands will serve as a long standing legacy and a reference dataset for the microwave and submillimeter emission in both intensity and polarization over the full sky. Astrophysicists will mine these maps for decades. In addition to the compelling science deliverables we know about today, even more exciting are all the discoveries buried in these maps, and that we can not yet imagine, nor describe. 

{\small 
\begin{table}[ht]
\begin{center}
{\footnotesize
\begin{tabular}{|c|c|c|c|c|c|c|c|c|c|c|c|}
\hline 
channel & beam  &  $N_{\rm det}$  &  $\Delta T$  &  $\Delta P$  & $\Delta I$& $\Delta I$  & $\Delta y\times 10^6$  &  PS  ($5\sigma$) \\
GHz     &	arcmin &    &  \SI{}{\micro\kelvin}$\cdot$arcmin &  \SI{}{\micro\kelvin}$\cdot$arcmin &  \SI{}{\micro\kelvin^{}_{RJ}}$\cdot$arcmin  & \SI{}{kJy/sr}$\cdot$arcmin &  $y^{}_{\rm SZ}\cdot$arcmin & mJy  \\
\hline 
\hline 

60      & 17.87   & 48      & 7.5          & 10.6         & 6.81          & 0.75         & -1.5         & 5.0                     \\ 
70      & 15.39   & 48      & 7.1          & 10           & 6.23          & 0.94         & -1.5         & 5.4              \\ 
80          & 13.52   & 48      & 6.8          & 9.6          & 5.76          & 1.13         & -1.5         & 5.7           \\ 
90         & 12.08   & 78      & 5.1          & 7.3          & 4.19          & 1.04         & -1.2         & 4.7            \\ 
100       & 10.92   & 78      & 5.0            & 7.1          & 3.90          & 1.2          & -1.2         & 4.9          \\ 
115       & 9.56    & 76      & 5.0            & 7.0            & 3.58          & 1.45         & -1.3         & 5.2         \\ 
130      & 8.51    & 124     & 3.9          & 5.5          & 2.55          & 1.32         & -1.2         & 4.2         \\ 
145      & 7.68    & 144     & 3.6          & 5.1          & 2.16          & 1.39         & -1.3         & 4.0             \\ 
160      & 7.01    & 144     & 3.7          & 5.2          & 1.98            & 1.55         & -1.6         & 4.1            \\ 
175        & 6.45    & 160     & 3.6          & 5.1          & 1.72          & 1.62         & -2.1         & 3.9         \\ 
195     & 5.84    & 192     & 3.5          & 4.9          & 1.41          & 1.65         & -3.8         & 3.6           \\ 
220      & 5.23    & 192     & 3.8          & 5.4          & 1.24          & 1.85         & -           & 3.6         \\ 
255      & 4.57    & 128     & 5.6          & 7.9          & 1.30          & 2.59         & 3.5          & 4.4         \\ 
295       & 3.99    & 128     & 7.4          & 10.5         & 1.12          & 3.01         & 2.2          & 4.5           \\ 
340      & 3.49    & 128     & 11.1         & 15.7         & 1.01            & 3.57         & 2.0            & 4.7             \\ 
390       & 3.06    & 96      & 22.0           & 31.1         & 1.08          & 5.05         & 2.8          & 5.8          \\ 
450      & 2.65    & 96      & 45.9         & 64.9         & 1.04            & 6.48         & 4.3          & 6.5           \\ 
520      & 2.29    & 96      & 116.6        & 164.8        & 1.03            & 8.56         & 8.3          & 7.4      \\ 
600       & 1.98    & 96      & 358.3        & 506.7        & 1.03            & 11.4         & 20.0           & 8.5            \\ 
\hline
\hline
Array         &              & 2100 & 1.2  & 1.7 &  &  & 0.41 &   \\
\hline 
\end{tabular}
}
\end{center}
\vspace{-\baselineskip}
\caption{\small  Proposed \coremfive\ frequency channels. The baseline aperture of the telescope is 1.2m in diameter. The detectors are cooled at 0.1K and their sensitivity is calculated assuming $\Delta \nu/\nu=30\%$ bandwidth, 60\% optical efficiency, total noise of twice the expected photon noise from the sky and the optics of the instrument at \SI{40}{K}. This configuration has 2100 detectors, about 45\% of which are located in CMB channels between 130 and \SI{220}{GHz}. Those six CMB channels yield an aggregated CMB polarization sensitivity of \SI{2}{\micro\kelvin}$\cdot$arcmin (\SI{1.7}{\micro\kelvin}$\cdot$arcmin for the full array). 
}
\label{tab:CORE-bands}
\end{table}
}

Other missions with similar targets have been proposed in the past. Apart from our previous proposals COrE \cite{2011arXiv1102.2181T} and PRISM \cite{2013arXiv1306.2259P}, a set of CMB polarization mission have been proposed in the USA (see e.g. \cite{2009arXiv0906.1188B}, \cite{2011JCAP...07..025K}). In Japan, the LiteBIRD proposal is currently in phase-A \cite{2012SPIE.8442E..19H}. This reflects the awareness of the scientific community that an unambiguous detection and characterizaiton of B-mode polarization from inflation cannot be achieved from ground-based measurements alone. There are two main problems: the need for a large sky coverage (the primordial B-mode signature in the Reionization peak is at the largest angular scales) and the need for a wide frequency coverage (to monitor overwhelming Galactic polarized foregrounds). Ground-based measurements suffer in both cases. Atmospheric noise, following a Kolmogorov statistics, is severe at large angular scales, and anisotropic ground pickup may seriously contaminate the detected signal at these scales. Atmospheric transmission and stability rapidly become a serious limitation at frequencies higher than 250 GHz, even in the best observing sites on Earth. Coordinated ground-based efforts can do extremely well in the well-known 40, 90, 140 GHz atmospheric windows and at intermediate and small angular scales \cite{2016arXiv161002743A}, but they have very serious limitations at higher frequencies (where the polarized dust foreground must be monitored) and at large angular scales.

The \coremfive \ instrument concept described in this paper has been conceived to minimize complexity and single-point criticalities, maximizing reliability in development, test, commissioning and operation. It uses a well developed, mirrors-only cross-Dragone telescope, with mirrors smaller than the mirrors used in Planck. It uses, as a baseline, simple single-frequency single-polarization detectors (dual-polarization is considered only for the lowest frequencies, as explained further on). It does not use a polarization modulator, avoiding critical cryogenic mechanisms. Exploiting the good angular resolution provided by the telescope, it is possible to synthesize in the analysis perfectly circular observation beams, so that the polarization of the sky can be measured simply by rotating the entire instrument. With a smaller aperture, this would not be possible, and the added complexity of a polarization modulator would be required to separate real sky polarization and signal produced by the rotation of an elliptical beam. In addition, the angular resolution provided by the 1.2m reflector allows for internal delensing of the data, a real plus when looking for such a low level of B-mode polarization \cite{2012JCAP...06..014S}. 

In Table~\ref{tab:CORE-bands} we anticipate the baseline configuration for the multi-band polarimeter, and its performance.

The rest of the paper is organized as follows: in the second section we describe the different telescope options which have been considered, taking into account the size constraints coming from the launcher, the need for sufficient angular resolution, the need for a wide, high Strehl ratio, and possibly planar and telecentric focal plane. In the third section we discuss advantages and disadvantages of a mechanical polarization modulator, our baseline choice of avoiding its use, and our approach in case it is found to be required in the phase-A study. In the fourth section we describe the general design of the focal plane array and its optimization in terms of size, number of detectors, weight of the different bands. In the fifth section we describe the selected detectors technology (Kinetic Inductance Detectors) and its implementation in the different frequency bands; the optical coupling and the readout electronics. In the sixth section we describe the cryogenic system. In the seventh section we describe the plan to calibrate the instrument. In section eight we describe a downscoped configuration of the instrument, with reduced aperture of the telescope (0.8m), reduced number of detectors (900), covering a reduced frequency range (100 to \SI{600}{GHz}). We conclude comparing our baseline instrument to other proposals, in terms of figure of merit for the measurement of CMB B-modes.

\section{Telescope}
\label{sec:telescope}



A combination of science requirements and practical constraints drive the choice of the telescope. The science requirements 
are (i) a focal surface with sufficiently large diffraction limited field of view (DLFOV) to accommodate the 2100 detectors, (ii) 
entrance aperture size of \SI{1.2}{m} to give a resolution between 7' and 8'  at \SI{145}{GHz}, and (iii) low instrumental polarization. 
Cost, the space mission implementation, and experience with previous mm-wave telescopes suggest a compact, 
low mass telescope, and reflectors that are sufficiently small to be manufactured as a single segment of silicon carbide (silicon 
carbide has been space proven with the Herschel and Aladin missions). A quantification of the level of instrumental polarization 
tolerable with \coremfive\ is awaiting a more comprehensive study of systematic uncertainties. 

The baseline detector technology for \coremfive\ also imposes constraints on the optical design.  \coremfive\
will implement arrays of lens-coupled kinetic inductance detectors fabricated on flat wafers. Therefore, 
the focal surface should be locally flat and telecentric. The optical design needs to provide a focal surface that 
is either planar, or can be tiled with flat detector wafers.  

We investigated two designs: a two-mirror Gregorian, and a crossed-Dragone. Both designs have rich heritage with 
CMB experiments \cite{hanany_niemack_page_review}. Most recently, WMAP and Planck used off-axis Gregorian 
telescopes. Tran et al. \cite{tran08} showed that with equal aperture sizes a crossed-Dragone system can provide larger DLFOV
compared to a Gregorian. (We follow standard convention designating an area in the focal surface `diffraction limited' when 
the Strehl ratio is larger than 0.8.) 

\subsection{Gregorian Design}
\label{sec:greg_designs}

\subsubsection{All Reflective Design}
\label{sec:greg}

\begin{figure}[htpb]
	\centering
        \includegraphics[width=0.45\textwidth]{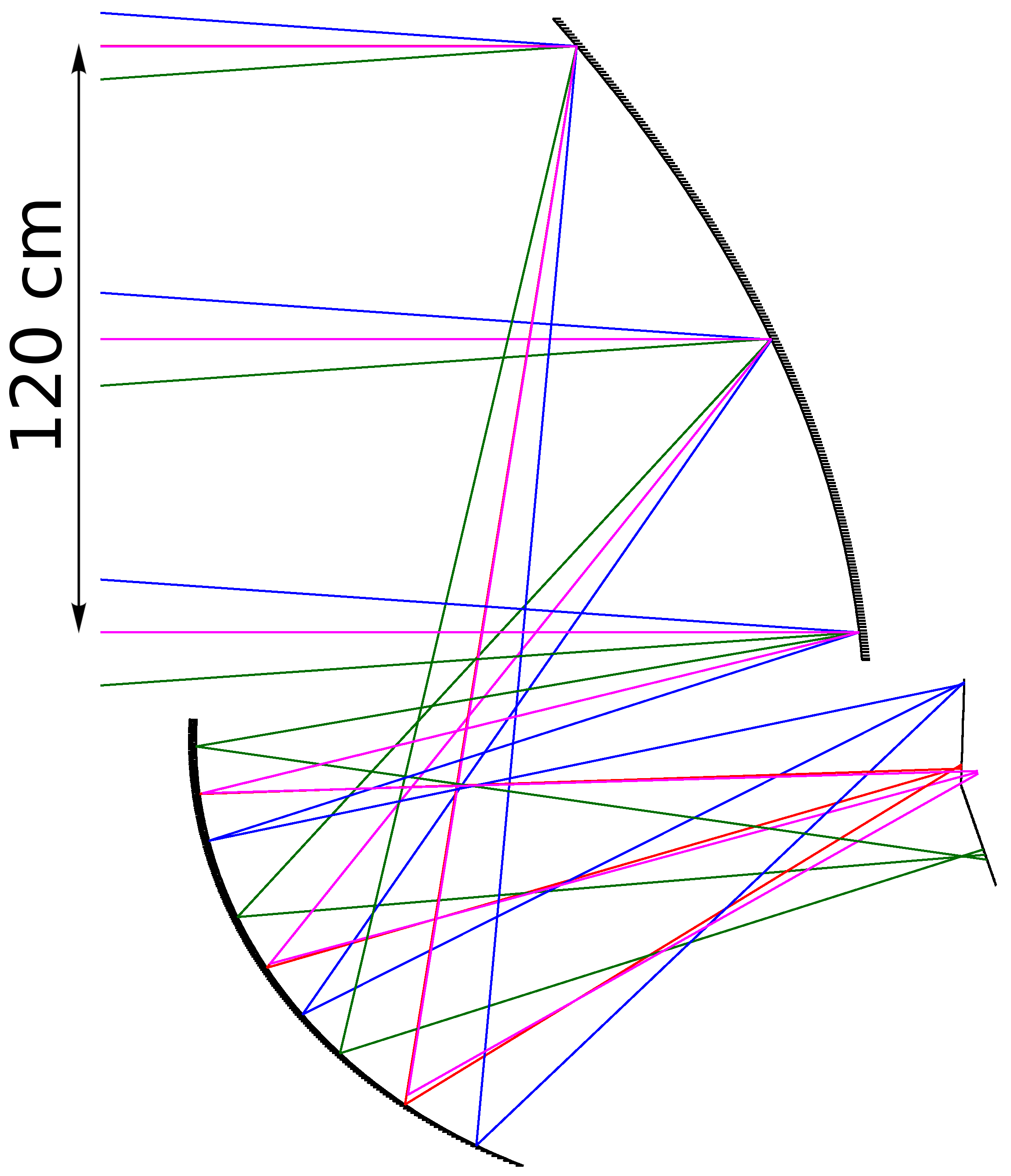}
        \qquad
        \includegraphics[width=0.41\textwidth]{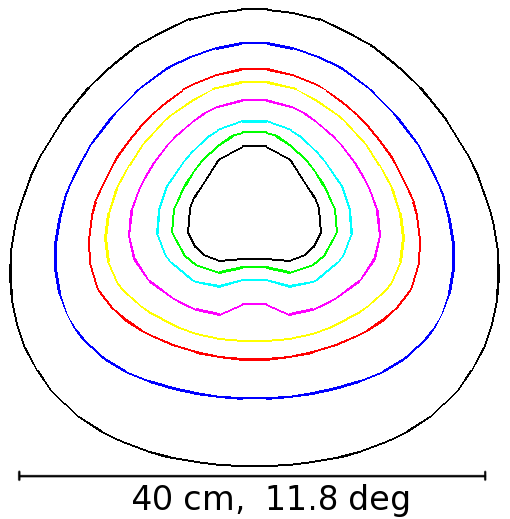}
	\caption{\small \SI{1.2}{m} Gregorian design and DLFOV contours for \SI{60}{GHz} (black, outer), \SI{90}{GHz} (blue), \SI{130}{GHz} (red),
         \SI{160}{GHz} (yellow), \SI{220}{GHz} (magenta), \SI{340}{GHz} (cyan), \SI{450}{GHz} (green), and \SI{600}{GHz} (black, inner). 
	\label{fig:gregschem} }
\end{figure}

The basic two-mirror Gregorian design has a parabolic primary and an ellipsoidal secondary. In Gregorian-Dragone
designs, astigmatism or both astigmatism and coma are canceled to first order \cite{hanany_niemack_page_review}. We began our 
analysis with a Gregorian-Dragone design in which astigmatism is canceled to first order \cite{dragone1978}
and used the optimization features available with a commercial ray tracing program\footnote{CodeV by Synopsys Inc.} 
to increase the DLFOV over the frequency range between 60 and \SI{600}{GHz}. Each of the reflector surfaces
was described as a figure of revolution $z(r)$ by  
\begin{equation}
z=\frac{cr^2}{1+\sqrt{1-\left(1+k\right)c^2r^2}}+\sum\limits_{i=1}^{n}a_ir^{2i}, 
\label{eq:zemaxasphere}
\end{equation}
where $k$ is the conic constant and $c$ is the radius of curvature.
Surfaces for which $a_{i} = 0$ for all $i$ are conics of revolution. 
During the numerical optimization we let all the parameters in Equation~\ref{eq:zemaxasphere} vary, as well 
as the parameters defining the relative orientations of the surfaces. The parameters of a system that gave a sufficiently large DLFOV are 
given in table~\ref{tab:greg_no_lens}. The system is shown in figure~\ref{fig:gregschem}. We note that the focal 
surface is not flat. It is a shallow cone. We could not find a solution with a sufficiently large DLFOV and flat focal surface. 

\begin{table}[htpb]
  \centering
  \begin{tabular}{|ll|l|ll|}
    \hline
         \multicolumn{2}{|c|}{Primary mirror} & Secondary mirror &  \multicolumn{2}{c|}{Telescope geometry} \\
    \hline
        & \SI{136}{cm} $\times$ \SI{120}{cm} & \SI{97}{cm} $\times$ \SI{67}{cm}       &  $D_{m}$    & \SI{120}{cm} \\
    $c$ & -$5.94\times10^{-3}$ &  0.0127                      &  $L_m$      & \SI{165}{cm} \\      
    $k$ & -0.981              &  -0.319                       &  $L_s$      & \SI{148}{cm} \\      
    $a_{2}$ &  $2.45\times10^{-9}$   &  $1.06\times10^{-9}$   &  $h$        & \SI{80.3}{cm} \\        
    $a_{3}$ &  $3.67\times10^{-13}$  &  -$7.06\times10^{-13}$ &  $\alpha$   & \SI{11.7}{\degree} \\  
    $a_{4}$ &  -$3.35\times10^{-17}$ &  $1.20\times10^{-15}$  &  $\beta$    & \SI{3}{\degree} \\  
    $a_{5}$ &  $1.40\times10^{-21}$  & -$2.72\times10^{-19}$  &  $\theta_0$ & \SI{51.2}{\degree} \\ 
    $a_{6}$ &  -$2.25\times10^{-26}$ &  $2.36\times10^{-23}$  &  &  \\
    \hline \hline
    \multicolumn{2}{|c|}{Focal ratio, F}   & \multicolumn{3}{c|}{1.6 at center, increases by 7~\% at focal plane edges}  \\
    \multicolumn{2}{|c|}{\SI{160}{GHz} DLFOV }    & \multicolumn{3}{c|}{Az $\times$ El = \SI{7.6}{\degree} $\times$ \SI{7.8}{\degree}, 85 F$\lambda \times 73 \text{~F}\lambda$} \\ 
    \hline
  \end{tabular}
    \vspace{0.15cm}
    \begin{tabular}{|c|llll|}
    \hline
    Focal Surface Location         & Center     & $\pm$ Az. edge & + El. edge &  $-$ El. edge \\
    Instrumental polarization  (\%) &  0.02   & 0.02        &  0.02    &  0.03     \\
    Polarization rotation  (\dg)    &  0        & $\pm 6.4$  &  0         &  0           \\
    \hline
    \end{tabular}
  \caption{\small Parameters for the \coremfive\ Gregorian design. Surface parameters refer to 
  Equation~\ref{eq:zemaxasphere}; parameters determining the overall telescope geometry are defined in 
  Granet~2002 \cite{granet2002}. 
  The parameters describing the focal plane dimensions are given at 160~GHz in 
  degrees and in units of F$\lambda$ ($\lambda=\SI{1.875}{mm}$) 
  to facilitate comparison between different telescope configurations.  The variation in the focal ratio $F$ across the 
  focal surface gives a \SI{160}{GHz} DLFOV that is larger 
  in degrees in the elevation direction, but smaller in F$\lambda$.}
  \label{tab:greg_no_lens}
\end{table}

This design has a \SI{12}{\degree}, \SI{40}{cm} field of view at \SI{60}{GHz}. The configuration is compact, and the system 
is conducive to strong baffling.  The instrumental polarization across the entire focal plane, calculated at \SI{145}{GHz} by 
assuming $n, k = 1445, 1455$ for aluminum \cite{ordal1983}, 
are below 0.05~\%, but polarization rotation exceeds \SI{6}{\degree} at the azimuth edge of the focal surface. There is no polarization rotation 
along the symmetry plane of the system; see table~\ref{tab:greg_no_lens}. 
A significant drawback for this design is the non-flat, non-telecentric focal surface. 
We concluded that if a Gregorian design is to be used, it requires additional image correction. 

\subsubsection{Reflective/Refractive Design}
\label{sec:greg_nolens}

\begin{figure}[htpb]
	\centering
	\includegraphics[width=0.36\textwidth]{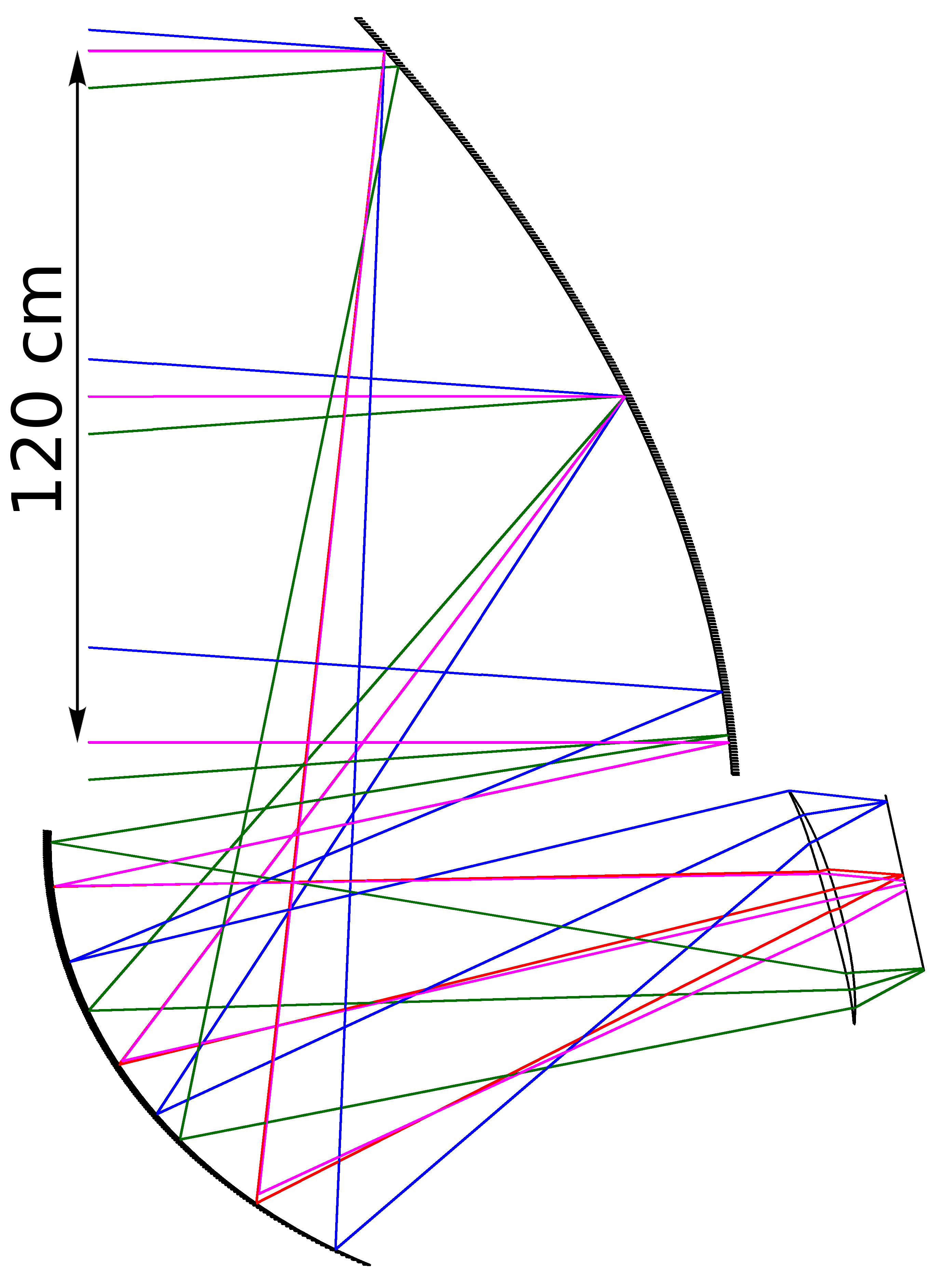}
        \qquad
        \includegraphics[width=0.39\textwidth]{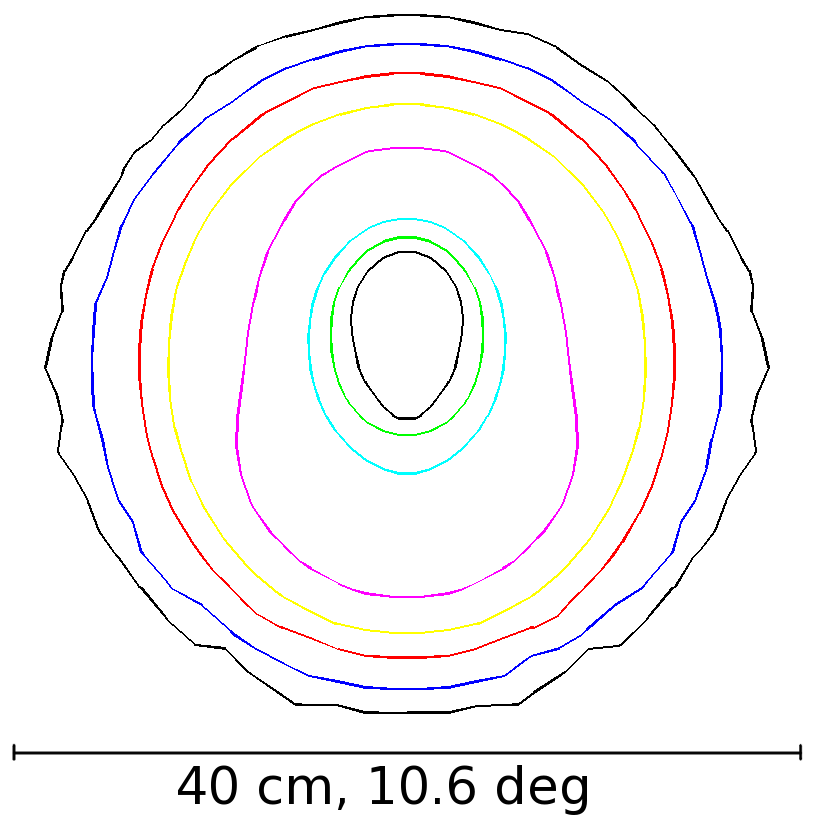}
	\caption{\small \SI{1.2}{m} Gregorian design with an alumina lens
	and DLFOV contours for \SI{60}{GHz} (black, outer), \SI{90}{GHz} (blue), \SI{130}{GHz} (red),
         \SI{160}{GHz} (yellow), \SI{220}{GHz} (magenta), \SI{340}{GHz} (cyan), \SI{450}{GHz} (green), and \SI{600}{GHz} (black, inner).
          \label{fig:greg_lens} }
\end{figure}

Figure~\ref{fig:greg_lens} shows a Gregorian system with one additional alumina lens ($n=3.1$). 
Table~\ref{tab:greg_lens} gives the parameter of the system. It has been optimized in a similar fashion to the one 
described in Section~\ref{sec:greg}. The focal surface is flat and nearly telecentric; the largest deviation from telecentricity 
is at the negative elevation edge (the lower part) of the FOV at a level of 4.1 degrees.
This design provides a \SI{9.2}{\degree} $\times$ \SI{9.5}{\degree}, \SI{36}{cm} $\times$ \SI{36}{cm}, DLFOV at \SI{60}{GHz}. 
These values are smaller than those for the all-reflective system. A more robust comparison that takes 
the focal ratios into account also shows that this system's performance falls short of the all-reflective one; see 
the lines giving the number of  $F\lambda$ at \SI{160}{GHz} in tables~\ref{tab:greg_no_lens} and~\ref{tab:greg_lens}. 
We could not fit all 2100 of CORE's detectors in the focal plane. 

The lens is \SI{44}{cm} in diameter with 
an optical aperture diameter of \SI{42}{cm} in diameter, and a mass of \SI{6}{kg}. 
The system is compact and the aperture of the lens causes vignetting for fields near the edge 
of the focal plane, as can be seen in figure~\ref{fig:greg_lens}.   
Assuming tophat illumination of the primary, the throughput for this edge of the focal plane is reduced by 7~\%.  
Vignetting reduces the throughput for the fields at both azimuth edges of the focal plane by 6~\%.

The anti-reflection 
coating (ARC) of the lens presents a challenge.  Broad, $\Delta \nu / \nu = 160$~\% ARC are required
with differential reflection sufficiently small to give low instrumental polarization. Although the development of broad-band
ARC is an active area of research \cite{datta2013, matsumura2016, plasma_spray_ar}, the development of this bandwidth 
ARC on alumina has not been demonstrated, let alone with low instrumental polarization. In table~\ref{tab:greg_lens} 
we give expected levels of instrumental polarization assuming an uncoated alumina lenses. The levels are 
substantial but are likely over-estimates. 
The magnitude of polarization rotation in this system is similar to that in the all-reflective system. 

In summary, the optical performance of the reflective/refractive system falls short of the requirements 
and the ARC presents a technical risk. 

\begin{table}[htpb]
	\centering
    \begin{tabular}{|ll|l|l|ll|}		
    \hline
        & Primary mirror & Secondary mirror & Alumina Lens & & Telescope geometry \\
    \hline
           & \SI{150}{cm} $\times$ \SI{130}{cm} & \SI{100}{cm} $\times$ \SI{70}{cm} & \SI{44}{cm}                      &  $D_{m}$ & \SI{120}{cm} \\
    $c$ & -$6.47\times10^{-3}$ &  0.0154  &  -$2.67\times10^{-3}$                     &  $L_m$ & \SI{145}{cm} \\      
    $k$ & -0.974              &  -0.333  &  0                                         &  $L_s$ & \SI{140}{cm} \\      
    $a_{2}$ &  -$1.02\times10^{-9}$  &  $6.08\times10^{-10}$  & $1.76\times10^{-6}$   &  $h$ & \SI{76.3}{cm} \\        
    $a_{3}$ &  $2.35\times10^{-14}$  &  -$1.91\times10^{-12}$ & -$2.69\times10^{-8}$  &  $\alpha$  & \SI{23.6}{\degree} \\  
    $a_{4}$ &  $5.73\times10^{-19}$  &  $1.27\times10^{-16}$  & $2.90\times10^{-11}$  &  $\beta$   & \SI{4}{\degree} \\  
    $a_{5}$ &  -$3.50\times10^{-23}$ &  $5.34\times10^{-20}$  & $4.64\times10^{-14}$  &  $\theta_0$ & \SI{55.9}{\degree} \\ 
    $a_{6}$ &  $5.65\times10^{-28}$  & -$1.35\times10^{-23}$  & -$7.98\times10^{-17}$ &  &  \\
    \hline \hline
    \multicolumn{2}{|c|}{Focal ratio, F}   & \multicolumn{4}{c|}{1.88 at centre, varies by 17~\% across focal plane}  \\
    \multicolumn{2}{|c|}{\SI{160}{GHz} DLFOV}    & \multicolumn{4}{c|}{Az $\times$ El = \SI{6.4}{\degree} $\times$ \SI{7.2}{\degree}, 
                                               68 F$\lambda \times 75 \text{~F}\lambda$} \\ 
    \hline
    \end{tabular}
    \vspace{0.15cm}
    \begin{tabular}{|c|llll|}
    \hline
    Focal Surface Location  & Center     & $\pm$ Az. edge & + El. edge &  $-$ El. edge \\
    Instrumental polarization  (\%) &  0.26  & 2.7       &  3.6   &  1.8     \\
    Polarization rotation  (\dg)   &  0        & $\pm 5.7$  &  0         &  0           \\
    \hline
    \end{tabular}
    \caption{\small Same as table~\ref{tab:greg_no_lens} but for the Gregorian design with an alumina lens.
	\label{tab:greg_lens} }
\end{table}

\subsection{Crossed Dragone}

The crossed-Dragone configuration more naturally provides a flat, telecentric focal plane and therefore it is a good 
match to focal planes with arrays of detectors that are micro-fabricated on flat silicon wafers \cite{tran08}. 
The configuration has heritage with the QUIET and ABS CMB polarization instruments \cite{quiet13, abs10}, 
and is currently the baseline for the JAXA-led LiteBIRD mission \cite{matsumura14}. It is being 
considered for the CMB-S4 project \cite{niemack16}. 

\begin{figure}[htpb]
	\centering
        \includegraphics[width=0.45\textwidth]{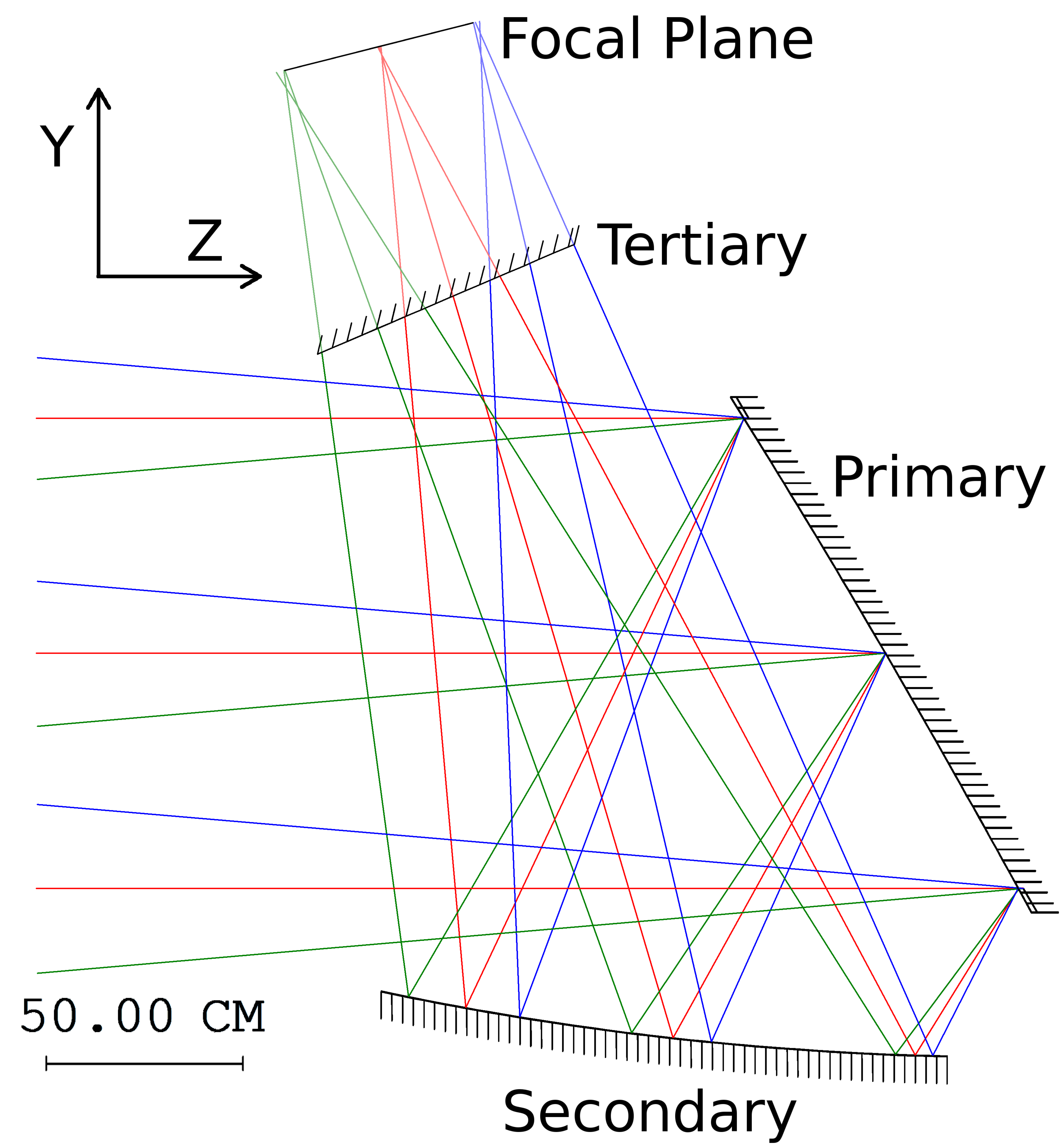}
	\includegraphics[width=0.50\textwidth]{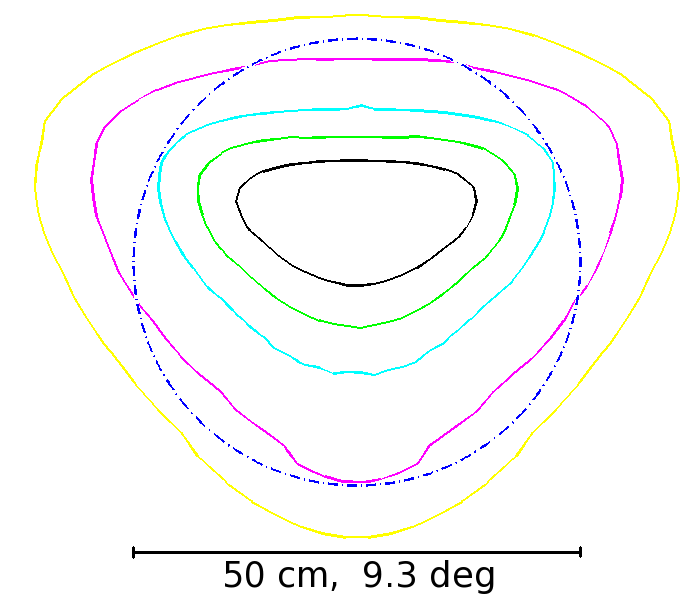} 
	\caption{\small \SI{1.2}{m} aperture \coremfive\ telescope and DLFOV 
	at \SI{160}{GHz} (yellow), \SI{220}{GHz} (magenta), \SI{340}{GHz} (cyan), 
	\SI{450}{GHz} (green), and \SI{600}{GHz} (black). The DLFOV at lower frequencies 
	is larger than that shown for \SI{160}{GHz}. The \SI{50}{cm} diameter usable FOV (dash dot blue) is limited by baffles, 
	not optical performance. 
	\label{fig:dragone_ray} }
\end{figure}

For \coremfive\ we started with the \SI{40}{cm} LiteBIRD telescope, scaled it up to \SI{1.2}{m} aperture, and re-optimized 
in a process similar to that described in Section~\ref{sec:greg}. The design included anamorphic surfaces, that is, surfaces 
with different radii of curvature in two orthogonal directions. 
The equation describing the surface $z(r)$ is 
\begin{equation}
\label{eqn:crossed}
z = \frac{\frac{x^2}{R_x^2} + \frac{y^2}{R_y^2}}
{ 1 + \sqrt{1 - (1+k_x) \frac{x^2}{R_x^2} - (1+k_y) \frac{y^2}{R_y^2}}}
+ A_{n,r} ( (1-A_{n,p})x^2 + (1 + A_{n,p})y^2)^n,
\end{equation}
where $R_x$ and $R_y$ are the radii of curvature, $k_x$ and $k_y$ are the conic coefficients, and $A_{n,r}$, $A_{n,p}$
define the higher order deformation coefficients with $n = 2,3,4,5$. 
When the higher order coefficients are zero this type of surface is called biconic. 
Figure~\ref{fig:dragone_ray} shows the system and Table~\ref{tab:xdragone} gives its parameters. 

\begin{figure}[htpb]
\centering
       \includegraphics[width=0.42\textwidth]{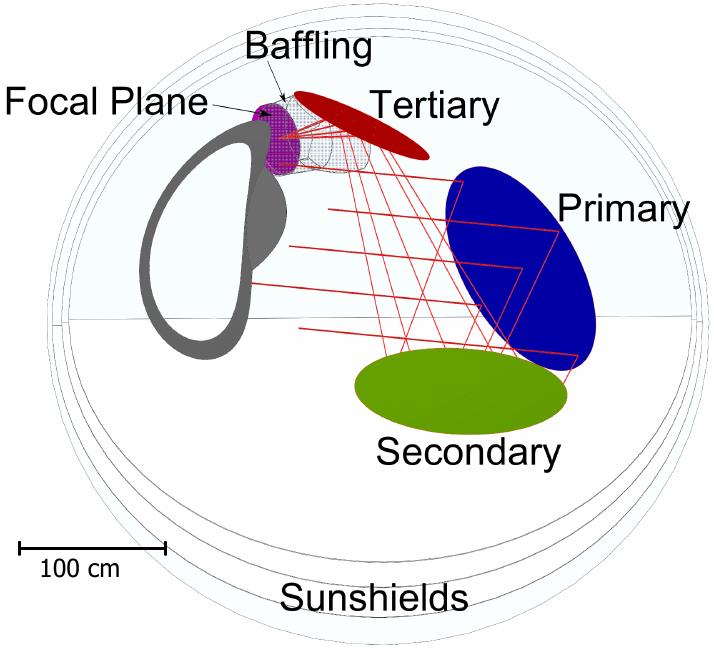}
       \includegraphics[width=0.4\textwidth]{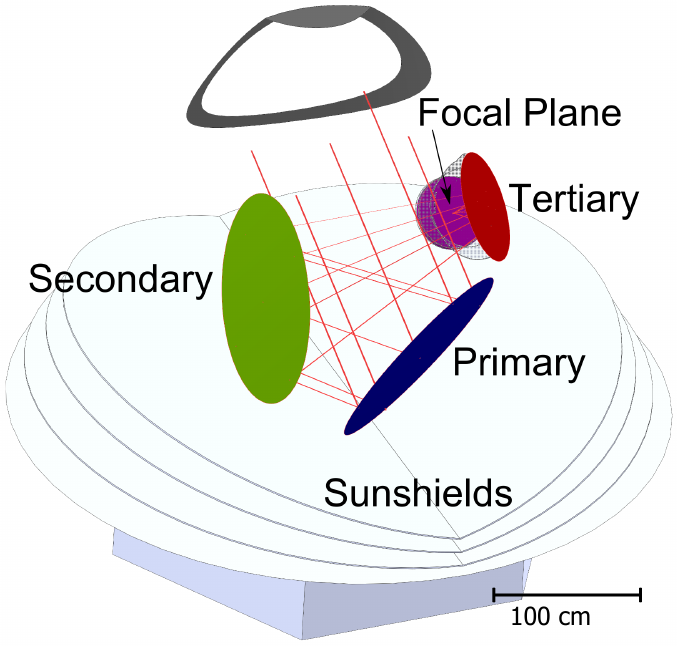}
       \caption{\small Perspective views of the telescope and focal plane relative to the spacecraft's 
       sunshields and bus. Only the edge of the entrance aperture of the enclosure is shown for clarity (gray).
\label{fig:dragone_cad} }
\end{figure}

We found that with only two mirrors the system was too big to fit within the satellite envelope. Therefore we 
added a flat tertiary fold mirror to make the system more compact. This flat mirror 
could be replaced by a reflective polarization modulator, if polarization modulation is deemed necessary. 
Figure \ref{fig:dragone_cad} gives solid model views of the telescope 
integrated within the payload module and sunshields. 
The telescope was rotated about the boresight to put the focal plane close to the main satellite body, reducing the need for 
additional supports. This was also the only orientation in which the telescope was completely shadowed by the sun-shields. 

The focal plane is flat with a DLFOV greater than 12 degrees across at \SI{160}{GHz}. 
It gives approximately 4 times the number of $F \lambda$ units at \SI{160}{GHz} compared to the all reflective 
Gregorian system, and a larger factor when compared to the partially refractive Gregorian.  
It is telecentric to within \SI{3.5}{\degree} over the entire FOV, and within \SI{2}{\degree}, which is 10~\% of the beam divergence 
angle at the focal plane, at field angles of less than \SI{2.8}{\degree} . 
Experience indicates that the system can be further optimized for even stronger telecentricity at the expense 
of DLFOV area at the edges of the focal plane that are currently not used. The instrumental polarization induced by the reflectors is 
a factor of 2-3 larger than the reflective Gregorian design, but still below 0.1\%. There is significantly lower polarization 
rotation, a consequence of the flatter reflecting surfaces. 

\begin{table}[htpb]
	\centering
	\begin{tabular}{|ll|l|ll|}
	\hline
        & Primary mirror & Secondary mirror &  & Telescope geometry \\
    \hline
        & \SI{131}{cm} $\times$ \SI{152}{cm} & \SI{125}{cm} $\times$ \SI{146}{cm} &  $D_{m}$ & \SI{120}{cm} \\
    $R_x$ & -$1.15\times10^{3}$ &  $8.29\times10^{2}$  &  $L_m$ & \SI{112}{cm} \\ 
    $R_y$ & -$7.14\times10^{2}$ &  $8.33\times10^{2}$  &  $L_s$ & \SI{264}{cm} \\ 
    $k_x$ &  2.97              &  -0.574               & $h$ & \SI{765}{cm} \\ 
    $k_y$ & -3.55              &  -7.31                & $\alpha$  & \SI{13.8}{\degree} \\ 
    $A_{2,r}$ &  $3.02\times10^{-10}$  &   $1.31\times10^{-9}$  & $\beta$   & \SI{90.2}{\degree} \\
    $A_{3,r}$ &  $9.54\times10^{-18}$  &   $3.05\times10^{-14}$   & $\theta_0$ & \SI{104}{\degree} \\
    $A_{4,r}$ & -$1.63\times10^{-22}$  &  -$8.17\times10^{-19}$   &  &  \\
    $A_{5,r}$ & -$3.49\times10^{-34}$  &   $2.36\times10^{-24}$  &  &  \\
    $A_{2,p}$ & -$1.027$                &  $0.183           $   &  &  \\
    $A_{3,p}$ & -$0.255$                &  $0.0671          $   &  &  \\
    $A_{4,p}$ & -$0.281$                &  $0.259           $   &  &  \\
    $A_{5,p}$ & -$0.914$                &  $0.669           $   &  &  \\
	\hline \hline
        \multicolumn{2}{|c|}{Tertiary mirror}         & \multicolumn{3}{c|}{\SI{104}{cm} $\times$ \SI{74}{cm} }  \\
	\multicolumn{2}{|c|}{Focal ratio, F}          & \multicolumn{3}{c|}{2.54 at centre, varies by 5~\% across focal plane}  \\
	\multicolumn{2}{|c|}{\SI{160}{GHz} DLFOV}    &  \multicolumn{3}{c|}{Az $\times$ El =\SI{14.0}{\degree} $\times$ \SI{12.9}{\degree}, 
	                          159 F$\lambda \times 140 \text{~F}\lambda$} \\ 
	\hline		
	\end{tabular}
    \begin{tabular}{|c|llll|}
    \hline
    Focal Surface Location   & Center     & $\pm$ Az. edge & + El. edge &  $-$ El. edge \\
    Instrumental polarization (\%)  &  0.06  & 0.07, 0.05       &  0.07   &  0.05     \\
    Polarization rotation   (\dg)   &  0        & $\pm 0.6$  &  0         &  0           \\
    \hline
    \end{tabular}
    \caption{\small Parameters for the \coremfive\ Crossed Dragone design.  Surface parameters refer to 
        Equation~\ref{eqn:crossed}; parameters determining the overall telescope geometry are defined in 
        Granet~2001 \cite{granet2001}. 
	\label{tab:xdragone} }
\end{table}

An initial challenge with the crossed Dragone system was baffling of the focal plane to reject stray light.  
To improve baffling we embedded the portion of the DLFOV that is used for detectors inside a `bucket', and a 
collar was added around the entrance aperture of the system; the bucket and collar are shown in grey 
figure~\ref{fig:dragone_cad}.
The default is all the paylod, including the collar, at \SI{40}{K}.


With these additions there is no 
direct view from the focal plane to the sky. The focal plane area populated by detectors is bounded by the 
blue dash-dot line in Figure~\ref{fig:dragone_ray} and is smaller than the total available DLFOV.

We analyzed the crossed-Dragone design using physical optics to determine the far field
beam shape at \SI{145}{GHz} for various positions on the focal plane. We assumed a Gaussian beam 
propagating from the focal plane outward. The Gaussian beam had a waist
$w_0=0.216\sqrt{T_e\left(dB\right)}\text{F}\lambda=\SI{5.14}{mm}$ with an 
edge taper of \SI{-20.5}{dB} on the primary mirror and $\text{F}=2.54$. For this first round of analysis 
we did not include the baffling structures.

\begin{figure}[htpb]
	\centering
	\includegraphics[width=0.48\textwidth]{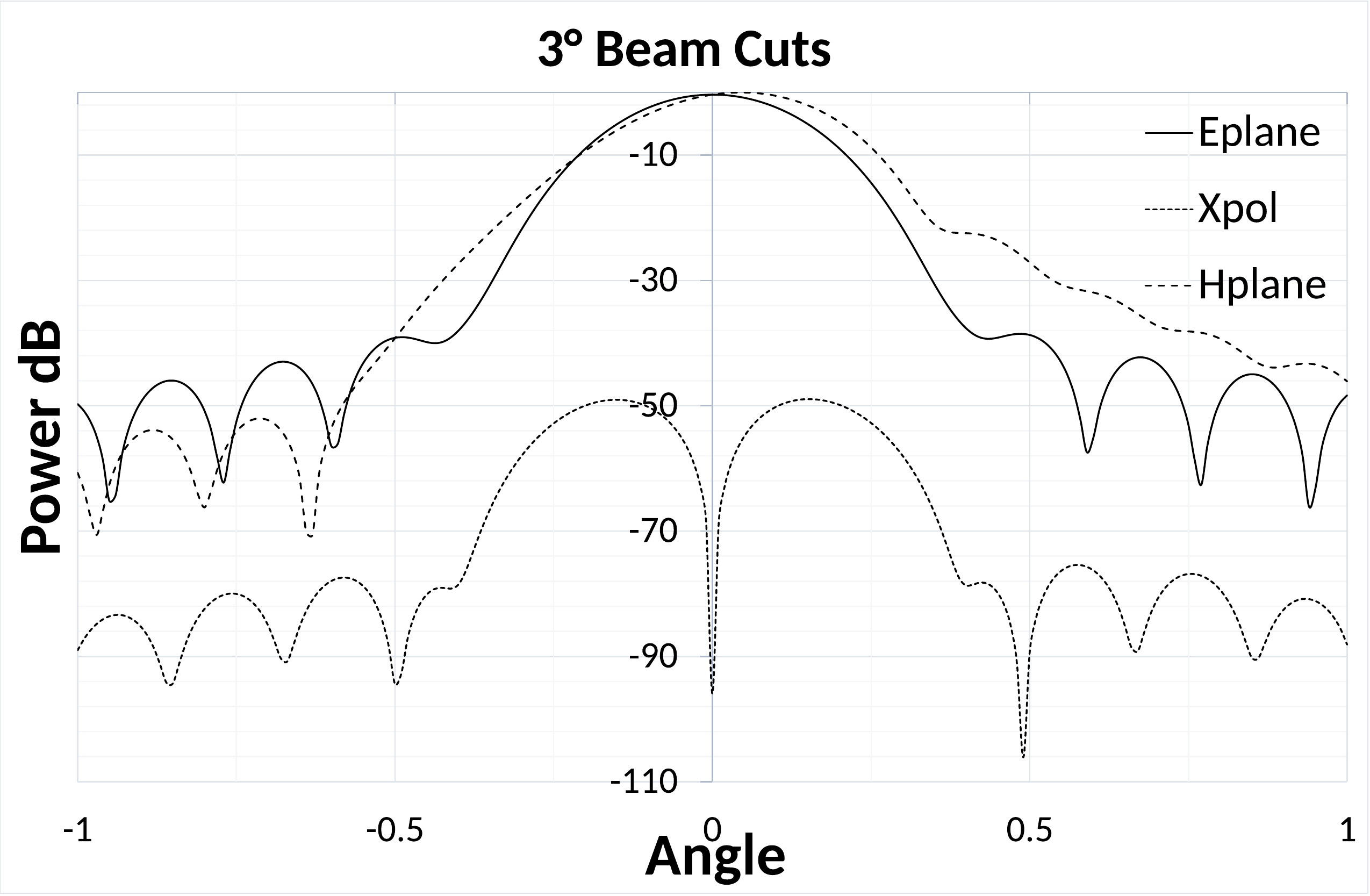}
	\quad
	\includegraphics[width=0.48\textwidth]{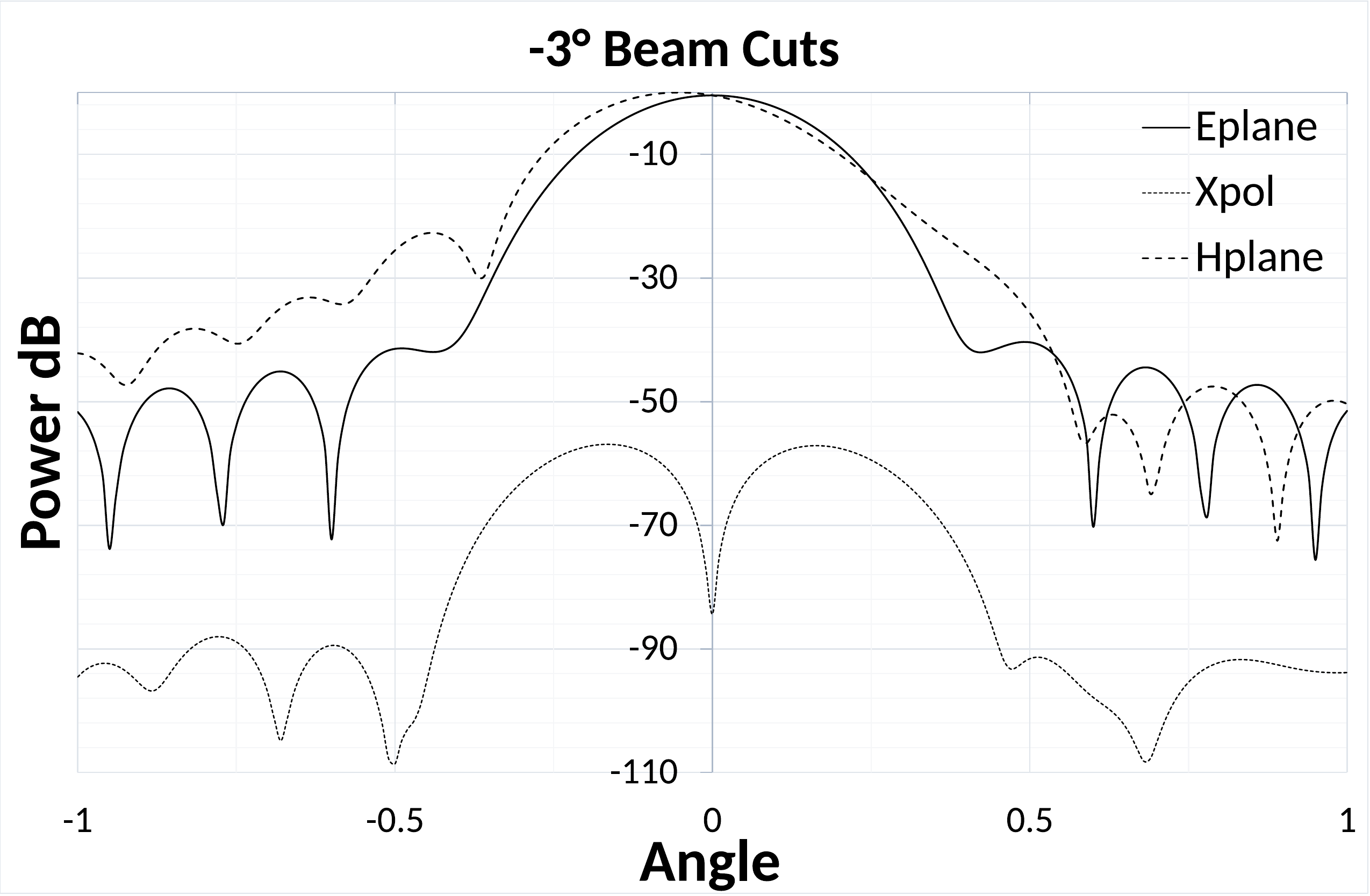} 
	\caption{\small Orthogonal co-pol and cross-pol cuts of the \SI{145}{GHz} far field beams at off-axis angles 
                 of $\pm 3^{\circ}$ along the $x$ direction of the telescope, as indicated in 
                 figure~\ref{fig:dragone_ray}.
	\label{fig:PO145GHz_cuts} }
\end{figure}

Figure~\ref{fig:PO145GHz_cuts} shows representative orthogonal cuts of the far-field beams 
at $\pm 3^{\circ}$ off-axis along the $x$ axis of the telescope (see figure~\ref{fig:dragone_ray}). 
These field positions correspond to locations that are 
\SI{15}{cm} from the center of the focal plane in the horizontal directions in the 
right panel of figure~\ref{fig:dragone_ray}. 
We present results for only one polarization. The orthogonal polarization, which was also propagated to the 
sky, is co-located and the differential gain between the two polarizations is \SI{-41}{dB} over 
the main beam.

We fit two-dimensional Gaussian to the far-field beam patterns to extract full-width
at half-maxima (FWHM) from which we calculate beam ellipticities. 
Ellipticity is defined as  $1-b/a$, 
where $a$ and $b$ represent the larger and smaller FWHM, respectively.
Cross-polarization 
levels are low, \SI{-48}{dB} at a maximum for the beams at $\pm 3^{\circ} $, but the co-polar components 
of the orthogonal polarizations show some aberration 
and have ellipticities of 6 and 8~\% for the -3$^{\circ}$ and +3$^{\circ}$ 
directions, respectively. 
A summary of the characteristics of the beams at several 
field locations is given in table~\ref{tab:POresults_cDragone}. 

The physical optics study of the telescope performance for a variety of focal plane positions 
and frequencies is ongoing. There is also initial work currently underway to simulate the far 
sidelobes and examine the impact of the shielding on the optical performance. 

\begin{table}[htpb]
\centering
\begin{tabular}{|l|l|l|l|l|l|}
  \hline
   Beam & FWHM$_{\theta}$ & FWHM$_{\phi}$ & $\theta_0$ & $\phi_0$ & Ellipticity \\
                 & (arcmin)      &  (arcmin) &                &              &                   \\
  \hline
    Y~$=3^{\circ}$~-~X-Polarization       & 7.16     & 7.80        &  0.0001       &  0.0195   &      0.082 \\
    Y~$=3^{\circ}$~-~Y-Polarization       & 7.16     & 7.80       &  0.0001       &    0.0195    &   0.082  \\
    Y~$=-3^{\circ}$~-~X-Polarization      & 7.24     & 7.73         &  0           & -0.022   &       0.063  \\
    Y~$=-3^{\circ}$~-~Y-Polarization      & 7.24     & 7.73        &  0            &   -0.022   &     0.063  \\
    Y~$=4^{\circ}$~-~X-Polarization       & 7.14     & 8.10         &  0.0002      &  0.0252   &      0.118  \\
    Y~$=4^{\circ}$~-~Y-Polarization       & 7.14     & 8.10          &  0.0002      &    0.0252    &   0.118 \\
    Y~$=-4^{\circ}$~-~X-Polarization      & 7.26     & 7.92         &  0.0001     & -0.0285   &       0.083  \\
    Y~$=-4^{\circ}$~-~Y-Polarization      & 7.26     & 7.92         &  0.0001    &   -0.0285   &     0.083   \\
  \hline
\end{tabular}
\caption{\small Far field beam parameters for two orthogonal polarizations of Gaussian beam inputs located at off-axis 
         angles of $\pm \SI{3}{\degree}$ and $\pm \SI{4}{\degree}$ along the $x$ direction of the telescope (as illustrated 
         in figure~\ref{fig:dragone_ray}), at \SI{145}{GHz}. 
         The coordinates $\theta_{0}$ and $\phi_{0}$ give the offsets
         of the beam centroids relative to a ray traced along the center of the input Gaussian. }
\label{tab:POresults_cDragone}
\end{table}

\subsection{Telescope Summary}
\label{sec:telescopesummary}

WMAP and Planck used off-axis Gregorian telescopes. However, their useable DLFOV was smaller than that of \coremfive\ and 
their detector technology is no longer suitable for modern instruments that use thousands of focal plane elements.  
Such instruments require a large, flat, and telecentric focal plane. The two mirror Gregorian falls short off the requirements. 
We attempted to populate this design with \SI{10}{cm} edge-to-edge flat tiles that were locally telecentric at the center 
of the tile. However, the optical performance at the edge of these tiles failed to be diffraction limited. 

The two-mirror Gregorian with the alumina lens gives an optical performance that is close 
to the requirements but the anti-reflection coating introduces a technology risk, and a spurious polarization risk. 
We selected the crossed-Dragone as the baseline because it gives a large, flat, telecentric 
DLFOV, and it fits within the satellite envelope. A full analysis of the system is ongoing. 
Table~\ref{tab:summary} summarizes the advantages and disadvantages of the systems we analyzed. 

\begin{table}[htpb]
\centering
\begin{tabular}{ |l|l|l| }
  \hline
  & \textbf{Gregorian} & \textbf{Crossed Dragone} \\
   \hline
  \multirow{3}{*}{\textbf{Pros}} & Small reflectors & Large, flat, telecentric DLFOV \\
                                                & No lens: high TRL    &  Low instrumental and low \\
                                                & Easy to baffle   &   \hspace{0.3in} polarization rotation \\
   \hline
   \multirow{2}{*}{\textbf{Cons}} & No lens : Smaller, non-flat DLFOV & Larger reflectors \\
                                                  & With lens: low TRL; insufficient DLFOV & Baffling more challenging \\
  \hline
\end{tabular}
\caption{\small Advantages (Pros) and disadvantages (Cons) of the optical systems considered for \coremfive .}
\label{tab:summary}
\end{table}

\section{Polarization Modulator}
\label{sec:polmod}
In \coremfive\ polarization modulation is achieved by means of an optimized telescope pointing strategy, leading to frequent scans of the same sky pixel with different orientations of the polarization sensitive detectors, as described in the Mission paper of this series \cite{missionpaper}. This approach has the overwhelming advantage of simplicity and reliability, due to the absence of moving parts in the instrument. 

However, the technology of active CMB polarization modulators has significantly advanced in recent years, thanks to an ESA funded project (\emph{Large radii Half-Wave Plate (HWP) development}, ESA Ref. T207-035EE) focusing on developments for future CMB satellite missions. A novel type of Reflective Half Wave Plate (R-HWP) was successfully manufactured and tested. It has high polarization modulation efficiency across a 150~\% bandwidth at incidence angles up to \SI{45}{\degree} \cite{pisanohwp16, pisanohwp14}. The design can be further improved to achieve the 164~\% bandwidth required for \coremfive. The R-HWP should be inserted in the beam path of polarization-sensitive detectors and rotated around an axis orthogonal to its plane to provide polarization modulation. The position of the flat tertiary mirror in the \coremfive\ optical design is a natural place to locate a rotating R-HWP, should it be deemed necessary to include polarization modulation. The target of the aforementioned ESA  project is develop facilities capable of manufacturing metre-sized devices. This will allow the manufacturing of a R-HWP fitting the size of the \coremfive\ tertiary mirror (\SI{1.0}{m} $\times$ \SI{0.7}{m}). 

A cryogenic rotation mechanism is required to operate such a modulator. Two broad classes of devices have been developed: steppers and spinners. In the former, the HWP is stepped across a set of angular positions. Each angular position is held for some integration time (e.g. for one full spin of the instrument) before stepping to the next one. In the latter, the HWP is continuously spun, encoding the polarization information at high frequency (4$\times$ the mechanical rotation frequency). 

A rotator of the first type was described in \cite{salatino11} and was flown on the \textit{PILOT} balloon-borne experiment \cite{2016ExA42199B}. The average dissipation of the device, due to friction in the cryogenic bearings, is a few mW. A rotator of the second type, based on magnetic levitation of the HWP, was described in \cite{bryan16}, has been flown on the \textit{EBEX} balloon-borne experiment \cite{2011SPIE.8150E..04K}, on the SPIDER experiment \cite{2016RScI...87a4501B}, and a similar system represents the baseline for the LiteBIRD satellite \cite{2016SPIE.9904E..0XI}. 

The advantage of a continuously spun HWP is the high modulation frequency (order of \SI{10}{Hz}) of the polarization information, far from the frequency region affected by 1/f noise. The disadvantage is that eddy currents can heat-up the HWP: this heat can only be radiated away, quite inefficiently at low temperatures. In the case of the stepped HWP the modulation frequency is much lower (order of 0.01-\SI{0.1}{Hz}), but the HWP can be thermally connected to the cold reference temperature by means of flexible copper straps. 

Simplicity and reliability considerations drove the decision of not using a polarization modulator in \coremfive. However, end-to-end simulations including systematic effects from the HWP on one side, and from the spin/precession scan strategy on the other, will be performed anyway during phase-A, to confirm quantitatively this choice.


\section{Focal Plane}
\label{sec:focalplane}
\subsection{Mission Constraints}
\label{missconst}

The \coremfive\ instrument is optimized to maximize its mapping speed by means of the widest possible Focal Plane Array (FPA) of diffraction-limited detectors. The total number of detectors is limited by several heterogeneous factors: 

$\smallskip \noindent \bullet$ the diffraction limited field of view of the telescope (which in turn depends on the size and configuration of the telescope, as described in Section \ref{sec:telescope}); 

$\smallskip \noindent \bullet$ the heat load on the cryogenic system (the larger/heavier the FPA, the stronger and conductive the supports connecting it to the higher temperature stage of the cryogenic system, and the larger the integrated radiative load); 

$\smallskip \noindent \bullet$ the maximum allowed data rate;

$\smallskip \noindent \bullet$ the electrical power dissipated on the readout electronics, both at room temperature (impacting on the power budget) and in the cryogenic section (impacting on the power lift budget of the cryo system). 

For these reasons, optimization of the FPA is a nontrivial procedure. Starting from the available bay size in the space carrier for a medium-size mission, we have targeted the optimization to a diameter of the FPA of $\sim$ \SI{0.5}{m} at \SI{0.1}{K}, and verified that we are at the edge of what can be obtained in terms of telescope and cryogenic system (cfr Section \ref{sec:telescope} and \ref{sec:cryo}).

\begin{figure}[ht]
\centering
\includegraphics[width=0.9\textwidth, trim={0.5cm 0.5cm 0cm 0.5cm},clip]{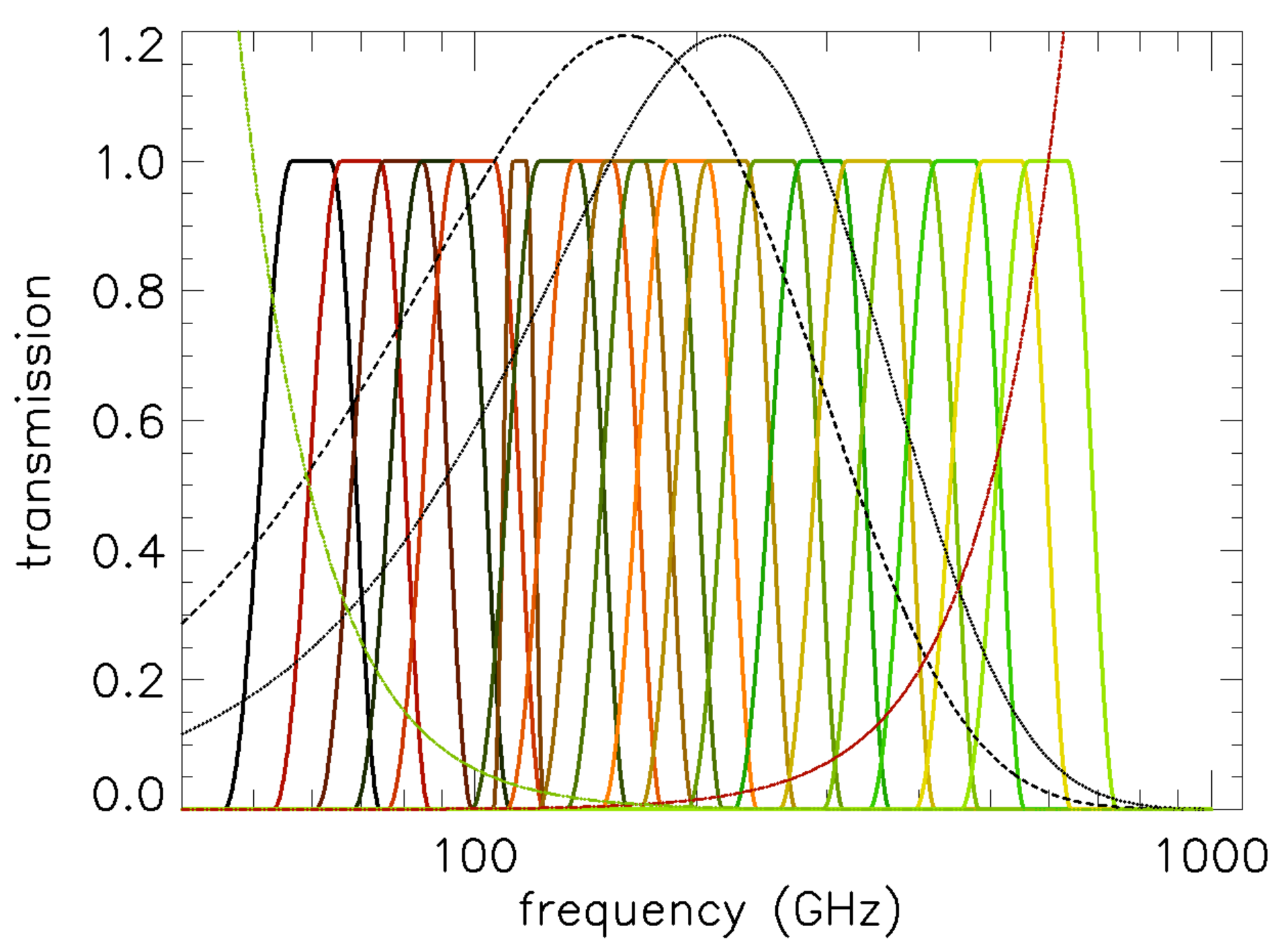}
\caption{\small 
\small 
Frequency coverage of the \coremfive\ focal plane (19 partially overlapping wide bands, plotted as red to light blue continuous lines), compared to arbitrarily normalized spectra of the CMB (dashed line), CMB anisotropy (dotted line), high-latitude dust (dot-dashed line), diffuse synchrotron (dash + 3 dots). 
}
\label{fig:FPU1}
\end{figure}

The next step of the optimization is the division of the available FPA area into the different frequency bands of observation. Covering a wide frequency range is the key to separate CMB polarization from a number of polarized foregrounds. The range \SI{60}{GHz} to \SI{600}{GHz} covers both low frequencies where diffuse synchrotron radiation from our Galaxy is dominant, CMB frequencies where the CMB signal is maximum, and high frequencies, where diffuse emission from interstellar dust 
is dominant. The \coremfive\ FPA covers this range with 19 frequency bands. The number of bands was selected to have a number of independent channels larger that the number of parameters required in a first-order description of all the relevant foregrounds. The optimization of the fractions of the FPA area assigned to the different channels has been carried out under the assumption that the detectors of a given frequency channel are sensitive to a wide band ($\Delta \nu / \nu \sim 30\%$, see Figure \ref{fig:FPU1}), and are limited by the photon noise associated with the incoming background power. This corresponds to a few to tens of aW/$\sqrt{\rm Hz}$, depending on the channel considered. The figure of merit of the optimization is the survey sentivity for CMB polarization signals, in $\SI{}{\micro\kelvin}\times$ arcmin, once the foreground removal procedure has been carried out. 

\begin{figure}[ht]
\centering
\includegraphics[width=0.9\textwidth, trim={0.cm 0.cm 0.cm 0.cm},clip]{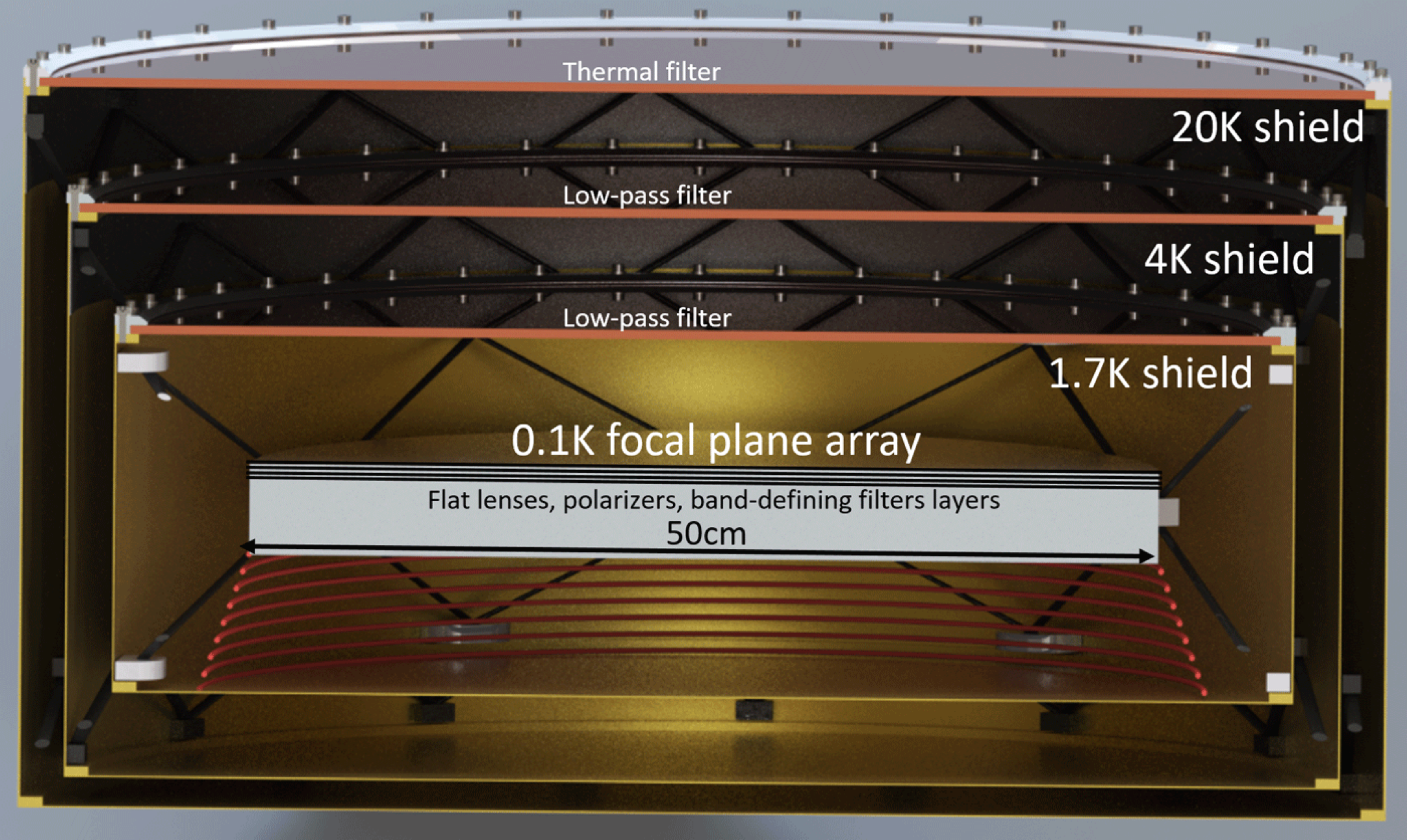}
\caption{\small 
\small 
The focal plane unit of \coremfive\ .
}
\label{fig:FPU2}
\end{figure}

\begin{figure}[ht]
\centering
\includegraphics[width=0.9\textwidth, trim={0.cm 0.cm 0.cm 0.cm},clip]{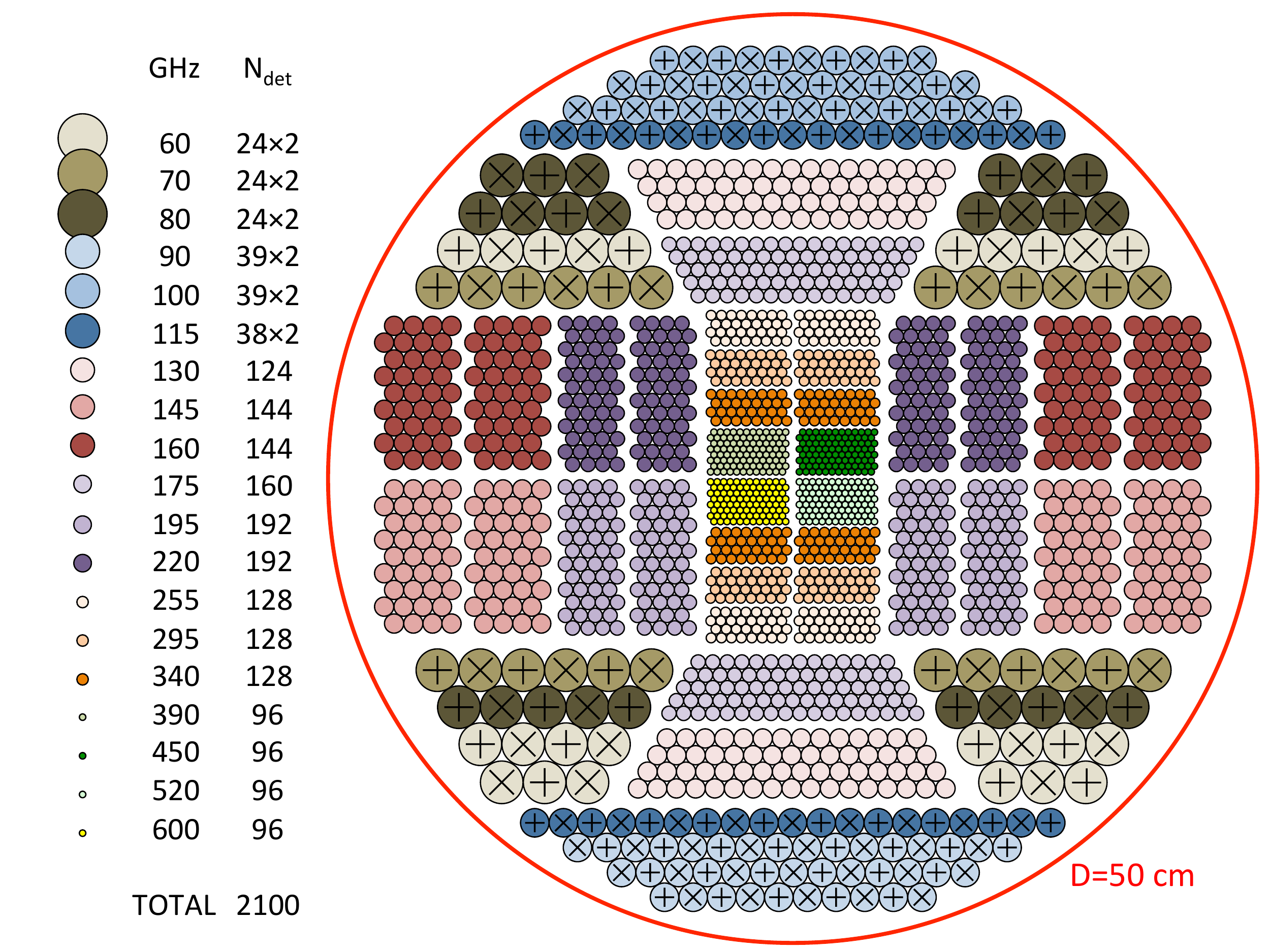}
\caption{\small 
\small 
Arrangement of 2100 single-band, single-polarization detectors, in the focal plane of the \coremfive\ telescope. The scan direction is horizontal.
}
\label{fig:FPU3}
\end{figure}

The outcome of these optimizations is that the \coremfive\ instrument will be equipped with a total of 2100 detectors (cfr Table \ref{tab:CORE-bands}), whose noise will not exceed the photon noise associated with the incoming background power. In the baseline configuration high-TRL single-band single-polarization pixels will be adopted, for all but the lowest frequencies ($\nu \leq$ \SI{115}{GHz}), where the use of dual-polarization sensitive pixels is envisaged. The coupling optics and the detectors for all of the pixels occupy a cylindrical volume, \SI{50}{cm} in diameter and \SI{5}{cm} tall, which is cooled at \SI{0.1}{K} for optimal operation of the detectors (see Figure \ref{fig:FPU2}). Detectors in the focal plane are arranged in tiles, each one including sensors operating in the same frequency band, and placing the highest frequency band tile in the center of the focal plane, where the optical quality is best. The general arrangement is shown in Figure \ref{fig:FPU3}.

\subsection{Radiation Coupling}

Techniques to achieve the desired radiation coupling while respecting the mass and volume constraints imposed by the mission framework described in Section \ref{missconst} are currently available and well developed within the \coremfive Consortium.

\begin{figure}[ht]
\centering
\includegraphics[width=0.7\textwidth]{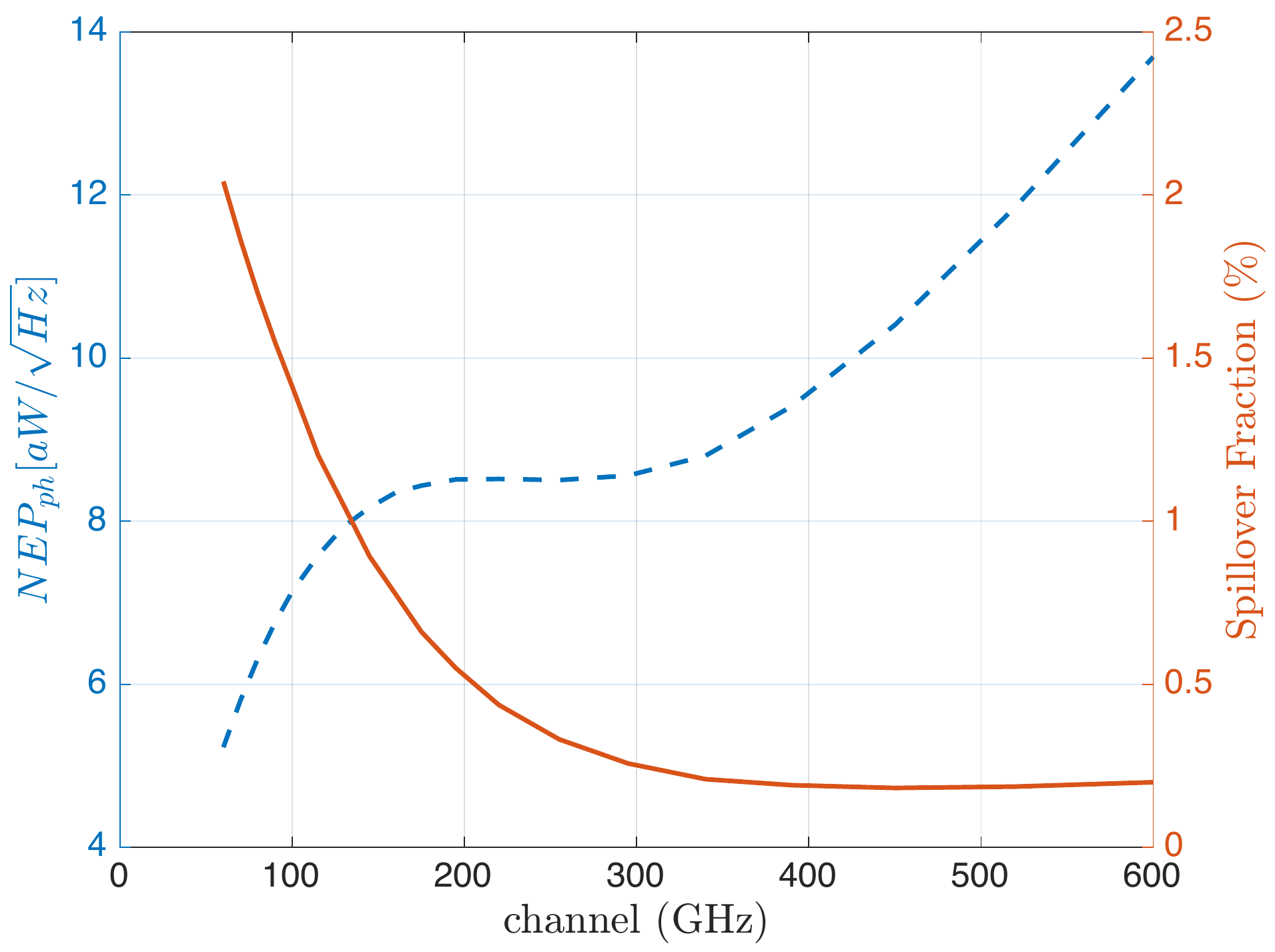}
\caption{\small 
\small 
In the orange curve we show the Spillover Fraction ($10^{-T_e/10}$)  that allows to reach the corresponding overall photon NEP, the plotted blue curve. 
}
\label{fig:NEP}
\end{figure}

One of the fundamental constraints driving the dimensioning of the focal plane is the acceptable spillover fraction, which in turn might have a different impact depending on the temperature of the telescope enclosure. This quantity must be small enough in such a way that the photon noise from the black payload alone is less than half of the contribution coming from the sky and from Planck-type mirrors together.
The sky brightness is a superposition of CMB, Galactic Dust (a greybody at \SI{18}{K}, with $\beta=1.7$, corresponding to the cleanest 90$\%$ of the \SI{857}{GHz} Planck sky map), Far-Infrared Background (a greybody at \SI{17}{K}, with $\beta=0.96$ and amplitude \SI{0.8}{MJ/sr} at the pivot frequency of \SI{1.87}{THz}) and Zodiacal light (a greybody at \SI{200}{K}, with $\beta=0.43$ and amplitude \SI{23.4}{MJ/sr} at the pivot frequency of \SI{5}{THz}) contributions. The telescope, at a temperature of \SI{40}{K}, is assumed to have a frequency dependent emissivity scaling as $\sqrt{\nu}$. 
For a \SI{40}{K} black payload, the optimised edge taper ranges from about \SI{17}{dB} at \SI{60}{GHz} to about \SI{27}{dB} at \SI{600}{GHz} (assuming a Gaussian telescope illumination), and its values for the different channels are plotted in Figure \ref{fig:NEP}, together with the corresponding overall photon noise. The illumination beam waist, for a given edge taper level $T_e({\rm dB})$ and telescope focal number ($F$), is given by $\omega_0=0.216\sqrt{T_e({\rm dB})}F\lambda$ \cite{goldsmith98}. For our optimised $T_e$ values, the diameter of the diffractive spot is in the range $1.8-2.2 F \lambda$.

\begin{figure}[ht]
\centering
\includegraphics[width=0.2\textwidth]{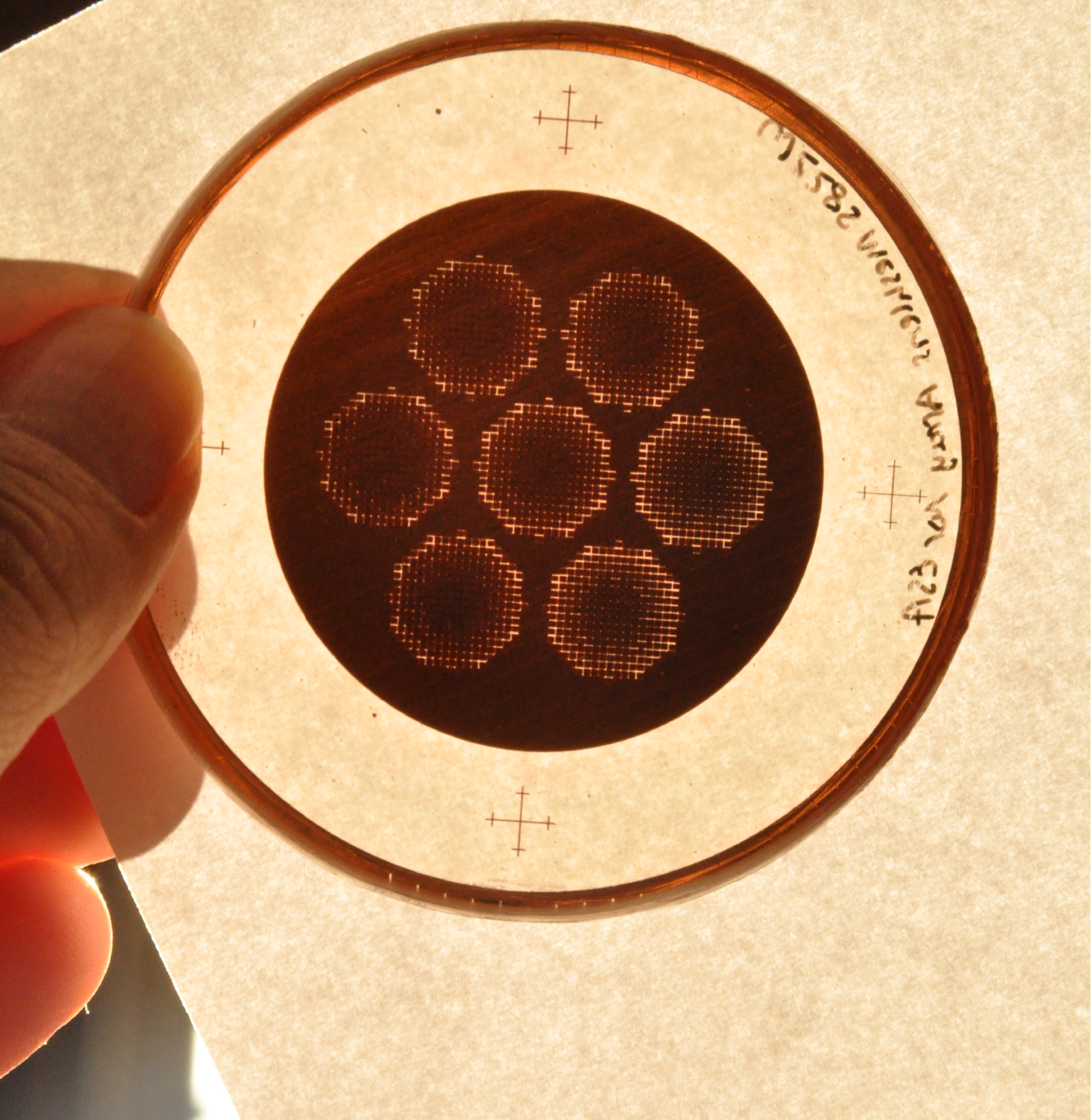} 
\includegraphics[width=0.7\textwidth]{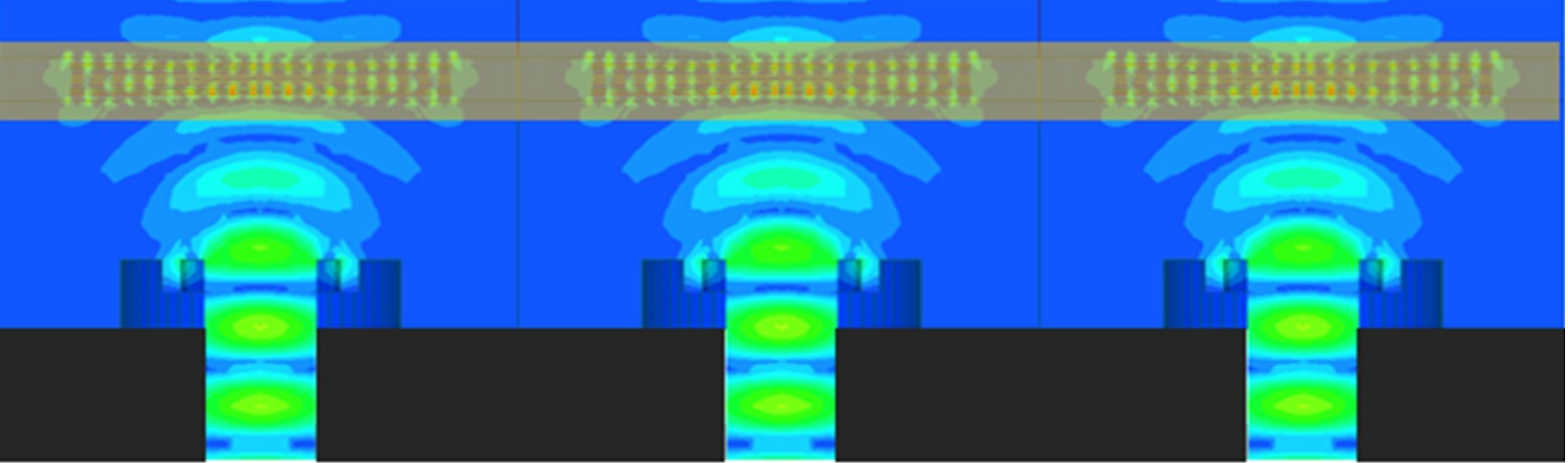}

\caption{\small 
\small 
Left Panel: Lenslet Array: fabrication detail. Right Panel: the action of planar lenslets on the beam radiated by a waveguide section.
}
\label{fig:Planar Lenslets}
\end{figure}

\noindent The beam-forming elements we will use to illuminate the telescope according to the above criteria are of two different kinds, depending on the frequency:

\begin{itemize}
\item At frequencies $\leq$ \SI{220}{GHz}, planar lenslets based on a metamaterial concept (see Figure \ref{fig:Planar Lenslets}) are a promising solution in terms of optical performance, and being based on the successful metal-mesh technology well consolidated for filter production, they rely on a solid know-how \cite{Pisano:13}. Since their thickness is comparable to $\lambda$ and their density is the one of polypropylene, their mass is slightly lower than 1 kg to cover all these bands. These planar lenslets are illuminated by a waveguide section which feeds a planar OMT for the Low Frequencies. 

\item The higher-frequencies (\SI{255}{GHz} and above) are based on antenna-coupled MKIDs, where the beam forming is achieved by means of a slot antenna endowed with a lenslet. These devices are fabricated on a monolithic Si substrate to which we mount, using a proven wafer bonding technique, a commercially available Si lens array fabricated using laser ablation. A parylene-C coating is to be applied to minimize reflection losses. Note that the lens-antenna coupling, including the AR coating, used for these MKIDs was used as well for the Herschel-HIFI band 5 and 6 mixers \cite{deLange08}. 
\end{itemize}



\section{Detectors}
\label{sec:detectors}
In order to minimize the complexity of the system, one single detector technology will be chosen to cover the full 60-\SI{600}{GHz} frequency range. Microwave Kinetic Inductance Detectors (MKIDs) are currently the most advanced solution at the European level in terms of TRL. They also have distinct advantages that make them stand out from other concurrent technologies. These include their intrinsic frequency domanin multiplexing (FDM) and the fast response time, not limited by thermal constraints. Furthermore, different tiles of MKIDs, for example containing the detectors of different sub-bands, can be fabricated separately and easily interconnected during the assembly of the FPA. This represents an additional advantage of the chosen technology, as it makes it possible to share the development and fabrication effort among different institutions.  MKIDs have therefore been adopted as the baseline detectors for \coremfive. To limit the radiative background on the resonators, from the telescope and its environment, high performance feed optics are used.

\subsection{Detection Technology}	
	
Over the last years, these detectors have been successfully used in many ground-based EU-lead experiments. The NIKA camera (\cite{nika1_a}~\cite{nika1_b}), installed at the IRAM \SI{30}{m} telescope in Spain, has been the first KID-based instrument to conduct on-the-sky observations. NIKA showed state-of-the-art performances using Aluminum (Al) Lumped Element KIDs (LEKIDs). Its follow-up instrument, NIKA2 (\cite{nika2_a}~\cite{nika2_b}), has a total of more than 3,000 detectors over two bands, covering the range 100-\SI{300}{GHz}, and is currently undergoing the final commissioning phases. Polarization sensitivity in the band centered at \SI{240}{GHz} is achieved using a wire-grid polarizer and two separate arrays, one for each polarization. At higher frequencies, the A-MKID\footnote{http://www3.mpifr-bonn.mpg.de/div/submmtech/bolometer/A-MKID/a-mkidmain.html} project is being commissioned at the focal plane of the APEX telescope, in Chile. This camera has a total of 25,000 KIDs split between two bands, centered at 350 and \SI{850}{GHz}. In parallel to these ground-based missions, the FP7 project  SPACEKIDS has been carried out, with the aim of optimizing KIDs for space applications. The results of this project, and of various laboratory measurements, confirm that MKIDs can meet the requirements of the \coremfive\ mission, resumed in Table \ref{table:COREdetsNeeds}. In particular, photon noise limited performance has been shown under optical loads representative of the mission~\cite{jannsen}~\cite{mauskopf_LTD15}. Furthermore, the tests conducted irradiating MKIDs with ionizing particles have demonstrated their very low susceptibility to Cosmic Rays hits~\cite{monfardinispace16}.

An overview of the current maturity level of MKIDs in the three main frequency ranges composing the FPA (Low Frequencies, CMB Frequencies, and High Frequencies) is given in the dedicated sections. The details of the design and materials to be adopted in each case will be determined by trade-off studies to be carried out during Phase A.

\begin{table}[htbp]
\begin{center}
\begin{tabular}{|l|c|c|c|c|}
\hline
 & Detector noise & Absorption efficiency & Yield & CR induced data loss \\
 \hline
 \coremfive\ goal &  5-\SI{30}{aW}/$\sqrt{\mathrm{Hz}}$ & $>$50$\,\%$ & $>$90$\,\%$ & $<$10$\,\%$ \\
 \hline
\end{tabular}
\caption{\small 
\small 
Summary of the main requirements in terms of performance for the \coremfive\ detectors. 
}
\label{table:COREdetsNeeds}
\end{center}
\end{table}





\subsection{Low Frequencies (channels: 60$\,$GHz - 115$\,$GHz):}
In MKIDs, the lowest energy that can be detected is determined by the superconducting gap of the material used for realizing the resonator. Thin Al films have proven to be the best choice for the frequency band between 110 and \SI{850}{GHz} (see NIKA2 on the IRAM \SI{30}{m} and AMKID camera on APEX), and have been undergoing intense developments over the last years. However, the use of Al results in a superconducting cut-off at around \SI{110}{GHz} and therefore is not suitable for the \coremfive\ bands covering the 60 to \SI{110}{GHz} frequency band. 

As a consequence, these bands are at present the less mature ones, and will need the strongest development effort during the phase A. In particular, new materials or configurations have to be adopted. The main route for achieving sensitivity in the low frequency bands will be the use of LEKIDs made of superconducting multilayers: the proximity effect\cite{Usadel1970} allows in this case to tune the critical temperature, and thus the cut-off frequency, by appropriately choosing the materials composing the multilayer and their thickness.  The limited number of pixels available for the LF bands also poses a challenge, as on-chip polarization sensitivity must be achieved, implying a further development effort.

\begin{figure}[!h]
\centering
\includegraphics[width=\textwidth, trim={0.1cm 0 0 0},clip]{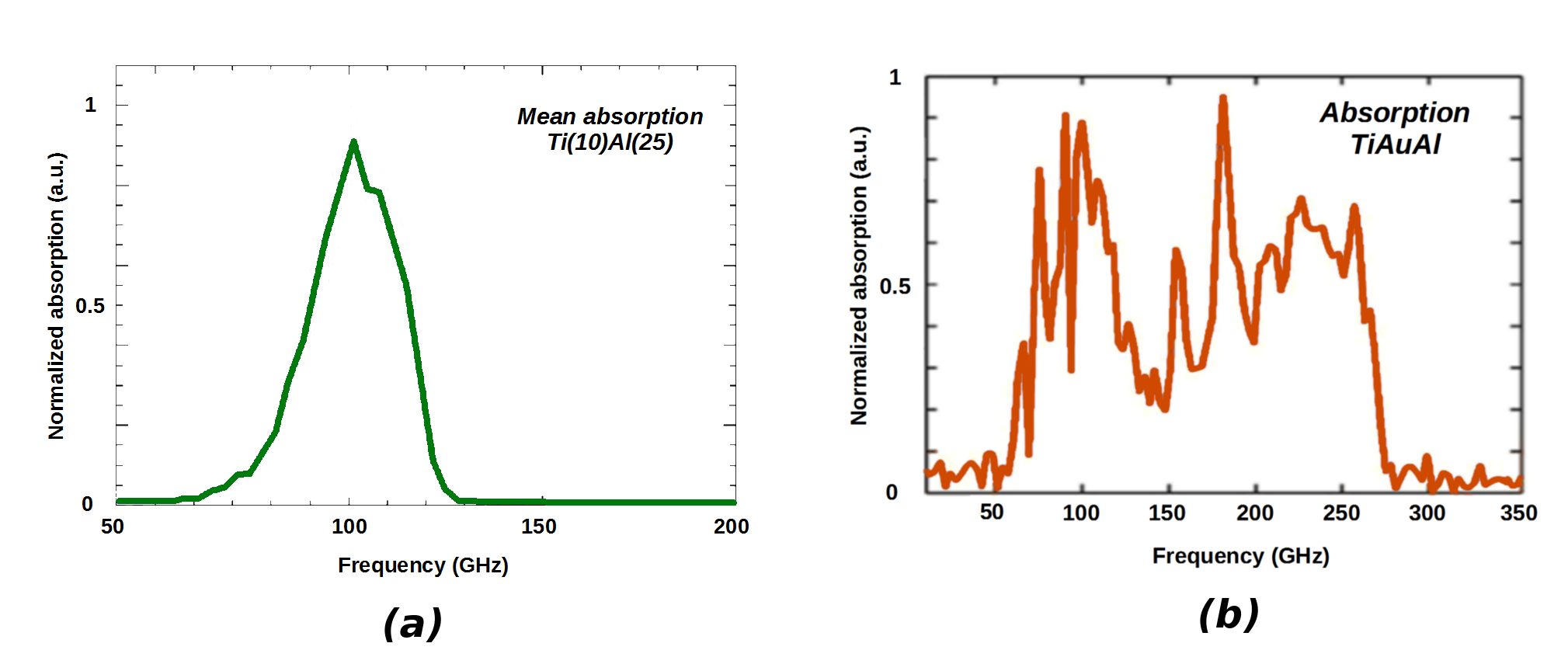}
\caption{\small 
\small
Left: absorption spectrum of LEKID realized using a Ti-Al bilayer, as reported in \cite{CatalanoGoupy}. These KID already demonstrated a good sensitivity in the band 80-\SI{120}{GHz}. Right: to reach even lower frequencies, different thicknesses or materials must be used. The preliminary measurements carried out using an Al-Ti-Au tri-layer  show for example a cutoff at \SI{60}{GHz} (credits: H. Le Sueur, A. Monfardini).
}
\label{fig:bilayer_trilayer}
\end{figure}

As far as the band coverage is concerned, the first measurements conducted on a Ti/Al bilayer have given promising results ~\cite{CatalanoGoupy}\cite{CatalanoSpace}. As shown in the left panel of figure~\ref{fig:bilayer_trilayer}, a simple \SI{10}{nm} Ti/\SI{25}{nm} Al bilayer already shows an absorption well adapted to the 80-\SI{120}{GHz} range \cite{CatalanoGoupy} \cite{Paiella2016}. In the measured array the mean pixel noise under a \SI{300}{fW} optical load was around \SI{20}{aW}/$\sqrt{\mathrm{Hz}}$, which is already within a factor 4 of the photon noise level expected at these frequencies. The end-to-end optical efficiency of the system, including the pixel absorption efficiency, was estimated to be above 30$\%$.

\begin{figure}[ht]
\centering
\includegraphics[width=\textwidth, trim={0.1cm 0 0 0},clip]{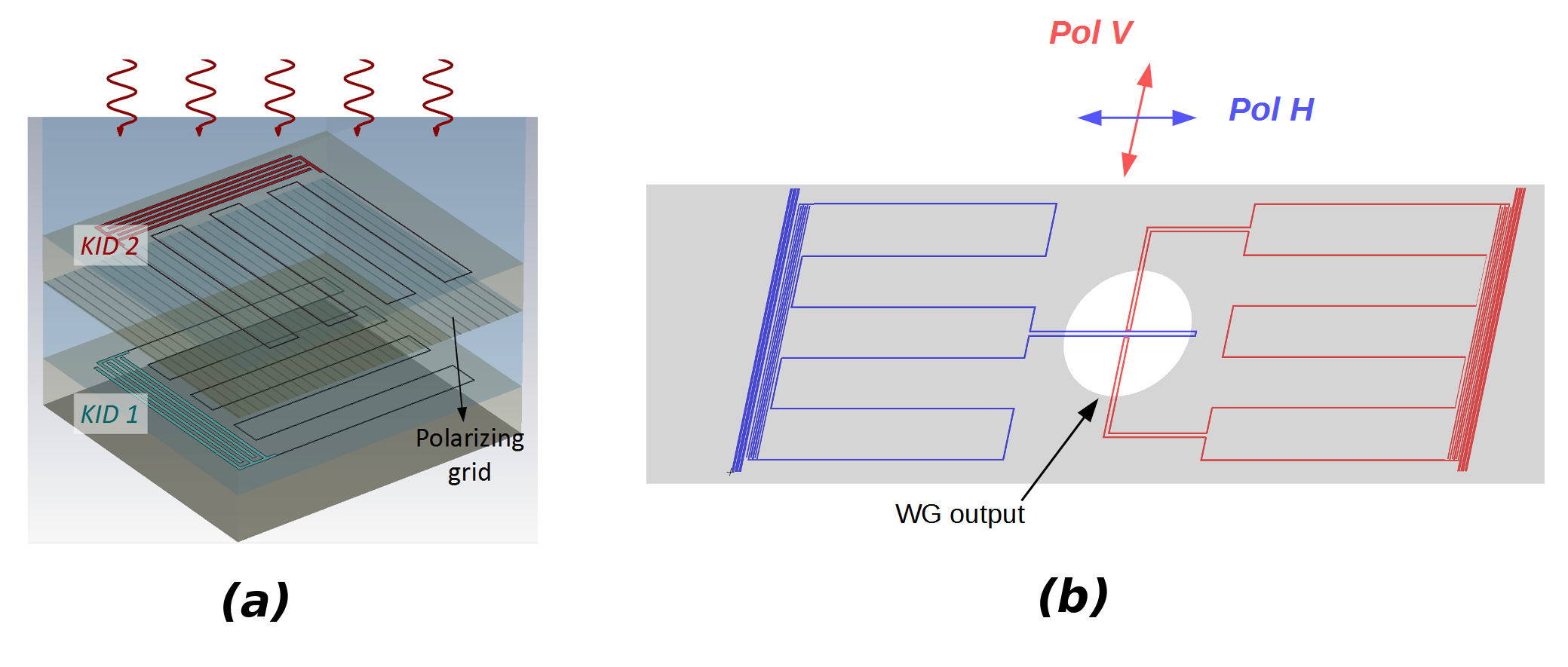}
\caption{\small 
\small
Comparison of two possible solution that can be adopted to achieve polarization sensitive KIDs. Left: a stack of two pixels separated by a wire grid. The first pixel is mainly sensitive to the polarization parallel to its meanders, but has also a considerable amount of cross polarization ($\sim$10$\%$). This effect can be accounted for and corrected thanks to the second pixel which, lying behind a wire-grid polarizer, is sensitive only to the polarization orthogonal to the grid. Right: two LEKIDs used in a planar OMT like configuration. The meanders extend to go sense the two polarization conveyed by an appropriate waveguide. This kind of approach has been proposed for example in \cite{doi:10.1117/12.2231830} and is expected to give very good results in terms of polarization purity. On the other hand, its need for a waveguide just in front of the pixels poses stricter requirements on the design of the radiation coupling section.
}
\label{fig:BiKID_vs_OMT-KID}
\end{figure}

Different paths are open in order to meet the \coremfive\ requirements and will be investigated during phase A:

\begin{itemize}
\item \emph{Band Coverage}: to obtain lower cut-off frequencies and thus be able to cover the full \coremfive\ LF bands, different Ti/Al thicknesses, as well as trilayers such as Ti/Au/Al, will be explored. The latter has already demonstrated a cut-off at around \SI{60}{GHz}, as required by \coremfive\ (right panel of Figure~\ref{fig:bilayer_trilayer}). We note here that the decrease in the Tc of the LEKID is associated with a decrease of their working temperature, and these detectors will most probably need to be operated at around \SI{120}{mK} or below. However, it is worth noting that lower Tc will also result in an increase in responsivity. 

\item \emph{Detector Noise}: the values reported in \cite{CatalanoGoupy} have been obtained using pixels that had not been conceived for low optical backgrounds. Furthermore, also the optical configuration was sub-optimal. Thus, an optimization of the pixel design (for example in terms of film thickness and coupling to the readout line) and the maximization of the optical coupling should easily allow to fill the gap between the current performance and the desired goal.

\item \emph{Polarization Sensitivity}: To obtain on-chip polarization sensitivity, different solutions will be tested. These include: \emph{a)} a bi-array structure, in which two arrays are superpoposed and separated by a polarizing grid. The first array will be sensitive to one polarization and the second to the orthogonal one (figure~\ref{fig:BiKID_vs_OMT-KID}, left). \emph{b)} On-wafer planar Ortho-Mode Transducer (OMT) designs coupled to LEKIDs, to benefit from their very low expected cross polarization level (Figure~\ref{fig:BiKID_vs_OMT-KID}, right). \emph{c)} Optical separation of the two polarizations thanks to the development of dedicated multi-mesh flat lenses, capable of focalizing the two orthogonal polarization on different spots of the focal plane.  

\end{itemize}

\subsection{CMB Frequencies (channels: 130 GHz - 220 GHz):}
The frequencies near the peak of the CMB polarization can be detected exploiting Aluminum LEKIDs, a technology consolidated by the developments carried out in several laboratories in Europe and already demonstrated on the field thanks to the the NIKA and NIKA2 instruments, installed at the IRAM 30-m telescope \cite{nika1_a}\cite{nika2_a}. The pixel design adopted for NIKA2 ~\cite{goupy} is based on a particular implementation of the LEKID, in which the meandering inductor has an Hilbert fractal shape ~\cite{roesch}. This solution is ideally suited for full-power measurements, as the Hilbert antenna efficiently absorbs both polarizations.

The NIKA2 pixels have shown excellent performance for the typical conditions of a ground-based experiment, and are already photon-noise limited under medium optical backgrounds \cite{mauskopf_LTD15}. The very same pixels, operated under optical loads representative of a CMB mission ($\sim$500 fW/pixel), have shown a mean noise level of 30$\,$aW/$\sqrt{\mathrm{Hz}}$ and an end-to-end optical efficiency of 30$\%$ ~\cite{CatalanoSpace}. As in the case of the low frequency bands, the noise level is already within a factor $\sim$4 of the \coremfive\ requirements, and an important margin for improvement is granted by the different design parameters that can be optimized to account for the lower optical load.

\begin{figure}[!h]
\centering
\includegraphics[width=0.9\textwidth]{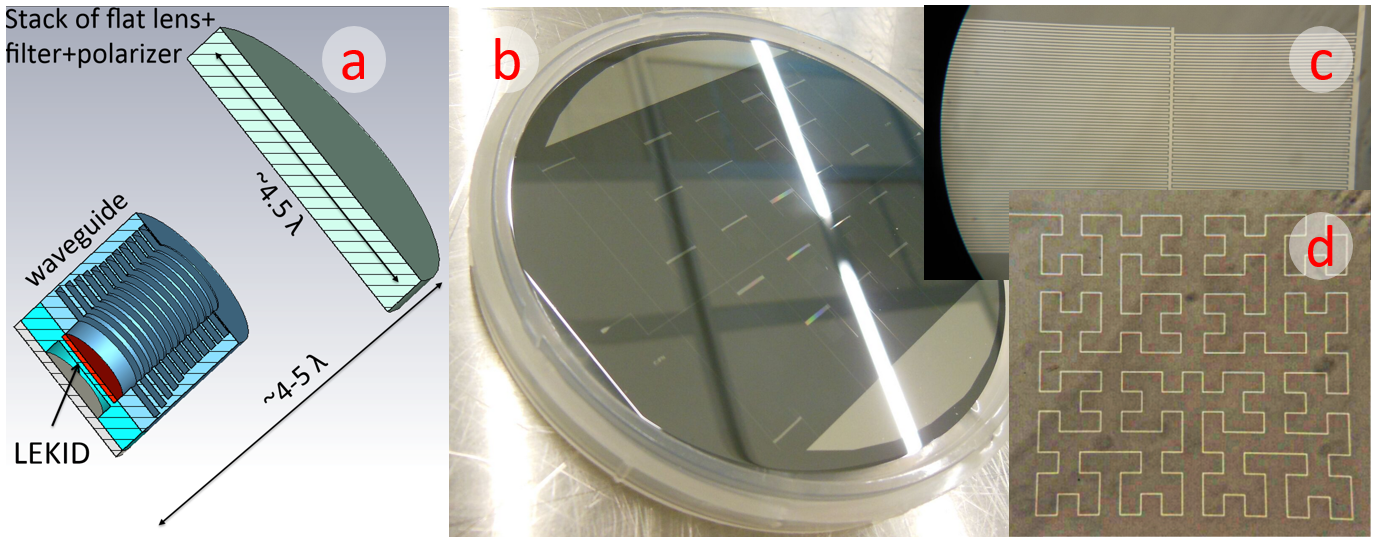}
\caption {\small a) Light-weight coupling optics proposed for the \coremfive\ focal plane. A waveguide selecting the HE11 mode is used to transfer on the detector absorber radiation focused by a flat metal mesh lenslet located in the focal plane. All the resonators at the same frequency are litographed on the same wafer; the lenslets, the polarizers and the band-defining filters are also produced on a continuous layer covering the entire array. b) Sample wafer of 150$\,$GHz LEKIDs produced in Rome (IFN-CNR \& Sapienza) c) detail of the capacitive part of the resonator d) detail of the inductor/absorber part of the resonator.  
}
\label{fig:CMB-f}
\end{figure}

In the implementation foreseen for the \coremfive\ mission, the resonators are coupled to incoming radiation through a combination of a plastic-embedded-mesh flat lenses (in the focal plane) and a short section of a waveguide, opening in a cavity surrounding the resonator (see Figure~\ref{fig:CMB-f}). This arrangement allows for efficient coupling with a lighter optical system with respect to horn-coupled LEKIDs. A polarization sensitivity upgrade can be envisaged by means of a layer of plastic-embedded metal-wire grids located in the focal plane, stacked on the band-defining embedded-mesh filters layer and on the flat lenses layer.

\begin{figure}[h]
\centering
\includegraphics[width=1\textwidth]{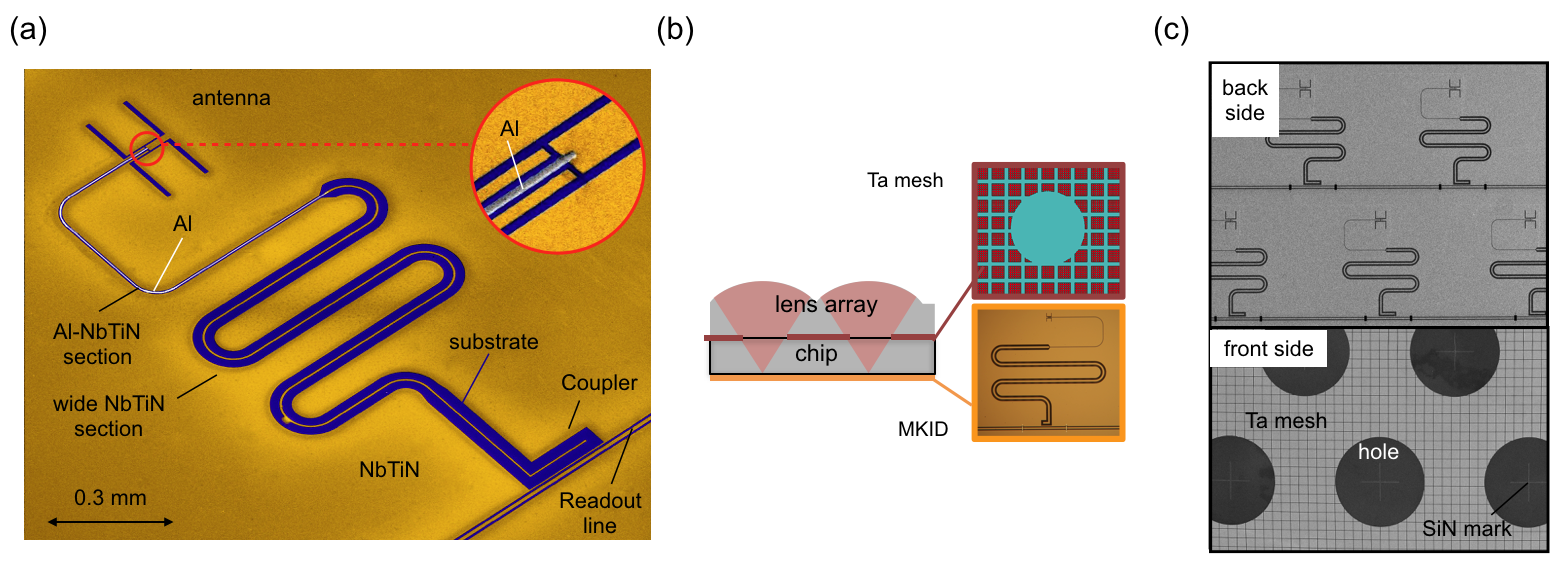}
\caption{\small (a) Scanning electron micrograph of a single hybrid lens-antenna coupled MKID, with its important structures labelled.  (b) Cross-sectional drawing of the MKID chip-lens array assembly. The backside of the chip contains the detectors, mounted to the chip frontside is a monolithical array of microlenses, made from Si. The wafer frontside contains a Beta-phase Ta mesh layer, designed to absorb stray radiation,. Holes co-aligned with the lenses allow the antenna beams to couple efficiently. (c) Front and backside image of the detector wafer prior to assembly, showing the mesh layer, Ta mesh and alignment marks.}
\label{fig:HighF1}
\end{figure}

\begin{figure}[h]
\centering
\includegraphics[width=1\textwidth]{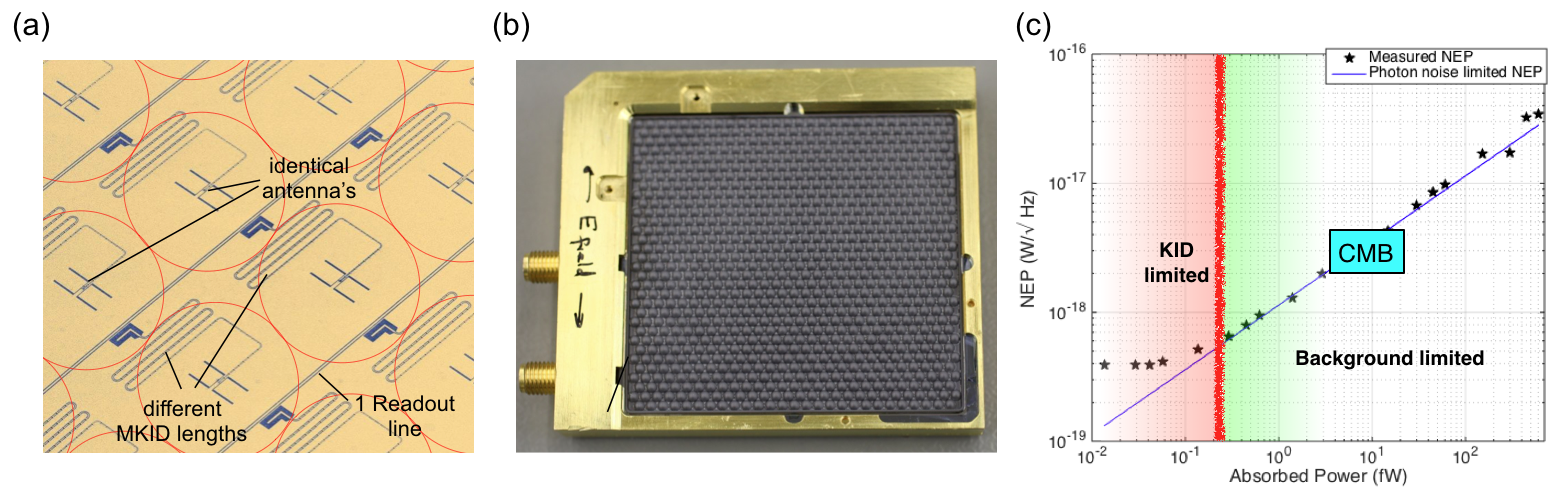}
\caption{\small (a) Scanning electron micrograph of a subset of a large array of lens-coupled hybrid MKIDs.  (b) Assembly of a 350 GHz imaging array of 880 detectors coupled to 2mm diameter lenses. (c) Measured performance of a single pixel as a function of power absorbed in the detector, the blue line is the background limited performance given by the photon noise of the source, the black points the measured data. \coremfive \ will operate around 100 fW/pixel, firmly in the region where the device performance is fully background limited.}
\label{fig:HighF2}
\end{figure}

\subsection{High Frequencies (channels: 255 GHz - 600 GHz):}
The high frequency detectors for \coremfive\ are envisioned to be made from lens-antenna coupled hybrid MKIDs. The device consists of a meandering coplanar waveguide (CPW) $\lambda/4$ resonator made of NbTiN on a high resistivity  Si substrate. The resonator has a wide section coupled weakly to a common readout line and a narrow section in which the central line is made of aluminum, which is shorted to ground at the resonator end. The length of the resonator sets the readout frequency of the MKID, the radiation coupled to the device is determined by the antenna located at the shorted end of the resonator. Radiation coupled to the antenna is injected in the Al-NbTiN narrow line, where it is absorbed in the aluminum at frequencies exceeding its gap frequency ($\nu >$\SI{90}{GHz}). There are no losses in the NbTiN ground plane since NbTiN has a gap frequency of \SI{1}{THz}, the use of NbTiN and Al therefore makes for an excellent, loss-free device at frequencies between \SI{90}{GHz} and \SI{1}{THz}. An image of a single hybrid MKID, with an antenna optimized for single polarization radiation coupling around \SI{350}{GHz}, is shown in Figure~\ref{fig:HighF1}(a). For CORE we envisage to use a monolithic array for each frequency band, shown schematically in Figure~\ref{fig:HighF2}(a). To allow efficient radiation coupling the detector array is coupled to a Si lens array using a well-tested bonding process. The cross section of this assembly is schematically shown in Figure~\ref{fig:HighF1}(b). Alignment is done using alignment marks etched in the SiN as shown in panel (c). Note that we use at the wafer frontside, i.e. in between the lens array and the detectors, a mesh layer of Beta-phase Ta, with a gap frequency of $\sim$\SI{50}{GHz}. This layer has two important functions: it reduces cosmic ray effects and absorbs stray radiation in the detector wafer. A 880 pixel assembly is shown in Figure~\ref{fig:HighF2}(b). These devices have shown background limited performance in combination with a high radiation coupling efficiency using a single-polarized, narrow-band twin slot antenna at \SI{350}{GHz}, see \cite{Janssen13}, and at \SI{850}{GHz}, see \cite{Baselmans16}. We reproduce in Figure~\ref{fig:HighF2}(c) the result from \cite{Janssen13}, making clear that we are comfortably background limited at \SI{350}{GHz} at the power levels expected for the instrument.

To adapt the current technology for the high frequency bands of \coremfive \ only very limited engineering effort is required. i) We need to make detectors for each frequency band, which requires a limited extrapolation to \SI{255}{GHz} from the demonstrated systems (operating at \SI{350}{GHz} and \SI{850}{GHz}) and interpolation for all other bands. ii) We need to optimize the lens-antenna design for minium spillover and maximum coupling, similar as done for the band 6 of Herschel-HIFI, which has used the same lens-antenna design principle as we will do. iii) We need to make lens arrays with larger lenses. Options here are laser machining, as we currently do, by an external commercial partner (Veldlaser) or mechanical machining using high speed diamond machining. It is to be noted that all frequency bands of \coremfive \ require significantly less pixels than demonstrated. In Ref. \cite{Baselmans16} we show that it is possible to multiplex up to 960 pixels using a single readout circuit using only a single cryogenic HEMT amplifier, which is Herschel-HIFI heritage, operated at \SI{4}{K} \cite{Lopez03}. On top of that we demonstrated that many relevant issues for space operation, such as yield, sensitivity, dynamic range and susceptibility to ionizing radiation, are very well under control.


\subsection{KID Susceptibility to Cosmic Rays}

Kinetic Inductance Detectors are pair-braking detectors, and are therefore insensitive to quanta of energy smaller than the superconducting gap 2$\Delta = 3.5 k_bT_c$. Thus, only photons or particles of energy larger than 2$\Delta$ can generate a measurable signal in the KID. This represents a key advantage with respect to bolometric detectors when it comes to the data loss induced by Cosmic Rays impacts. When a high energy particle hits a dielectric wafer, it generates a shower of high energy phonons, which then rapidly thermalize in the wafer. The KID are sensitive only to the first, non-thermal part of the event, while it is not effected by the final, thermal tail of the event during which the energy of the impact is released to the cold bath. Thus, the temporal evolution of a CR induced glitch in a KID is sensibly faster than the corresponding glitch in a bolometer. The recovery time of a KID is in fact of order of \SI{1}{ms} or less, depending on the details of the pixel and the material used. In a bolometer, the typical values are of hundreds of ms or longer, as outlined by the Planck experiment.

To further mitigate the data loss due to CR impacts, a solution has already been proposed \cite{monfardinispace16}. This is based on the deposition, on the same wafer where the KID arrays are fabricated, of layers of a superconducting material having a critical temperature lower than the one of the material used for the detectors themselves. This additional superconducting layer efficiently absorbs the energetic non-thermal phonons. The phonons that are then re-emitted have an energy just above the gap of this superconductor, and are therefore unable to induce a signal in the KIDs. This phonon mopping technique is very effective at limiting the area of the KID array that is affected by each event, and at shortening the timescale of each glitch. The typical CR induced data loss in KID arrays hardened this way has been estimated to be of only a few $\%$ ~\cite{CatalanoSpace}\cite{Baselmans16}, and thus fully compatible with the requirements of the \coremfive\ mission.

\subsection{Readout Electronics}
The readout system on board \coremfive\ will be based upon a space-qualified
version of the existing readout systems in use for: a) the NIKA2
instrument by Grenoble \cite{bourrion/etal:prep}; b) the system
developed by SRON for the SpaceKIDs project
\cite{vanrantwijk/etal:2016}; and c) the readout for the
A-MKID 
instrument by the Max Planck Institute for Radio Astronomy in Bonn. A
multiplexing factor of 1000 has already been achieved in the latter two systems,
and the former (based on the NIKEL readout using the DDC algorithm) allows for around 600. Any
proposed readout is independent of the exact KID architecture (absorber
or antenna coupled), and thus the particular KID architecture adopted
will not influence the readout electronics scheme.

\begin{figure}[!h]
\centering
\includegraphics[width=1.\textwidth]{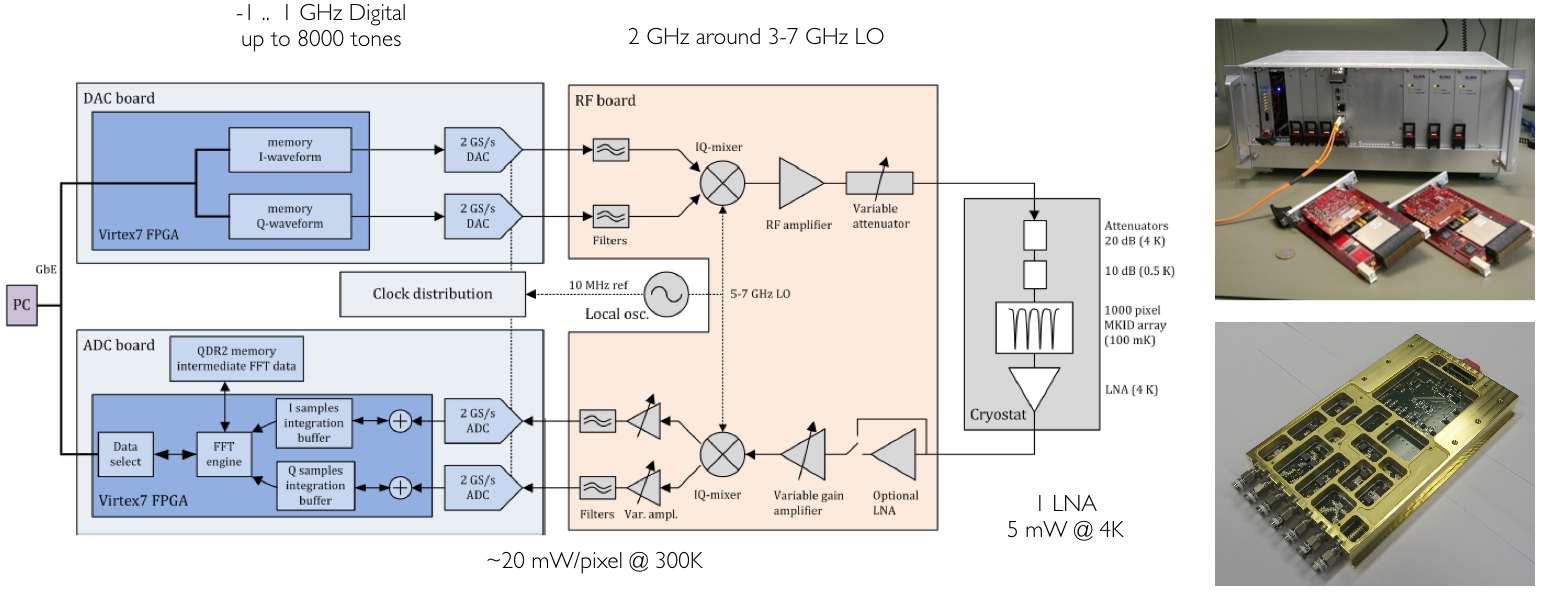}
\caption{\small \baselineskip=-0cm
\small 
Overview of the baseline readout system (left) and images of the ADC card (top right) and analog board (bottom right).}
\label{fig:readout}
\vspace{0.2cm}
\end{figure}

The baseline system architecture consists of the following components
(see figure~\ref{fig:readout}):
A digital carrier board, which can be loaded with the complex waveforms
to read-out up to 1000 MKIDs at the baseband frequency of 0-1$\,$GHz. The
complex waveform is converted to the analog world by two DAC's. The
signal is upconverted from the baseband to MKID frequencies in the analog
board using IQ upconversion and a separate LO, and the result is a
readout band of \SI{2}{GHz} around the LO frequency. The signal is
subsequently attenuated/amplified to the correct power level for the
MKID array. After being altered by the array, the signal is amplified by
a single LNA at \SI{15}{K} \cite{lopez-fernandez:2003} and down-converted back
to the baseband. We propose the Herchel-HIFI heritage TRL 9
Yebes 2-\SI{4}{GHz} LNA, requiring \SI{5.5}{mW} per amplifier.
The demodulation board demodulates the complex sampled
signal using a FFT and bin-selection algorithm. Depending on the exact
system design, QDR2 memory is needed in this board. Control of the
system is done using a dedicated Leon3 control processor, which controls
the digital boards, the clocks, and the RF attenuator. The clocks for
both the converters and the mixers are generated on a dedicated clocking
board, both of which are locked to the same reference source.

The estimated power consumption and mass per readout chain are
\SI{52}{W} and \SI{4.5}{kg}, respectively. Each chain can readout up to
1000 MKIDs. With a margin we will need 3 readout chains, and thus
the required total power and mass are \SI{156}{W} and \SI{13.5}{kg},
respectively. 
In this design, there is a margin but no redundancy. The
best option would be to make the system fully redundant, which would
require a doubling in mass and volume to be able to have a full readout
system as backup. On top of this it might be considered to make also the
detector system redundant, which would double the cold power dissipation
by the LNAs.

The entire readout system proposed here can be made from space-qualified
components, most importantly the VIRTEX-5 FPGA. A qualification program is not needed on the component level, but only for the integrated system.
\vspace{1.2cm}

\section{Cryogenic System}
\label{sec:cryo}

The sensitivity of \coremfive \ detectors should be limited by the intrinsic photon noise of the sky emission for diffraction-limited detectors observing in broad bands of $\Delta \nu / \nu \sim 0.3$. In the 60-\SI{600}{GHz} frequency range this corresponds to minimum background loads in the range of 100-\SI{200}{fW}. Achieving this sensitivity requires continuously cooling the detector array to $\sim$ \SI{100}{mK}, and the mirrors down to \SI{40}{K}-\SI{100}{K}. The CORE cryogenic chain leverages on the ultra-low temperature space missions heritage (Planck \cite{cryo1}, Hitomi) and ongoing developments (MIRI/JWST, XIFU/Athena) to succeed.

The \coremfive \ cryogenic architecture minimizes the development risk for a continuous low temperature cooling chain, and is open to potential evolutions of the instrument needs or configurations during the phase-A study. 

The cooling system implementation is based on the following assumptions:

\begin{itemize}

\item The system should work continuously for at least 5 years in space, and cool the focal plane at 100mK and the telescope at 40-100K.

\item All active coolers shall be redundant, except for the \SI{100}{mK} stage.

\item For the cooling power, a margin of more than 25\% at all temperature stages is required, except for the \SI{100}{mK} stage, where we require 100\% margin. 

\item Coolers from Europe shall be baselined.

\end{itemize}

\smallskip

\begin{figure}[ht]
\centering
\includegraphics[width=\textwidth]{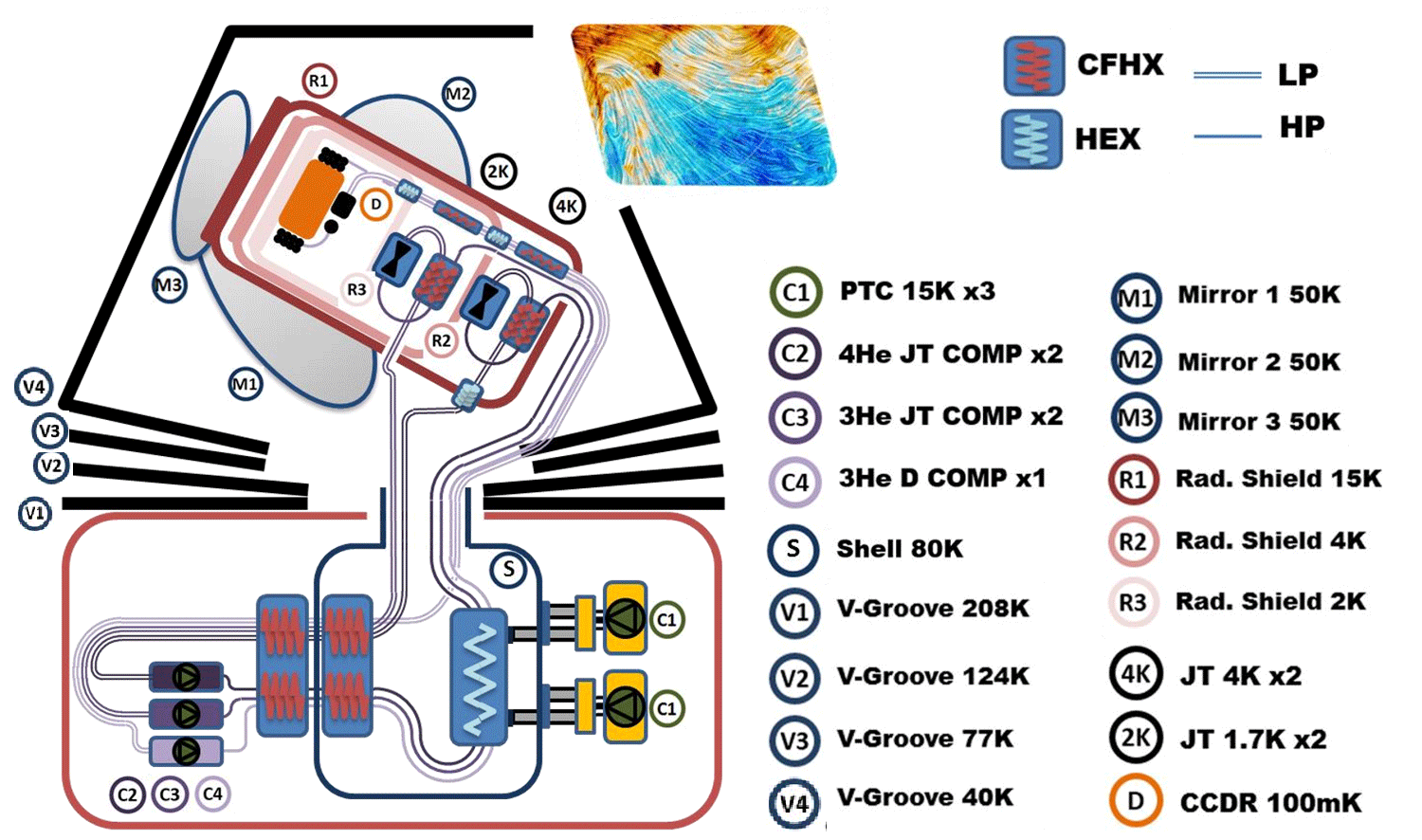}
\caption{\small 
\small 
Overall configuration of the cryogenic system for \coremfive \ .
}
\label{fig:overall-cryo}
\end{figure}

The cryo-system is basically divided into 3 sub-systems, as visible in Figure~\ref{fig:overall-cryo} (\cite{cryo2}). 

\begin{itemize}

\item The first sub-system is located in the service module (SVM) and includes all the compressors for the active coolers (described below) in order to limit micro-vibrations and EMI issues at payload module (PLM) level. It also includes a small cryostat thermally anchored to the pulse tube coolers (PTC). 

\item The second sub-system is the SVM to PLM connection. It consists in a set of V-Grooves and insulated lines to transfer $^4$He and $^3$He of the coolers from the SVM, where they are pre-cooled, to the PLM. This passive cooling system is made possible by the large surface area of the instrument baffle.  With our telescope configuration, V-Grooves similar to the ones proposed for the Planck mission \cite{cryo1}, or the SPICA \cite{cryo10} and Ariel studies, can be re-proposed. Given the impact of the scan strategy on the requirements for these V-grooves, their optimization is discussed in detail in the mission companion paper \cite{missionpaper}. Here we assume that a baseline temperature of 40-100 K is produced by this subsystem. 

\item The last sub-system consists of a series of concentric shells, going from $\sim$ 18K down to $\sim$ 1.7K, in order to limit the thermal loads on the lower temperature stages. It also includes the 0.1K focal plane unit cooled by the low temperature stages of the closed-cycle dilution refrigerator. The cool-down to around 18K of the PLM external envelope (R1) is achieved by using remote cooling from the SVM, with a concept similar to the one developed for MIRI in JWST.

\end{itemize}

With this configuration, the baseline system provides \SI{2.3}{\micro\watt} continuous at \SI{100}{mK} (for a requirement of \SI{1.1}{\micro\watt}). The total mass is $\sim$\SI{210}{kg} (excluding V-Grooves) with a total electrical power required $\sim$\SI{1300}{W}, including thermal control. The expected lifetime is 5 years, potentially extendable to 10 years. The budgets for this cryogenic system are summarized in Table~\ref{tab:CRYO-TABLE} below.


\begin{sidewaystable}[h]
\centering
\resizebox{\textwidth}{!}{%
\begin{tabular}{|lcclcclllclllll|lc|}
\hline
\multicolumn{17}{|l|}{\textbf{CORE Cryogenics: Engineering Budget}}                                                                                                                                                                                                                                                                                                                                                                                                                                                                                                                                                                                                                                                                                                                              \\ \hline
\multicolumn{4}{|l|}{}                                                                                                            & \multicolumn{2}{l}{Type of Cooler}                                                                 & PTC                                                      & JT \SI{4}{K}                                                   & JT \SI{2}{K}                                                   & \multicolumn{1}{l}{CCDR}                    & \multicolumn{5}{l}{}                                                                                                                                                                                                                                                  &                            & \multicolumn{1}{l|}{} \\
\multicolumn{4}{|l|}{}                                                                                                            & \multicolumn{2}{l}{Number of coolers}                                                              & \multicolumn{1}{c}{3}                                    & \multicolumn{1}{c}{2}                                   & \multicolumn{1}{c}{2}                                   & 1                                           & \multicolumn{5}{l}{}                                                                                                                                                                                                                                                  & \multicolumn{2}{c|}{\textbf{STAGES MASS}}          \\
\multicolumn{4}{|l|}{\multirow{-3}{*}{}}                                                                                          & \multicolumn{2}{l}{Nb of ON coolers}                                                               & \multicolumn{1}{c}{2}                                    & \multicolumn{1}{c}{1}                                   & \multicolumn{1}{c}{1}                                   & 1                                           & \multicolumn{5}{l}{\multirow{-3}{*}{}}                                                                                                                                                                                                                                &                            & \multicolumn{1}{l|}{} \\ \hline
                               & \multicolumn{1}{l}{} & \multicolumn{1}{l}{}                        & \multicolumn{1}{c}{Planck}  & \multicolumn{1}{l}{}                        & \multicolumn{1}{l}{}                                 &                                                          &                                                         &                                                         & \multicolumn{1}{l}{}                        & \multicolumn{1}{c}{OFF}                                  & \multicolumn{1}{c}{Detec}                                & \multicolumn{1}{c}{TOTAL}                                 &                                 & \multicolumn{1}{c|}{Resulting}                    & \multicolumn{1}{c}{Per}    & \multicolumn{1}{l|}{} \\
\multicolumn{1}{|c}{Reference} & Component            & T (K)                                       & \multicolumn{1}{c}{Cond.}   & Cond.                                       & Rad. In.                                             &                                                          &                                                         &                                                         & \multicolumn{1}{l}{}                        & Cooler                                                   & \multicolumn{1}{c}{tors}                                 & \multicolumn{1}{c}{Loss}                                  &                                 & \multicolumn{1}{c|}{Margins}                      & \multicolumn{1}{c}{Item}   & Cumulated             \\ \hline
                               & VG 1                 & {\color[HTML]{3531FF} \textbf{208}}         &                             & {\color[HTML]{FE0000} \textbf{32.8}}        & {\color[HTML]{FE0000} \textbf{40.7}}                 &                                                          &                                                         &                                                         & \multicolumn{1}{l}{}                        &                                                          &                                                          &                                                           & \multicolumn{1}{c}{\textbf{W}}  &                                                   &                            & \multicolumn{1}{l|}{} \\
\textbf{Passive}               & VG 2                 & {\color[HTML]{3531FF} \textbf{124}}         &                             & {\color[HTML]{FE0000} \textbf{4.2}}         & {\color[HTML]{FE0000} \textbf{12.4}}                 &                                                          &                                                         &                                                         & \multicolumn{1}{l}{}                        &                                                          &                                                          &                                                           & \multicolumn{1}{c}{\textbf{W}}  &                                                   & \multicolumn{1}{r}{29.9}   & \textbf{119 kg}       \\
\textbf{Cooling}               & VG 3                 & {\color[HTML]{3531FF} \textbf{77}}          &                             & {\color[HTML]{FE0000} \textbf{1.4}}         & {\color[HTML]{FE0000} \textbf{1.4}}                  &                                                          &                                                         &                                                         & \multicolumn{1}{l}{}                        &                                                          &                                                          &                                                           & \multicolumn{1}{c}{\textbf{W}}  &                                                   & \multicolumn{1}{r}{29.4}   & \multicolumn{1}{l|}{} \\
                               & VG 4 + Baffle        & {\color[HTML]{3531FF} \textbf{40,1}}        &                             & {\color[HTML]{FE0000} \textbf{0.6}}         & {\color[HTML]{FE0000} \textbf{0.6}}                  &                                                          &                                                         &                                                         & \multicolumn{1}{l}{}                        &                                                          &                                                          &                                                           & \multicolumn{1}{c}{\textbf{W}}  &                                                   & \multicolumn{1}{r}{59.3}   & \multicolumn{1}{l|}{} \\ \hline
\textbf{\SI{80}{K} Stage}             & 5 Shells             & {\color[HTML]{3531FF} \textbf{82}}          &                             & {\color[HTML]{FE0000} \textbf{120}}         & {\color[HTML]{FE0000} \textbf{475}}                  & \multicolumn{1}{c}{{\color[HTML]{00009B} \textbf{4059}}} & \multicolumn{1}{c}{{\color[HTML]{F56B00} \textbf{276}}} & \multicolumn{1}{c}{{\color[HTML]{F56B00} \textbf{200}}} & {\color[HTML]{F56B00} \textbf{0}}           & \multicolumn{1}{c}{{\color[HTML]{F56B00} \textbf{2052}}} & \multicolumn{1}{c}{{\color[HTML]{F56B00} \textbf{0}}}    & \multicolumn{1}{c}{{\color[HTML]{FE0000} \textbf{3123}}}  & \multicolumn{1}{c}{\textbf{mW}} & \multicolumn{1}{c|}{{\color[HTML]{009901} 30\%}}  & \multicolumn{1}{c}{+20\%}  & +20\%                 \\
PTC                            &                      & \multicolumn{1}{l}{{\color[HTML]{3531FF} }} &                             & \multicolumn{1}{l}{{\color[HTML]{FE0000} }} & \multicolumn{1}{l}{{\color[HTML]{FE0000} \textbf{}}} & {\color[HTML]{00009B} M=10\%}                            & \multicolumn{1}{c}{{\color[HTML]{F56B00} M=20\%}}       & \multicolumn{1}{c}{{\color[HTML]{F56B00} M=100\%}}      & \multicolumn{1}{l}{{\color[HTML]{F56B00} }} & \multicolumn{1}{c}{1 PT + 2 JT}                          &                                                          & {\color[HTML]{FE0000} }                                   &                                 & {\color[HTML]{009901} }                           & \multicolumn{1}{c}{Margin} & Margin                \\ \hline
\textbf{\SI{15}{K} Stage}             & PTC                  & {\color[HTML]{3531FF} \textbf{15}}          & \SI{39.2}{mW}                     & {\color[HTML]{FE0000} \textbf{12}}          & {\color[HTML]{FE0000} \textbf{60}}                   & \multicolumn{1}{c}{{\color[HTML]{00009B} \textbf{783}}}  & \multicolumn{1}{c}{{\color[HTML]{F56B00} \textbf{114}}} & \multicolumn{1}{c}{{\color[HTML]{F56B00} \textbf{90}}}  & {\color[HTML]{F56B00} \textbf{9}}           & \multicolumn{1}{c}{{\color[HTML]{F56B00} \textbf{188}}}  & \multicolumn{1}{c}{{\color[HTML]{F56B00} \textbf{30}}}   & \multicolumn{1}{c}{{\color[HTML]{FE0000} \textbf{503}}}   & \multicolumn{1}{c}{\textbf{mW}} & \multicolumn{1}{c|}{{\color[HTML]{009901} 56\%}}  & \multicolumn{1}{c}{10.4}   & \textbf{\SI{29.9}{kg}}      \\
PTC                            & + R1 Shield          & \multicolumn{1}{l}{{\color[HTML]{3531FF} }} & \multicolumn{1}{c}{\SI{97.5}{kg}} & Mass x 0.31                                 & From \SI{50}{K}                                             & \multicolumn{1}{c}{{\color[HTML]{00009B} M=10\%}}        & \multicolumn{1}{c}{{\color[HTML]{F56B00} M=20\%}}       & \multicolumn{1}{c}{{\color[HTML]{F56B00} M=100\%}}      & {\color[HTML]{F56B00} M=50\%}               & \multicolumn{1}{c}{1 PT + 2 JT}                          &                                                          & {\color[HTML]{FE0000} }                                   &                                 & {\color[HTML]{009901} }                           &                            & \multicolumn{1}{l|}{} \\ \hline
\textbf{\SI{4}{K} Stage}              & \SI{4}{K} JTC               & {\color[HTML]{3531FF} \textbf{4.5}}         & \SI{5.7}{mW}                      & {\color[HTML]{FE0000} \textbf{14.5}}        & {\color[HTML]{FE0000} \textbf{2.30}}                 &                                                          & \multicolumn{1}{c}{{\color[HTML]{00009B} \textbf{24}}}  &                                                         & {\color[HTML]{F56B00} \textbf{2.25}}        & \multicolumn{1}{c}{{\color[HTML]{F56B00} \textbf{0.02}}} & \multicolumn{1}{c}{{\color[HTML]{F56B00} \textbf{0.19}}} & \multicolumn{1}{c}{{\color[HTML]{FE0000} \textbf{19.22}}} & \multicolumn{1}{c}{\textbf{mW}} & \multicolumn{1}{c|}{{\color[HTML]{009901} 25\%}}  & \multicolumn{1}{c}{6.9}    & \textbf{\SI{19.5}{kg}}      \\
JT                             & + R2 Shield          & \multicolumn{1}{l}{{\color[HTML]{3531FF} }} & \multicolumn{1}{c}{7.63 kg} & Mass x 2.5                                  & From \SI{20}{K}                                             &                                                          & {\color[HTML]{00009B} M=20\%}                           &                                                         & {\color[HTML]{F56B00} M=50\%}               & {\color[HTML]{F56B00} }                                  & {\color[HTML]{F56B00} }                                  & {\color[HTML]{FE0000} }                                   &                                 & {\color[HTML]{009901} }                           &                            & \multicolumn{1}{l|}{} \\ \hline
\textbf{\SI{1.7}{K} Stage}            & \SI{1.7}{K} JTC             & {\color[HTML]{3531FF} \textbf{1.65}}        & \multicolumn{1}{c}{\SI{0.3}{mW}}  & {\color[HTML]{FE0000} \textbf{1.26}}        & {\color[HTML]{FE0000} \textbf{0.13}}                 &                                                          &                                                         & \multicolumn{1}{c}{{\color[HTML]{00009B} \textbf{10}}}  & {\color[HTML]{F56B00} \textbf{5.25}}        & \multicolumn{1}{c}{{\color[HTML]{F56B00} \textbf{0.02}}} & \multicolumn{1}{c}{{\color[HTML]{F56B00} \textbf{0.08}}} & \multicolumn{1}{c}{{\color[HTML]{FE0000} \textbf{6.72}}}  & \multicolumn{1}{c}{\textbf{mW}} & \multicolumn{1}{c|}{{\color[HTML]{009901} 49\%}}  & \multicolumn{1}{c}{4.6}    & \textbf{\SI{12.6}{kg}}      \\
JT                             & + R3 Shield          & \multicolumn{1}{l}{{\color[HTML]{3531FF} }} & \multicolumn{1}{c}{\SI{3.01}{kg}} & Mass x 4.2                                  & From R2                                              &                                                          &                                                         & \multicolumn{1}{c}{{\color[HTML]{00009B} M=33\%}}       & {\color[HTML]{F56B00} M=50\%}               & {\color[HTML]{F56B00} }                                  & {\color[HTML]{F56B00} }                                  & {\color[HTML]{FE0000} }                                   &                                 & {\color[HTML]{009901} }                           &                            & \multicolumn{1}{l|}{} \\ \hline
\textbf{\SI{0.1}{K} Stage}            & CCDR                 & {\color[HTML]{3531FF} \textbf{0.1}}         & \multicolumn{1}{c}{NA}      & {\color[HTML]{FE0000} \textbf{$\sim$0.0}}   & {\color[HTML]{FE0000} \textbf{$\sim$0.0}}            &                                                          &                                                         &                                                         & {\color[HTML]{00009B} \textbf{2.3}}         &                                                          & \multicolumn{1}{c}{{\color[HTML]{F56B00} \textbf{1.1}}}  & \multicolumn{1}{c}{{\color[HTML]{FE0000} \textbf{1.1}}}   & \SI{}{\micro\watt}                              & \multicolumn{1}{c|}{{\color[HTML]{009901} 112\%}} & \multicolumn{1}{c}{6.7}    & \textbf{\SI{8}{kg}}         \\
CCDR                           & \multicolumn{1}{l}{} & \multicolumn{1}{l}{{\color[HTML]{3531FF} }} &                             & Intercepted                                 & \multicolumn{1}{l}{From R3}                          &                                                          &                                                         &                                                         & {\color[HTML]{00009B} M=33\%}               &                                                          &                                                          &                                                           &                                 &                                                   &                            & \multicolumn{1}{l|}{} \\ \hline
\multicolumn{6}{|l|}{\textbf{Electrical Power incl. CDE (W)}}                                                                                                                                                                          & \multicolumn{1}{c}{\textbf{900}}                         & \multicolumn{1}{c}{\textbf{120}}                        & \multicolumn{1}{c}{\textbf{120}}                        & \textbf{50}                                 & \multicolumn{5}{l|}{\textbf{TOTAL POWER (Coolers)+\SI{100}{W} Thermal Control}}                                                                                                                                                                                              &                            & \textbf{\SI{1290}{W}}       \\ \hline
\multicolumn{6}{|l|}{\textbf{Cooler Mass incl. CDE}}                                                                                                                                                                                   & \multicolumn{1}{c}{\textbf{84}}                          & \multicolumn{1}{c}{\textbf{40}}                         & \multicolumn{1}{c}{\textbf{54}}                         & \textbf{31}                                 & \multicolumn{5}{l|}{\textbf{TOTAL MASS (Coolers)}}                                                                                                                                                                                                                    &                            & \textbf{\SI{209}{kg}}       \\ \hline
\multicolumn{6}{|l|}{\textbf{Cooler Mass with V-Grooves and Baffle}}                                                                                                                                                                   &                                                          &                                                         &                                                         & \multicolumn{1}{l}{}                        & \multicolumn{5}{l|}{\textbf{TOTAL MASS WITH V-GROOVES}}                                                                                                                                                                                                               &                            & \textbf{\SI{328}{kg}}       \\ \hline
\end{tabular}%
}
\caption{\small Mass, Power and Heat budgets for the \coremfive \ cryogenic system.}
\label{tab:CRYO-TABLE}
\end{sidewaystable}

In the following we give additional information on key sub-systems of the cryogenic system of \coremfive.

\clearpage

\subsection{Coolers}

In addition to the V-Grooves, the cryo chain is composed of:

\begin{itemize}

\item 3 Pulse Tube Coolers (PTC): two systems ON running below their maximum power and one PTC kept OFF for the purpose of redundancy. The \SI{15}{K} PTC developed by ALAT/CEA/TCBV is baselined \cite{cryo3}. A prototype has recently been tested and has demonstrated \SI{435}{mW} at \SI{15}{K} on the lower end together with \SI{2}{W} at \SI{80}{K} on the first stage. The system is compact and weighs about \SI{18}{kg} without its drive electronics (CDE). It uses a maximum of \SI{300}{W} of electrical input at compressor level, i.e. $\sim$ \SI{450}{W} including the CDE. A development phase has recently brought this cooler to TRL 5.

\item 2 Joule Thomson (JT) Coolers operating at \SI{4.5}{K} (one cooler is ON and one is OFF for redundancy) \cite{cryo5}. The compressor will benefit from the ongoing \SI{2}{K} JT cooler development, but will use just 2 compression stages instead of 4. The cold piping and ancillary panels are similar to the \SI{4}{K} JT Planck cooler and do not need additional development. A higher cooling power (\SI{30}{mW} at \SI{4.5}{K}) than for Planck (\SI{20}{mW} at \SI{4}{K}) can be supplied by taking advantage of the lower \SI{15}{K} pre-cooling temperature available from the PTC, and by using an increased ($\sim$ 10 \%) mass-flow. 
 
\item 1 Closed Cycle Dilution Refrigerator (CCDR). This cooling technology is based on the Planck heritage, but includes some major evolutions such as the distillation and recirculation of $^3$He and $^4$He. It requires precooling at \SI{1.7}{K} (\SI{5.25}{mW} with 50\% margin). It is baselined because the \coremfive \ mission calls for continuous observation but also because the focal plane is heavy. Mechanical modeling studies show that to avoid a complex mechanical release mechanism as on Planck, 24 Carbon Fiber Reinforced Plastic tubes OD/ID \SI{5/3}{mm} are needed to support a \SI{500}{mm} diameter \SI{8}{kg} detector plate under \SI{120}{g} static acceleration and with a resonant frequency above \SI{200}{Hz}. The price to pay is a conduction load as high as \SI{20}{\micro\watt} from \SI{1.7}{K} to \SI{0.1}{K}. This subsystem is described in detail below. 

\end{itemize}

\subsection{Closed Cycle Dilution Refrigerator}

\label{sec:org6ef6356}

Gravity plays an important role in the operation of \He{3}-\He{4} dilution
refrigerators on Earth since it localizes the interfaces among a relatively dense \He{3}-poor liquid phase, a less dense
\He{3}-rich liquid phase, and an even less dense \He{3}-rich vapor phase in
convenient positions.

The dilution refrigerator on the Planck satellite has a different design
where capillary forces play the role of gravity. This dilution refrigerator
consists of three capillaries soldered in parallel to form a heat exchanger.
The capillaries form an Y-junction at the cold end of the dilution
refrigerator. Liquid \He{3} and liquid \He{4} flow separately through
two of those capillaries to the Y-junction, where mixing produces cooling.
The mixture returns through the third capillary and expands in a \SI{1.6}{\K}
Joule-Thompson cooler \SI{1.5}{\K} before being pumped by space.
Therefore, the Planck dilution refrigerator is called an Open Cycle Dilution
Refrigerator (OCDR).

The cooling power and the life time of this refrigerator are limited by the
amounts of \He{3} and \He{4} stocked on the spacecraft. In order to increase
the cooling power, the N\'{e}el Institute (in collaboration with Air Liquide) has
developed an \He{3}-\He{4} isotope separator working at
\SIrange{0.6}{1.2}{\K} to close the cycle.

The Closed Cycle Dilution Refrigerator (CCDR) has three disadvantages with
respect to the OCDR:
\begin{enumerate}
\item It requires a \SI{1.7}{\K} cooler, because the isotope separator in the
CCDR takes the place of the Joule-Thompson cooler in the Planck
refrigerator.
\item It requires to localize a liquid-vapor interface in zero gravity, because
the extraction of \He{3} uses the difference in vapor pressures of \He{3}
and \He{4} in the phase separator.
\item It requires a \He{3} circulation pump.
\end{enumerate}

\begin{figure}[t!h]
  \centering
  \includegraphics[width=10cm]{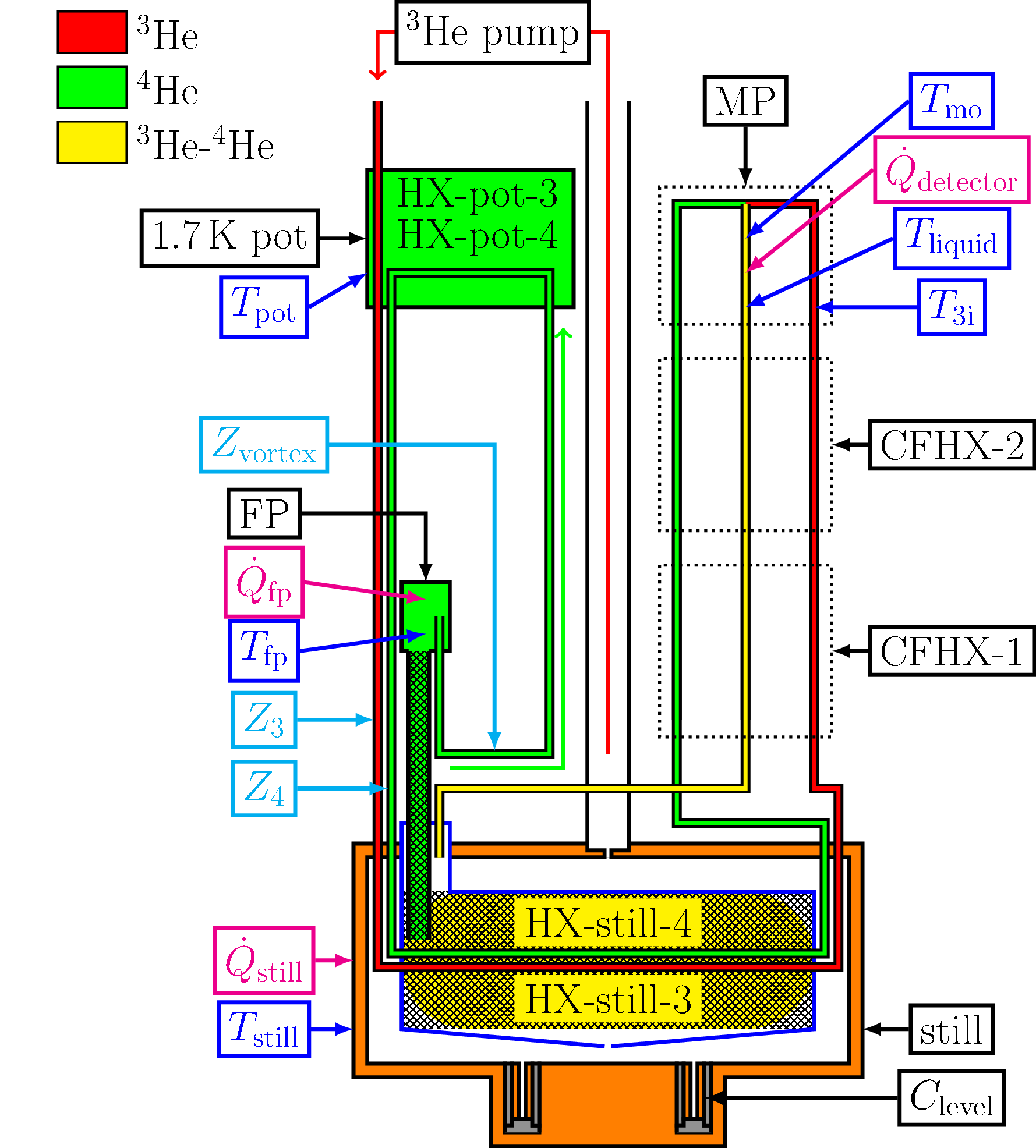} 
  \caption[Diagram of the heart of the CCDR setup]{\small
    Diagram of the upside-down closed cycle dilution refrigerator inside
    the \SI{4.2}{\K} vacuum can.
    The still contains a sponge pot (outlined in blue) filled with a Procelit
    P160 sponge (black crosshatch) to confine the liquid mixture (yellow).
A \He{3} pump extracts almost pure \He{3} through the pumping line, the
    \SI{1}{\mm} \diameter{} orifice at the bottom of the pumping line, and the
    \SI{1}{\mm} \diameter{} orifice at the sponge pot bottom out of the
    Procelit P160 sponge.
    A fountain pump FP extracts pure \He{4} (green) from the Procelit P160
    sponge.
    The almost pure \He{3} (red) and pure \He{4} (green) flow back into the
    dilution refrigerator heat exchanger (sections CFHX-1 and CFHX-2) after
    cooling in the heat exchangers in the \SI{1.7}{\K} pot and the still.
We show some of the heaters (magenta labels), some of the thermometers (blue
    labels), and the level gauge \PQ{C}{level} used to operate and characterize
    the CCDR.
    The three flow impedances (cyan labels) serve either to liquefy \He{3} gas
    (\PQ{Z}{3}), or to make the fountain pump work (\PQ{Z}{vortex}), or to
    diminish the heat load from the \SI{1.7}{\K} pot on the still (\PQ{Z}{4}) in
    case the fountain pump circulates no \He{4}.}
  \label{fig:schematic}
\end{figure}

We discuss the current state of the art by showing some results obtained with
an upside-down CCDR in combination with a 3-stage reciprocal compressor
developed by JAXA by modifying their 2-stage \SI{1.7}{\K} \He{3}
Joule-Thompson compressor.
We have built the upside-down CCDR shown in Figure~\ref{fig:schematic} to
demonstrate that liquid \He{3}-\He{4} can be retained by capillary forces in
a sponge under the conditions of negative gravity instead of zero gravity.
The CCDR consists of 3 subsystems:
\begin{enumerate}
\item A Counter Flow Heat eXchanger (divided in sections CFHX-1 and CFHX-2 in
Figure~\ref{fig:schematic}) and a Mixing Part (MP). We have scaled up
the diameters of the capillaries in this subsystem with respect to
Planck's OCDR to provide more cooling power at lower temperatures.
The CCDR will cool the focal plane through the \He{3}-\He{4} mixture
return capillary indicated by the heater \dQ{detector}.
\item An isotope separator (still) allowing to pump \He{3}-rich vapor (\(\>
       \SI{90}{\percent}\)) from  the liquid \He{3}-\He{4} mixture confined in a
sponge in the still by means of a \He{3} pump as well as superfluid \He{4}
by means of a fountain pump (FP).
Heat exchangers (HX-still-3 and HX-still-4) cool the \He{3} and \He{4}
streams before they enter into the CFHX.
\item A \SI{1.7}{\K} pot (future \He{3} Joule-Thompson expansion provided by
RAL) to liquefy the \He{3} gas returning from the \He{3} pump and to cool
the \He{4} liquid coming from the fountain pump.
\end{enumerate}

We have designed the heat exchanger above the sponge in the still
to make the capillary confinement harder because of the hydrostatic pressure
exercised by the liquid in heat exchanger.
In fact, the MP is sitting \SI{150}{\mm} above the still while we estimate
the capillary height in the Procelit P160 sponge to be \SIrange{30}{40}{mm}.
Indeed, the sponge leaks in case of zero \He{3} and \He{4} flow rates.
Nevertheless, the CCDR works well without any liquid leaking from the sponge
as monitored by the capacitance liquid level gauge \PQ{C}{level}.
We explain the successful confinement by the viscous pressure drops in the
capillaries of the CFHX when the CCDR is circulating.

\begin{figure}[t!]
  \centering
  \includegraphics[width=10cm]{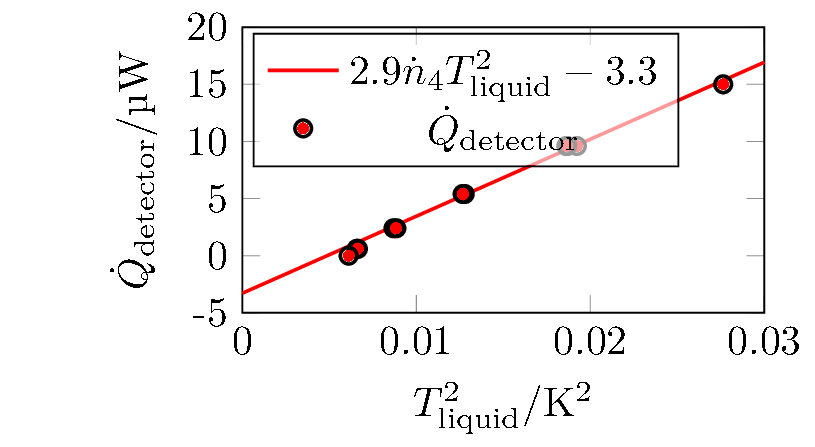}
  \caption[CCDR cooling power]{\small
    CCDR focal plane cooling power data obtained by applying heat to
    \dQ{detector} and measuring the temperature \PQ{T}{liquid} of the
    outgoing liquid \He{3}-\He{4} mixture between \dQ{detector} and the
    CFHX.}
  \label{fig:cooling-power}
\end{figure}

Figure~\ref{fig:cooling-power} shows cooling power data obtained by applying
heat to \dQ{detector} and measuring the temperature \PQ{T}{liquid} which is
the temperature of the liquid \He{3}-\He{4} mixture downstream of
\dQ{detector} and upstream of the CFHX (see Figure~\ref{fig:schematic}).
We have fitted the data by the approximate relation
\(\dQ{detector}=A\dn{4}T_{\text{liquid}}^2-\dQ{leak}\), where
\(\dn{4}=\SI{235}{\micro\mole\per\second}\) obtained from the relation
\(\dQ{fp}=\dn{4}\PQ{T}{fp}\PQ{s}{40}(\PQ{T}{fp})\) which is based on the
second law of thermodynamics and the molar entropy (\PQ{s}{40}) data of
liquid \He{4}, \(A=\SI{2.9}{\joule\per\mole\kelvin\squared}\) which is in
very good  agreement with thermodynamic data of \He{3}-\He{4} mixtures, and
\(\dQ{leak}=\SI{3.3}{\micro\watt}\).
Note that \(\dQ{detector}=\SI{3.5}{\uW}\) at \(\PQ{T}{liquid}=\SI{0.1}{\K}\)
even in the presence of a heat leak of \SI{3.3}{\micro\watt}.

\subsection{Focal Plane Suspension with Thermal Intercepts}
\label{sec:org814ae0d}

\begin{figure}[t!]
  \centering
  \includegraphics[height=80mm]{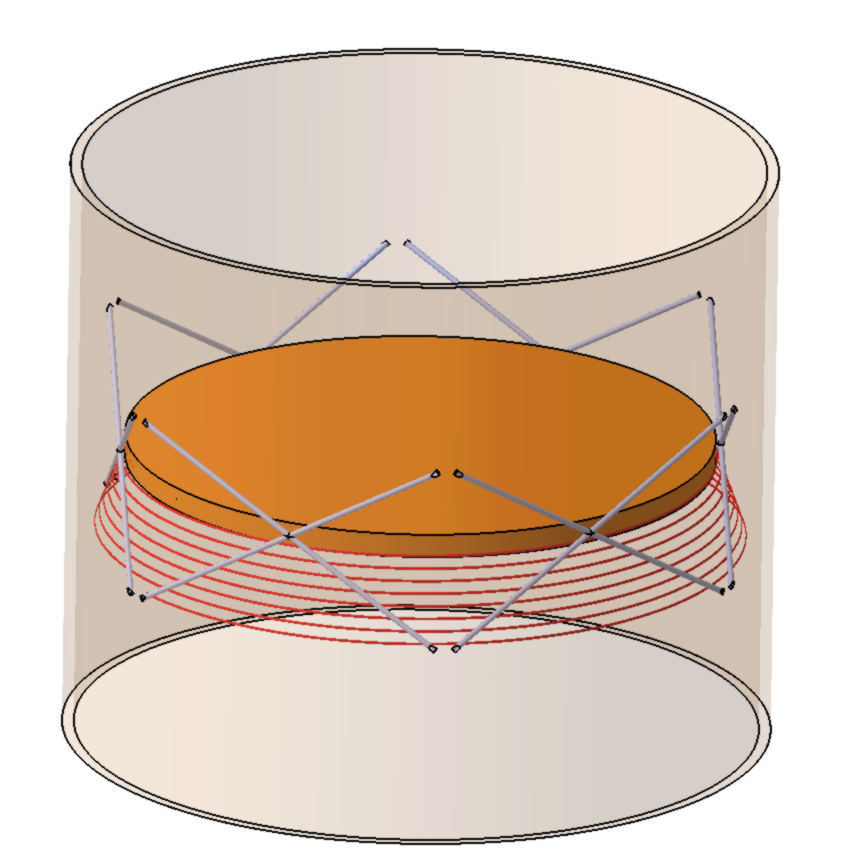}
  \caption[Focal plane suspension with heat intercepts]{\small
    The target design model for the preliminary mechanical and thermal
    calculations consists (1) of a \SI{0.1}{\K} focal plane (rigid
    cylindrical plate with a nominal mass of \SI{8}{\kg} and a nominal
    diameter of \SI{500}{\mm}), (2) of 24 CFRP suspension rods between
    the focal plane and the \SI{1.7}{\K} radiation screen (nominal
    diameter of \SI{580}{\mm}), and (3) the dilution refrigerator CFHX in
    the form of a spiral below the focal plane to intercept heat currents
    through the 12 CFRP struts below the focal plane.
    In addition, vertical high thermal conductivity metal links (not
    shown) between the upper struts and CFHX interception points on the
    lower struts will intercept the heat currents through the 12 CFRP
    struts above the focal plane.}
  \label{fig:focal-plane-suspension}
\end{figure}

Figure~\ref{fig:focal-plane-suspension} shows a simplified model of the
focal plane suspension with thermal intercepts to check the feasibility of a
suspension system meeting the mechanical requirements of a satellite launch
as well as the thermal requirements imposed by the operation of the focal
plane at \SI{0.1}{\K}.
We have choosen Carbon Fiber Reinforced ePoxy (CFRP) rods because of its
high Young's modulus, its high yield strength, and its low thermal
conductivity.
The low thermal contraction of CFRP with respect to metals may be a
disadvantage, since the cooling of the struts may impose addition strain.
However, in case the satellite is being launched with the focal plain at
\SI{300}{\K} this disadvantage plays no role.
The geometry of the suspension system in Figure~\ref{fig:focal-plane-suspension} is an advantage, since it has the optimal
position of the center of mass of the focal plane.

The results of the mechanical study are still preliminary but look
promising.
We summarize here the results of calculations done to find the optimal
angles for 24 CFRP tubes with a fixed length of \SI{195}{\mm}, a fixed inner
diameter of \SI{3}{\mm} and a fixed outer diameter of \SI{5}{\mm}.
The effect of varying the angles is that (1) the vertical distance between
the strut fixations on the \SI{1.7}{\K} envelope below and above the focal
plane can change and that (2) the diameter of the \SI{1.7}{\K} envelope can
change.
The vertical distance has been varied from \SI{120}{\mm} to \SI{220}{\mm}
and the diameter of the \SI{1.7}{\K} envelope has been varied from
\SI{680}{\mm} to \SI{740}{\mm}.
The lowest resonant frequencies vary from \SI{182}{\Hz} to \SI{265}{\Hz} for
those configurations while the struts do not buckle for accelerations of 120
times gravity on earth.
In order to reduce the diameter of the \SI{1.7}{\K} to \SI{580}{\mm}, the
length and the cross-section area of the tubes have to be reduced by the
same factor to keep the same resonance frequencies.
In this case, we expect the buckling threshold to increase.
Those results are very promising, but we point out that the vibration modes
of the focal plane should be handled together with its suspension structure.

In order to have a design aid for the focal plane suspension system, we have
written a program to solve differential equations for the enthalpy balance
in the CFHX using a collocation method for boundary value problems.
The differential equations take into account
(1) the flows of enthalpy due the incoming \He{3} stream and the outgoing
\He{3}-\He{4} mixture stream,
(2) the heat transfer from the incoming \He{3} stream to the outgoing
\He{3}-\He{4} mixture stream through the walls the CFHX (the dominant
thermal resistance is the boundary resistance between the helium and the
CuNi capillaries forming the CFHX), and
(3) viscous dissipation due to Poiseuille flow.
The program is flexible enough to handle a CFHX consisting of an arbitrary
number of sections with different lengths and diameters as shown in Table~\ref{tab:cfhx-temperature-profile-input}.
Table~\ref{tab:cfhx-temperature-profile-input} also shows that is possible
(1) to short-circuit the CFHX sections with sections of CRFP struts and
(2) to apply heat loads to the outgoing \He{3}-\He{4} mixture between the
CFHX.
Other parameters are for instance the \He{3} circulation rate \dn{3}, the
\He{4} circulation rate \dn{4}, the still temperature \PQ{T}{still}, and the
diameter \PQ{D}{mo} and the length \PQ{L}{mo} of the \He{3}-\He{4} mixing part
capillary (the yellow capillary inside the dashed MP box in Figure~\ref{fig:schematic}).

We have compared solutions of the differential equations with the
performance of the upside-down CCDR and our best upside-up CCDR with
\(\PQ{T}{detector}=\SI{51.4}{\mK}\) and \(\PQ{T}{liquid}=\SI{44.0}{\mK}\) for
\(\dQ{detector}=\SI{1}{\uW}\).
We have seen that the temperature \PQ{T}{liquid} of the solutions agrees
within \SI{5}{\percent} with the experimental data.

\begin{table}[t!]
  \centering
  \begin{tabular}{rrrrrrr}
    Section & \PQ{L}{cfhx} & \PQ{D}{i} & \PQ{D}{o} & \dQ{load} & \PQ{L}{cfrp} & \PQ{A}{cfrp}/\PQ{L}{cfrp}\\
            & \si{m} & \si{mm} & \si{mm} & \si{\uW} & \si{mm} & \si{mm}\\
    \hline
    0 & 0.1 & 0.2 & 0.2 & 0.0 &      &\\
    1 & 1.9 & 0.2 & 0.2 & 0.0 & 80.0 & 3.77\\
    2 & 3.0 & 0.4 & 0.4 & 0.0 & 40.0 & 7.54\\
    3 & 3.0 & 0.4 & 0.4 & 0.0 & 40.0 & 7.54\\
    4 & 3.0 & 0.4 & 0.4 & 7.0 & 40.0 & 7.54\\
  \end{tabular}
  \caption[CFHX temperature profile input data]{\small
    CFHX section input data for the calculation of the CFHX temperature
    profiles shown in Figure~\ref{fig:cfhx-temperature-profile}.
    The data consists of the section number,
    the length \PQ{L}{cfhx} of the section,
    the diameter of the incoming liquid \He{3} stream \PQ{D}{i},
    the diameter of the outgoing liquid \He{3}-\He{4} mixture stream
    \PQ{D}{o},
    an externally applied heat load \dQ{load} to a thermal anchoring
    point at the cold end of the heat exchanger section with the
    exception of the last section where \(\dQ{load}=\dQ{detector}\),
    the distance \PQ{L}{cfrp} between two thermal anchoring points,
    and the ratio \PQ{A}{cfrp}/\PQ{L}{cfrp} where \PQ{A}{cfrp} is the cross
    section area of the 24 CRFP struts in parallel.
    The total length of the CFRP struts is \SI{200}{\mm}, the distance
    between the thermal anchoring points on thermal screen at \SI{1.7}{\K}
    and the cold end of section 1 equals \SI{80}{\mm}, and \PQ{L}{cfrp} in
    section 4 is the distance between the thermal anchoring point at the
    cold end of section 3 and the focal plane.
    The CFRP struts are tubes with an outer diameter of \SI{5}{\mm} and an
    inner diameter of \SI{3}{\mm}.}
  \label{tab:cfhx-temperature-profile-input}
\end{table}

\begin{figure}[t!]
  \centering
  \includegraphics[width=10cm]{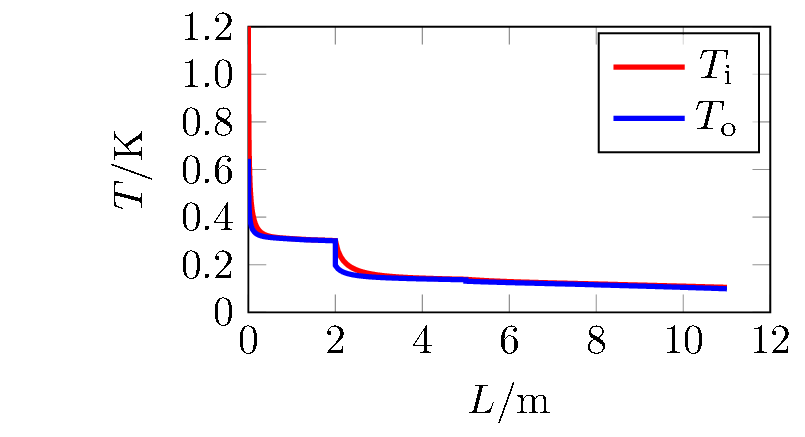}
  \caption[CFHX temperature profile]{\small
    Temperature profiles of the incoming almost pure \He{3} stream \PQ{T}{i}
    and the outgoing \He{3}-\He{4} mixture stream along the CFHX between the
    still at \SI{1.2}{K} and the mixing part sligthly below \SI{0.1}{\K}
    cooling the focal plane. In fact, \(\PQ{T}{o}\) at \(L=\SI{11}{\m}\) --
    the equivalent of \PQ{T}{liquid} in Figure~\ref{fig:schematic} -- equals
    \SI{99}{\mK}.}
  \label{fig:cfhx-temperature-profile}
\end{figure}

Table~\ref{tab:cfhx-temperature-profile-output} and Figure~\ref{fig:cfhx-temperature-profile} show the solution of the differential
equations for the input data shown in Figure~\ref{tab:cfhx-temperature-profile-input}.
Some of the other parameters are \dn{3}=\SI{35}{\micro\mole\per\second},
\dn{4}=\SI{205}{\micro\mole\per\second}, \PQ{T}{still}=\SI{1.2}{\K},
\PQ{D}{mo}=\SI{0.6}{\mm}, and \PQ{L}{mo}=\SI{1.5}{\m}.
The flow rates \dn{3} and \dn{4} are representative of the values during the
tests of the upside-down CCDR in combination with the space qualified JAXA
\He{3} circulation pump.

The results show that the thermal intercept at the cold end of CFHX section
1 evacuates \SI{43.3}{\uW} in total at \SI{0.3}{\K} or \SI{1.8}{\uW} per
strut.
Vertical thermal links between the heat intercepts on the struts below the
focal plane and the struts above the focal plane have to intercept the heat
current through those struts.
The thermal conductivity of high-quality copper or silver is
\(1000T\;\si{\W\per\K\per\m}\).

\begin{table}[t!]
  \centering
  \begin{tabular}{rrrrrrr}
    Section & \dQ{cfrp,cold} & \dQ{load,cold} & \PQ{T}{i,warm} & \PQ{T}{i,cold} & \PQ{T}{o,warm} & \PQ{T}{o,cold}\\
            & \si{\uW} & \si{\uW} & \si{K} & \si{K} & \si{K} & \si{K}\\
    \hline
    0 &        & 0.000 & 1.200 & 0.413 & 0.644 & 0.347\\
    1 & 43.303 & 0.000 & 0.413 & 0.301 & 0.347 & 0.300\\
    2 &  1.410 & 0.000 & 0.301 & 0.139 & 0.197 & 0.137\\
    3 &  0.071 & 0.000 & 0.139 & 0.119 & 0.131 & 0.116\\
    4 &  0.041 & 7.000 & 0.119 & 0.105 & 0.115 & 0.099\\
  \end{tabular}
  \caption[CFHX calculation output data]{\small
    CFHX calculation output data for the input data shown in Table~\ref{tab:cfhx-temperature-profile-input}.
    The data consists of the section number, the heat current \dQ{cfrp,cold}
    through the CRFP struts and intercepted by a thermal anchor at the cold
    end of the heat exchanger section (for the last section it is the heat
    current to the focal plane),
    the externally applied heat load \PQ{T}{load,cold} to a thermal
    anchoring point at the cold end of the heat exchanger section with the
    exception of the last section  where \(\dQ{load}=\dQ{detector}\),
    the temperature \PQ{T}{i,warm} of the incoming liquid \He{3} stream at
    the warm end of the heat exchanger section,
    the temperature \PQ{T}{i,cold} of the incoming liquid \He{3} stream at
    the cold end of the heat exchanger section,
    the temperature \PQ{T}{o,warm} of the outgoing liquid \He{3}-\He{4}
    mixture stream at the warm end of the heat exchanger section, and
    the temperature \PQ{T}{o,cold} of the outgoing liquid \He{3}-\He{4}
    mixture stream at the cold end of the heat exchanger section.}
  \label{tab:cfhx-temperature-profile-output}
\end{table}

In this case, a thermal link for a heat current of \SI{1.8}{\uW} with a
cross-section of \SI{0.1}{\mm\squared} and a length of \SI{0.1}{\m} will
have a temperature difference of \SI{6}{\mK} over the length of the link.

The thermal intercept at the cold end of section 3 is only \SI{1.4}{\uW} at
\SI{0.14}{\K} and is easier to handle and the intercept at the cold end of
section 3 is not very useful, since it evacuates only \SI{1}{\percent} of
the heat load on the mixer part.

We point out that the total length of the heat exchanger should be at least
\SI{15}{\m} in order to intercept heat on the struts at three different
temperatures. 
However, prolonging the heat exchanger to for instance \SI{20}{\m} not
affect the equivalent of \PQ{T}{liquid} in Figure~\ref{fig:schematic}, since
it is entirely controlled by \dQ{detector}.

Finally, we point out that we did not discuss the mechanical suspension of
the still, but we think it should be treated separately from the CFHX and
focal plane assembly and we expect it to be straightforward because the
still is much lighter and smaller than the focal plane assembly.

\subsection{\SI{1.7}{\K} \He{3} Joule-Thompson expansion cooler requirements}
\label{sec:org13429c1}

The \He{3} Joule-Thompson expansion cooler serves (1) to cool the liquid
\He{4} coming from the fountain pump to \SI{1.7}{\K} and to cool and liquefy
the \He{3} gas coming from the \He{3} pump.
Therefore, we may write the heat load on the \He{3} Joule-Thompson expansion
as the sum of two contributions
\(\dQ{JT}(\dn{3})=\dn{3}\left(\PQ{L}{30}(T=\SI{1.7}{\K})+2.5R(\PQ{T}{gas}-\SI{1.7}{\K})\right)\)
and
\(\dQ{JT}(\dn{4})=\dn{4}\left(\PQ{H}{40}(\PQ{T}{fp})-\PQ{H}{40}(\SI{1.7}{\K})\right)\),
where \PQ{L}{30} is the latent heat of liquefaction of \He{3}, \PQ{H}{40}
the enthalpy of liquid \He{4}, and \PQ{T}{gas} the temperature of the \He{3}
gas before liquefaction.
In all our experiments up to now \(\PQ{T}{gas}=\SI{4.2}{\K}\), but we plan
to reduce it to below \SI{2.0} by means of a heat exchanger between the
incoming and outgoing \He{3} gas. In this case the value
\dQ{JT}(\dn{3}=\SI{35}{\micro\mole\per\second}) will decrease from
\SI{3.4}{\mW} to below \SI{1.8}{\mW}.
In case of \PQ{T}{fp}=\SI{2.15}{\K},
\(\dQ{JT}(\dn{4}=\SI{205}{\micro\mole\per\second})=\SI{1.7}{\mW}\).
Since \(\PQ{H}{40}(\PQ{T}{})\) is a steep function of \PQ{T}{}, (\PQ{H}{40}
decreases by about \SI{20}{\percent} when \PQ{T}{} decreases from
\SI{2.15}{\K} to \SI{2.1}{\K}), the design of the fountain pump has to make
it work at the lowest temperature possible.
In summary, in case \dn{3}=\SI{35}{\micro\mole\per\second} and
\dn{4}=\SI{205}{\micro\mole\per\second}, the current configuration requires
a heat lift of \SI{5.1}{\mW} from the \He{3} Joule-Thompson expansion
cooler, but the requirement will decrease to below \SI{3.5}{\mW} with a heat
exchanger between the incoming and outgoing \He{3} gas.

\section{Instrument Calibration}
\label{sec:calibration}


By instrument calibration we intend the measurement of all instrument characteristics which
impact the data analysis and hence the science products of the mission. The experience of \emph{Planck} shows that the ultimate data quality of extreme sensitivity measurements is likely
to be limited not by white noise but by residual systematic effects, which mainly reflect the level of accuracy achieved in the calibration of the instrument.
This experience also shows that non-idealities can be corrected for in the data processing as long as they are known with sufficient precision. 
A broad requirement for calibration accuracy is that uncertainties in the
measurements of all the instrument parameters give rise to a level of
systematics which is significantly less than the statistical noise. By
statistical noise we mean the white noise in the final product maps
arising from detector sensitivity averaged over the entire range of angular
scales of interest. However, the way in which systematic effects impact
the science products of the mission is far from trivial (e.g. 
\cite{Planck2013_A3}). In general it is necessary to simulate
systematic effects and propagate them through the data analysis pipeline
up to the cosmological results, with an accuracy and a level of detail
that depends on the significance of the specific effect being evaluated.

Defining requirements is essential in the calibration plan and requires close interaction with instrument design and mission strategy (see \cite{missionpaper}). We propose here two main stages in the calibration plan: 1) ground-based system level characterization and 2) in-flight measurements using astronomical sources. The \coremfive\ calibration plan will include a combination of tests on ground and in flight for each of the calibration parameters to be measured. The main classes of instrumental parameters to be characterized are: 
\begin{enumerate}
\itemsep0em 
\item  Photometric (or absolute) calibration: conversion of the
product maps from generic telemetry units to physical units
(\SI{}{\micro\kelvin_{\rm CMB}}). Gain factors for each radiometer will be measured on the ground at several stages. 
The final calibration will be performed in-flight.
\item Relative calibration (signal stability): stability of the gain and zero-level. This will be simplified by the redundancy of the scanning strategy.
\item Thermal effects: systematics induced by thermal fluctuations in the
\SI{0.1}{K}, \SI{1.7}{K}, \SI{4}{K}, \SI{20}{K} and \SI{300}{K} stages; cooler induced microphonics. A set of
temperature sensors will monitor the  thermal configuration of the instrument and
stability. The necessary accuracy and resolution of thermal calibration will be based on simulations
requiring a detailed thermal model of the focal plane.
\item Detector chain non-idealities: detector time-response; non-linearity
of the detector response; non-linearity of ADC converters;  impact of cosmic
rays; sensitivity to microphonics.
\item Spectral calibration: detailed bandpass measurements on the ground.
In-flight verification of the measured bandpasses will be
possible through observation of diffuse and point sources with steep
spectra. 
\item Optical calibration: main beam determination, near sidelobes, far
sidelobes (both total intensity and polarization). Direct measurements of
the main beams and near lobes down to -35 $\slash$ -\SI{40}{dB} will be possible
in-flight exploiting signals from planets and strong polarized sources.
Detailed models of the far sidelobes will be constructed with
state-of-the-art physical optics codes (GRASP\footnote{www.ticra.com}) and convolved to models of
the full sky emission to propagate signatures from straylight. The optical
model will be validated by comparing simulation with direct measurements
of main beams and near lobes.
\item Polarization-specific calibration: polarization efficiency and 
polarization angle of each detector; and in the case of using a half-wave plate the systematics that it may induce.
These will be measured both on the ground and, possiblty, in-flight. 

\item Intrinsic noise characterization: detailed measurements of the noise
properties (noise power spectrum, $1/f$ noise, possible non-gaussianity)
and their time evolution are needed. These will be standard measurements
in ground testing and in-flight operation
\end{enumerate}

\subsection{Ground Calibration}
The experience gained from the \emph{Planck}-HFI has shown that while a great deal of effort was put in instrument ground calibration \cite{2010A&A...520A..10P}, some effects were not anticipated at the level observed (cosmic ray hits for instance) while some other parameters could have benefited from a better characterization. Therefore we will follow a similar but more complete instrument ground calibration and with higher accuracy in the determination of parameters that are crucial for polarization measurements. However, the difficulty will be to adapt this calibration strategy from tens of pixels for \emph{Planck} to thousands of pixels for \coremfive. It is likley that full characterization of each individual pixel with extremely high accuracy would be too costly and time consuming. It that case, the \coremfive\ strategy will include general tests conducted on all components and pixels
to check for anomalies, failures, and performance inhomogeneity, complemented with thorough testing on a subset of pixels across the frequency bands.

The \emph{Planck} data analysis has also highlighted the criticality of a detailed knowledge of beams and far-sidelobes, particularly for polarization.
Clearly, even the extensive optical testing carried out for \emph{Planck} \cite{tauber:etal2010}
would be significantly inadequate for \coremfive\, which
requires RF calibration to be more accurate by an order of magnitude. The \coremfive\ optical ground testing will include a highly representative focal plane system, comparable with the final flight model, and multiple in-band pattern measurements.

\subsection{In-flight calibration}
Following the \emph{Planck} strategy, photometric and
relative calibration will be based on the solar and orbital dipoles as
primary calibrators and cross-checked on planets. 
Using orbital modulation
\emph{Planck} has reached absolute calibration at few $\times 10^{-4}$ level,
which can be exploited by \coremfive\ in its first phase of data analysis.
However, \coremfive\ will be able to far improve over long time intervals by at least one order of magnitude.
The \emph{Planck} experience has shown the extreme importance of direct
monitoring of the instruments (subsystem temperatures, electrical bias
parameters, thermal stability, cooler performance) and regular processing
of the time-ordered data. Indeed, the cross-correlation between monitored
parameters and data processing have proved essential for improving
systematic effect removal algorithms.
Similarly, noise characteristics will be regularly measured in terms of
power spectrum, $1/f$ and other non-gaussian components.
Noise characteristics will be affected by cosmic rays, that show up
as glitches in the data. We will exploit the considerable experience,
gained in \emph{Planck}, of measurements of the time evolution and amplitude of
glitch signatures.
Detector time response and main beam characterization will be simultaneously performed on planets as they are strongly degenerate. This high level of accuracy will make use of multiple crossing observations (multiple scanning directions and angles) of various planets to achieve a high signal-to-noise well outside the main beam. As \coremfive\ 
will map about half sky with all of its detector every 4 days, planets will be scanned repeatedly over the full duration of the mission, allowing very precise sampling of the beams and continuos cross-check of the gain stability. These observations will be also used to measure instrumental polarization and scanning strategy-induced systematic effects (in-scan and cross-scan differences). 
Polarization efficiency and polarization angles can be measured on
well-known polarized astronomical sources, such as the Crab Nebula. \coremfive\ may benefit from future, high-precision, dedicated ground
observation campaigns of these polarized calibration sources.


\section{Downscoping Options}
\label{sec:downscoping}

The instrument configuration described above is robust enough to allow for downscoping, would cost/programmatic issues require it. The downscoping strategy would be to aim at maintaining extreme CMB polarization sensitivity and accuracy, while reducing the performance for measurements of: the polarized interstellar medium of our galaxy and the study of the galactic magnetic field; Cosmology with galaxy clusters; Extragalactic sources and cosmic infrared background. In practice, a \MiniCORE \ mission would be aimed at:

\begin{itemize}

\item Constraining inflationary B-modes at the level required to reach $\sigma (r) = 0.001$ with \MiniCORE \ alone, and $\sigma (r)=0.0003$ in combination with future ground-based measurements;

\item Observing dark matter structures by measuring lensing B-modes at S/N$>$3 per individual mode, with angular resolution $\lsim$ 3', in combination with future ground-based measurements;

\item Measuring polarisation E-modes with cosmic-variance dominated errors over $>30\%$ of the sky, in combination with future ground-based observatories.

\end{itemize}

The reduction of the requirement on $\sigma (r)$ by a factor of 3 (\MiniCORE \ alone vs \coremfive \ alone) translates in a reduction of the requirement on the sensitivity by a factor of $\sqrt{3}$, or $\Delta P = \SI{3}{\micro\kelvin}\cdot$arcmin (similar to LiteCORE-80 in the companion papers of this series). As foreground emission dominates over lensing B-modes for $\ell < 1000$, we require the CMB to be mapped from space up to $\ell=1000$, i.e. with an angular resolution of about 12 arcmin FWHM. To reach its goals in combination with future ground-based observatories, \MiniCORE \ should provide the capability to monitor high-frequency foreground emission at a level compatible with $\sigma (r) = 0.0003$.

The first change with respect to \coremfive \ is the reduction of the diameter of the telescope aperture to \SI{80}{cm}. This results in a number of savings, as detailed below. The second important change is the reduction of the number of detectors and the removal of all channels with frequency below \SI{100}{GHz} (15 frequency bands in place of 19). This will require teaming with ground-based surveys, which are possible and effective at low frequency. Angular scales to a few arcmin will be mapped from the ground with several-meters-aperture telescopes, up to 150 GHz: \MiniCORE \ will add the essential coverage of the same angular scales at frequencies up to \SI{600}{GHz}, where polarized dust emission becomes very strong. The proposed frequency channels of  \MiniCORE \ are outlined in Table~\ref{tab:downscope}. The aggregated CMB polarisation sensitivity is \SI{3.2}{\micro\kelvin}$\cdot$arcmin for a temperature of \SI{85}{K} of the telescope and its surroundings.

{\small 
\begin{table}[htb]
\begin{center}
{\footnotesize
\begin{tabular}{|c|c|c|c|c|}
\hline 
Channel & Beam   &  $N_{\rm det}$ &  $\Delta P$    & $\Delta I$                      \\
GHz     & arcmin &                 &  \SI{}{\micro\kelvin}$\cdot$arcmin & \SI{}{\micro\kelvin_{\rm RJ}}$\cdot$arcmin  \\
\hline 
\hline 

100     & 16.9   &  40    &  11.8   &  2    \\
115     & 14.8 	 &  40    &  11.7   &  2.41 \\         
130     & 13.2   &  40    &  11.7   &  2.82 \\
145     & 11.9   &  90    &   7.9   &  2.16 \\         
160     & 10.8   &  90    &   8.1   &  2.43 \\         
175     & 10.0   &  90    &   8.5   &  2.71 \\         
195     &  9.0   &  90    &   9.2   &  3.07 \\    
220     &  8.0   &  90    &  10.4   &  3.55 \\        
255     &  7.0   &  90    &  13.1   &  4.29 \\        
295     &  6.1   &  40    &  27.6   &  7.91 \\        
340     &  5.3   &  40    &  43.9   &  9.98 \\        
390     &  4.6   &  40    &  79     & 12.85 \\       
450     &  4.0   &  40    & 171     & 17.11 \\       
520     &  3.5   &  40    & 445.8   & 23.18 \\       
600     &  3.0   &  40    &1397.1   & 31.47 \\
\hline
\hline
\end{tabular}
}
\end{center}
\vspace{-\baselineskip}
\caption{\small  Proposed \MiniCORE \ frequency channels. The sensitivity is calculated assuming $\Delta \nu/\nu=30\%$ bandwidth, 60\% optical efficiency, total noise of twice the expected photon noise from the sky and the optics of the instrument at \SI{85}{K}. This configuration has 900 detectors. 
}
\label{tab:downscope}
\end{table}
}

Assuming a filling factor of 70\%, the focal plane radius is \SI{15.6}{cm} instead of \SI{25}{cm}. The area and mass of the sub-Kelvin focal plane are reduced by a factor 2.5, with 900 detectors in place of 2100. The \MiniCORE \ payload is passively cooled to \SI{85}{K} instead of \SI{40}{K}. This can be achieved with only two V-groove plates instead of three, or with MLI insulation between the SVM and the PLM. The mass of the telescope is reduced from $\sim$\SI{100}{kg} to $\sim$\SI{40}{kg}. The mass of the optical bench from $\sim$\SI{50}{kg} to $\sim$\SI{20}{kg}; the mass of the telescope structure from $\sim$\SI{20}{kg} to $\sim$\SI{5}{kg}. The total mass of the payload reduces from $\sim$\SI{200}{kg} to $\sim$\SI{80}{kg}. The combination of reduced number of detectors and reduced angular resolution results in reduced sampling requirements, reducing the continuous data rate from \SI{1.15}{Mbit/s} to \SI{330}{kbit/s}. The size and mass reduction of the payload allows for the use of only one pulse-tube instead of two; moreover with 900 detectors instead of 2100 the readout electronics is also reduced, so that the on-board power required is reduced from $\sim$\SI{2100}{W} to $\sim$\SI{1500}{W}. 

All these changes result in a very significant reduction of the cost of the instrument and of the mission, maintaining basically untouched its main scientific goals, provided that ground-based observatories can indeed meet their ambitious objectives and complement the space mission observations, in particular with noise-dominated small scale CMB maps over a large fraction of sky.

\section{Conclusions}
\label{sec:conclusions}

We have described a space-borne polarimeter optimized for a medium-size space mission within the Cosmic Vision Programme of the European Space Agency. 
The instrument uses the legacy of the successful Planck mission, and the most recent advances in the technology of panoramic detectors for mm-waves. The Technology Readiness Level (TRL) for the different parts of the instrument is summarized in Table \ref{instrumentTRL}, and is perfectly compatible with the requirements for the M5 call of ESA. With its wide frequency coverage (60-\SI{600}{GHz}), and large throughput this instrument will map CMB polarization with unprecedented sensitiviy and accuracy. The outstanding scientific performance of \coremfive \ is thoroughly discussed in companion papers of this series (\cite{2016arXiv161208270C}, \cite{2016arXiv161200021D}, \cite{Meli17}, \cite{missionpaper}). 
\begin{table}[htb]
\begin{center}
\scalebox{0.8}{\begin{tabular}{|l|c|l|c|}
  \hline
 {\bf Component}   &  {\bf Current} &  {\bf Heritage} & {\bf TRL @ end}    \\
                             &   {\bf TRL}     &                        & {\bf of phase-A}    \\
  \hline
\hline
telescope frame &  9    & Planck + Herschel heritage   &  9    \\
 \hline
telescope mirrors &  9    & SiC, D$<$1.2m, Herschel heritage  &  9    \\
 \hline
thermal filters &  4    &  used in current suborbital CMB  exp. &  6    \\
 \hline
plastic-embedded  metal-mesh flat lenses & 4 & demonstrated in the lab & 6 \\
\hline
plastic-embedded metal-grid polarizers & 4 & used in current balloon CMB  exp. & 6 \\
\hline
band-defining filters & 9 & Planck HFI heritage & 9 \\
\hline
waveguides & 9 & Planck HFI heritage & 9 \\
\hline
KIDs ($\nu < \SI{110}{GHz}$) & 3-4 & demonstrated in the lab & 6 \\
\hline
KIDs ($\nu > \SI{110}{GHz}$) & 5  & used at mm-wave  telescopes& 6 \\
 &   &  (IRAM / APEX)& \\
\hline
cryogenic LNA & 9 & Herschel-HIFI heritage & 9 \\
\hline
readout electronics & 5 & used at the telescope with  & 6 \\
 &  &  space qualified components &  \\
\hline
\hline
CCDR & 3-4 &   & 6 \\
\hline
\SI{1.7}{K} JT cooler & 4 &  & 6 \\
\hline
\SI{4}{K} JT cooler & 4 &  & 6 \\
\hline
\hline

\end{tabular}}

\caption{TRLs for the scientific instrument and main SVM systems
\label{instrumentTRL}}
\end{center}
\end{table}



\section{Acknowledgements}
\label{sec:acknowledgements}
PdB acknowledges support from University of Rome La Sapienza (project ``Cosmologia di Precisione 2015'') and from the Italian Space Agency (Agreement 2016-019-H.0 ``Kinetic Inductance Detectors for Space''). CJM is supported by an FCT Research Professorship, contract reference IF/00064/2012, funded by FCT/MCTES (Portugal) and POPH/FSE (EC). C.H.-M. acknowledges the financial support of the Spanish Ministry of Economy and Competitiveness via I+D project AYA-2015-66211-C2-2-P. GDZ acknowledges the financial support of ASI/INAF agreement n.~2014-024-R.1. J.G.N. acknowledges financial support from the Spanish MINECO for a ``Ramon y Cajal'' fellowship (RYC-2013-13256) and the I+D 2015 project AYA2015-65887-P (MINECO/FEDER). F.J.C., E.M.-G. and P.V. acknowledge support from the Spanish Ministerio de Econom\'{i}a y Competitividad project ESP2015-70646-C2-1-R co-financed with EU FEDER funds.

\bibliographystyle{JHEP}

\small
\bibliography{bib_CMB-f.bib,bib_detectors.bib,bib_high-f.bib,bib_telescope.bib,bib-cryo.bib,bib_filters.bib,bib_hwp.bib,bib_instrument.bib,bib_calibration.bib,bib_modulator.bib}
\clearpage
\newpage

%
%
%
%
%
%
%
%
%

\end{document}